\documentclass[twocolumn, showpacs, english, 10pt,preprintnumbers, amsmath, amssymb, superscriptaddress, aps, prl,longbibliography,nofootinbib]{revtex4-2}
\usepackage{silence}
\usepackage{xspace}

\usepackage[utf8]{inputenc}
\usepackage{graphicx}
\usepackage{grffile}
\usepackage[dvipsnames]{xcolor}
\usepackage{amsmath,bm}
\usepackage[normalem]{ulem}
\usepackage{appendix} 
\definecolor{orange}{rgb}{1,0.5,0}
\definecolor{goodgreen}{rgb}{0.1,0.5,0}
\definecolor{goodred}{rgb}{0.7,0,0}
\definecolor{goodblue}{rgb}{0,0,0.7}
\definecolor{goodyellow}{rgb}{0.7,0.65,0.1}

\renewcommand\vec{\boldsymbol}

\usepackage{tikz}
\makeatletter

\usepackage[colorlinks,urlcolor=goodgreen,citecolor=blue,linkcolor=goodred]{hyperref}

\newcommand{\orcid}[1]{\href{https://orcid.org/#1}{\includegraphics[width=8pt]{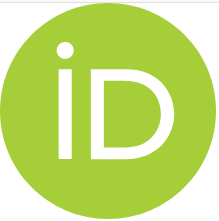}}}

\usepackage{color}

\usetikzlibrary{patterns.meta}
\usetikzlibrary{arrows.meta}

\usepackage{float}

\let\oldepsilon\epsilon \let\epsilon\varepsilon \let\varepsilon\oldepsilon

\makeatother

\usepackage{babel}

\begin{document}

\title{Shielded order}
\author{Tim Kokkeler \orcid{0000-0001-8681-3376}}
\email{tim.h.kokkeler@jyu.fi}
\affiliation{Department of Physics and Nanoscience Center, University of Jyväskylä, P.O. Box 35 (YFL), FI-40014 University of Jyväskylä, Finland}
\author{F. Sebastian Bergeret\orcid{0000-0001-6007-4878}}
\email{fs.bergeret@csic.es}
\affiliation{Centro de Física de Materiales (CFM-MPC) Centro Mixto CSIC-UPV/EHU,
E-20018 Donostia-San Sebastián, Spain}
\affiliation{Donostia International Physics Center (DIPC), 20018 Donostia--San
Sebastián, Spain}
\author{Tero T. Heikkilä \orcid{0000-0002-7732-691X}}
\email{tero.t.heikkilä@jyu.fi}
\affiliation{Department of Physics and Nanoscience Center, University of Jyväskylä, P.O. Box 35 (YFL), FI-40014 University of Jyväskylä, Finland}

\begin{abstract}
We introduce the concept of shielded order in correlated electron systems. Shielded orders appear in materials that have multiple electronic instabilities, such as spin-density waves and superconductivity. We consider instabilities that compete at the level of mean-field theory and show that their interplay changes dramatically in the presence of random inhomogeneities of the dominant order parameter phase. We develop a model to describe materials with such phase disorder and show that they are very sensitive to subdominant instabilities, but only to those that are preserved upon scattering off the phase disorder. The resulting long-range orders stemming from the subdominant interaction survive as long as the dominant order does, and thus have critical temperatures that are much larger than expected based on their own interaction strength. This is due to the excitation gap introduced by the remnant short-range correlations, which shields the subdominant order. We show that if the dominant order consists of spin-density waves, \textit{s}-wave superconductivity is obstructed by the local order, but a specific type of unconventional superconductivity, such as \textit{d}-wave or \textit{s}$_\pm$-wave superconductivity, is shielded. Shielded orders can be observed if they preserve long-range order while the dominant order does not. Candidate materials displaying shielded orders are cuprates, iron pnictides, graphene multilayers and transition metal dichalcogenides. Therefore, shielded orders provide a possible explanation for the high temperature unconventional superconductivity in these materials. Moreover, when no long-range order survives at all, our model predicts a phase in which the density of states is suppressed, reminiscent of the pseudogap phase.
\end{abstract}
\maketitle
\section{Introduction}
 
The usual paradigm in condensed matter physics is to identify which  interaction mechanisms lead to a given correlated electronic state that wins over other candidates. But can we turn this competition to a collaboration? Prominent materials with multiple electronic instabilities to different correlated states are the cuprate superconductors, where antiferromagnetism, pseudogap state, charge order and d-wave superconductivity are encountered, twisted multilayer  graphene or transition metal dichalcogenide (TMD) based systems, whose phase diagrams host correlated insulators, superconductivity and magnetism.
Superconductivity frequently is found  to compete with other correlations, such as charge \cite{peierls1955quantum,gruner1988dynamics} and spin density waves \cite{overhauser1960giant,overhauser1962spin,gruner1994dynamics}, which typically appear at considerably higher temperatures, and within the conventional picture, superconductivity emerges as these orders are suppressed by doping or gating.

This raises the question of whether, and how, correlated insulators can instead coexist with superconductivity, or even enhance its critical temperature.
When treated on the microscopic level, these orders tend to compete: the dominant instability gaps out the Fermi surface and suppresses the others, so coexistence is restricted to narrow regions of the phase diagram. This picture, however, becomes questionable in the materials introduced above. A common feature for all of them is that they are  (quasi-)two-dimensional. In two dimensions, quantum fluctuations are strongly enhanced, and the Mermin–Wagner theorem \cite{mermin1966absence,hohenberg1967existence,coleman1973there} forbids orders to extend over arbitrary range at any finite temperature.  At finite temperatures, two-dimensional orders can therefore be divided into  (quasi-) long-range orders containing correlations with algebraic (power-law) decay, and short-range orders with exponentially decaying correlations. The distinction between the two has been studied in detail for superconductors, in which they are separated by the Berezinskii-Kosterlitz-Thouless transition \cite{berezinskii1971destruction,berezinskii1972destruction,kosterlitz1973ordering}. A question that opens up is what happens to subdominant interactions once the dominant order loses its spatial phase coherence. The interplay between multiple orders and fluctuations in both space and time near quantum criticality has been discussed in Ref.~\cite{efetov2013pseudogap} and the competition between charge-density waves and superconductivity  in Refs. \cite{nie2015fluctuating,caplan2015long}, while the presence of different orders with comparable strength may result in intertwined orders \cite{fradkin2015colloquium}. Moreover, superconductivity may appear locally if a dominant order hosts defects at which it is suppressed \cite{yu2019fragile}. However, a generic framework for the interplay between a short-range dominant order  and a long-range subdominant order is lacking and their coexistence remains an open question which we address in this work.

\begin{figure}
\resizebox{0.5\textwidth}{!}{
    \begin{tikzpicture}
        \draw[-{Circle[sep=-.45pt,width=2.04pt,length=2pt,,black]Triangle[width=10pt,length=7pt,black]},ultra thick,black] (11,5.4)--(11,-1.4);
        \fill[] (11,0) circle (0.2cm);
        \fill[] (11,1) circle (0.2cm);
        \fill[] (11,4) circle (0.2cm);
        \fill[] (11,5) circle (0.2cm);
        \node[text width = 4.5cm] at (13.75,-0.25){\Large Short-range order with shielded order};
        \node[text width = 5cm] at (14.0,1.25){\Large Short-range order without shielded order};
        \node at (13.2,4){\Large Vulnerable order};
        \node at (12.35,5){\Large No order};
        \draw[] (2,3.5)--(2,6.5)--(7,6.5)--(7,3.5)--(2,3.5);
        \draw[] (2,1.5)--(2,-1.5)--(7,-1.5)--(7,1.5)--(2,1.5);
        \node at (8.3,5){\Large$\textcolor{goodred}{\langle c^{\dagger}Bc\rangle\neq 0}$};
        \node at (8.3,0){\Large$\textcolor{goodblue}{\langle c^{\dagger}Cc\rangle\neq 0}$};
        \node at (5.7,2.5){\Large$[\textcolor{goodred}{B},\textcolor{goodblue}{C}] = 0$};
        \draw[-{Circle[sep=-.45pt,width=1.04pt,length=1pt,black]Triangle[width=5pt,length=5pt,black]},thick,black](4.5,3.5)--(4.5,1.5);
        \draw[-{Circle[sep=-.9pt,width=5.04pt,length=2pt,goodblue!80]Triangle[width=5pt,length=5pt,goodblue]},ultra thick,goodblue](4,0)--(5, 0);
        \draw[](6.5,1)--(6.75,1)--(6.75,1.25)--(6.5,1.25)--(6.5,1);
        \draw[dashed](6.75,1.25)--(7,3.5);
        \draw[dashed](6.5,1.25)--(2,3.5);
        \draw[-{Circle[sep=-.9pt,width=5.03pt,length=2pt,goodred!80]Triangle[width=5pt,length=5pt,goodred]},ultra thick,goodred](2.500, 4.000)--(3.300, 4.600);
        \draw[-{Circle[sep=-.9pt,width=5.03pt,length=2pt,goodred!80]Triangle[width=5pt,length=5pt,goodred]},ultra thick,goodred]	(4.2, 5.0)--(4.8,5.8);
        \draw[-{Circle[sep=-.9pt,width=5.03pt,length=2pt,goodred!80]Triangle[width=5pt,length=5pt,goodred]},ultra thick,goodred]	(5.5,4.2)--(5.9,5.117);
        \draw[-{Circle[sep=-.9pt,width=5.03pt,length=2pt,goodred!80]Triangle[width=5pt,length=5pt,goodred]},ultra thick,goodred]	(3.6,5.2)--(3, 6.0);
        \draw[-{Circle[sep=-.9pt,width=5.03pt,length=2pt,goodred!80]Triangle[width=5pt,length=5pt,goodred]},ultra thick,goodred]	(6,5)--(6.8, 5.6);
        \draw[-{Circle[sep=-.9pt,width=5.03pt,length=2pt,goodred!80]Triangle[width=5pt,length=5pt,goodred]},ultra thick,goodred]	(2.8,5.2)--(3.6,4.6);
        \draw[-{Circle[sep=-.9pt,width=5.03pt,length=2pt,goodred!80]Triangle[width=5pt,length=5pt,goodred]},ultra thick,goodred]	(4.5, 3.8)--(5.1,4.6);
        \draw[-{Circle[sep=-.9pt,width=5.03pt,length=2pt,goodred!80]Triangle[width=5pt,length=5pt,goodred]},ultra thick,goodred]	(6.2, 4.5)--(6.6,3.583);
        \draw[-{Circle[sep=-.9pt,width=5.03pt,length=2pt,goodred!80]Triangle[width=5pt,length=5pt,goodred]},ultra thick,goodred]	(5.5, 6)--(6.4,5.436);
        \draw[-{Circle[sep=-.9pt,width=5.03pt,length=2pt,goodred!80]Triangle[width=5pt,length=5pt,goodred]},ultra thick,goodred]	(3.9, 3.8)--(2.9,3.8);
        \draw[-{Circle[sep=-.9pt,width=5.03pt,length=2pt,goodred!80]Triangle[width=5pt,length=5pt,goodred]},ultra thick,goodred]	(3.9, 6.2)--(4.9,6.2);
        \draw[-{Circle[sep=-.9pt,width=5.03pt,length=2pt,goodred!80]Triangle[width=5pt,length=5pt,goodred]},ultra thick,goodred]	(2.2, 5.5)--(2.7,6.2);
        \node at (11,5.7){\LARGE Free energy};
    \end{tikzpicture}}
\caption{Illustration of shielded order. \textbf{Left)} The system has a dominant order ($B$), with constant amplitude. The phase, illustrated using arrows, varies rapidly due to the phase disorder. On larger scales the system develops a long-range subdominant order ($C$). \textbf{Right)} The free energy of the shielded order is much lower than that of the normal state due to the short-range order. Without it, it would be close in energy to the normal state and thus be very vulnerable.}
    \label{fig:IllustrationShieldedOrder}
\end{figure}
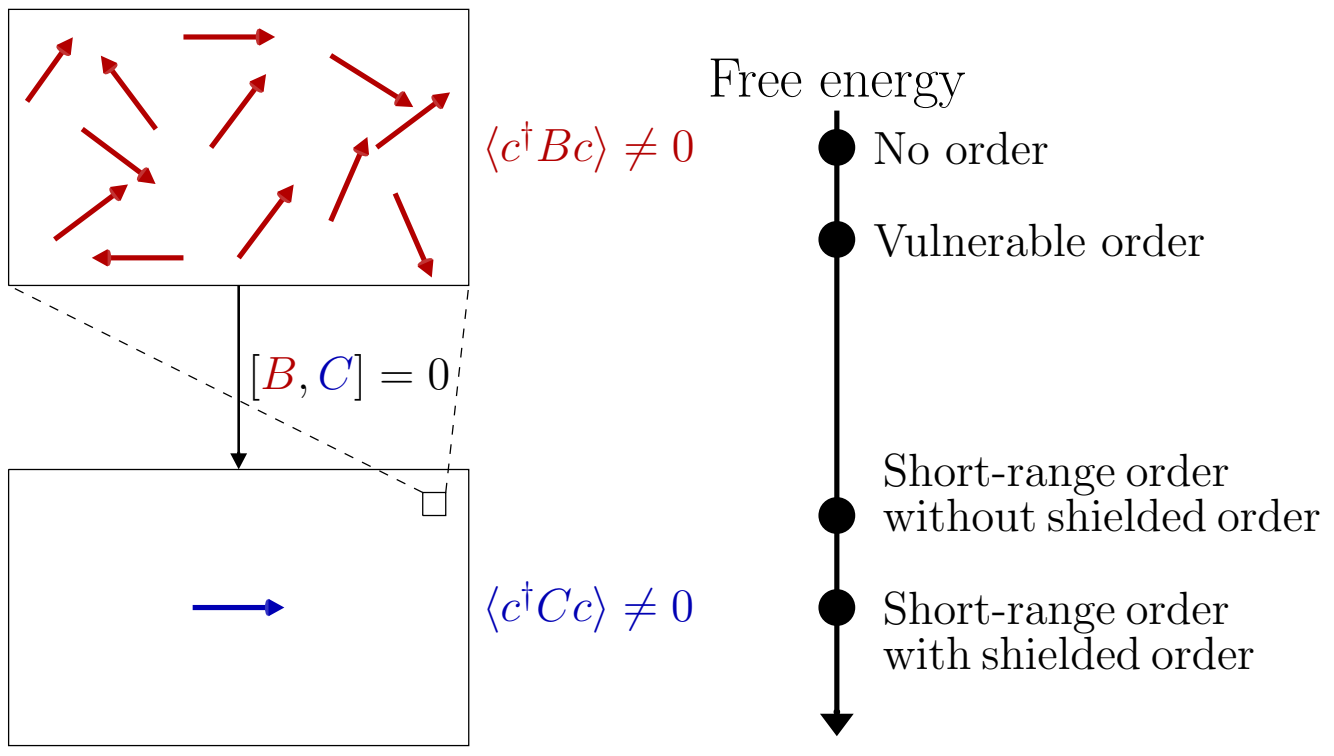

We show that if the dominant order loses its spatial coherence, its interplay with subdominant orders changes significantly, and competition turns into collaboration. Specifically, we show that in this case sub-dominant correlations that are preserved upon scattering off the phase disorder, \textit{i.e.}, that commute with the dominant order, can exhibit long-range order up to temperatures much larger than expected from their interaction strength, because the remnant short-range dominant order pushes quasiparticle excitations to high energies, as illustrated in Fig.~\ref{fig:IllustrationShieldedOrder}. The short-range dominant order thus shields the subdominant long-range order.

Shielded orders appear in materials that exhibit multiple types of correlation channels, of which the dominant type loses its spatial phase coherence more easily. We identify several candidate materials for shielded order. A particularly interesting candidate are the cuprate superconductors. Their phase diagram  shows that at zero doping, they exhibit an insulating phase, which can be described as a spin-density wave / antiferromagnetic  order or an antiferromagnetic Mott insulator \cite{lee2006doping,keimer2015quantum}. Because both orders can be described using an order parameter with spatially varying spin correlations, for the purpose of this manuscript we do not distinguish them and refer to them as SDW. The SDW state shows long-range order at low dopant concentrations. For large enough dopant concentrations, the SDW order is lost, and high-temperature \textit{d}-wave superconductivity \cite{tsuei2000pairing} prevails. We show that an SDW order without phase coherence can sustain high-temperature \textit{d}-wave superconductivity as a shielded order.

\section{Shielded orders}
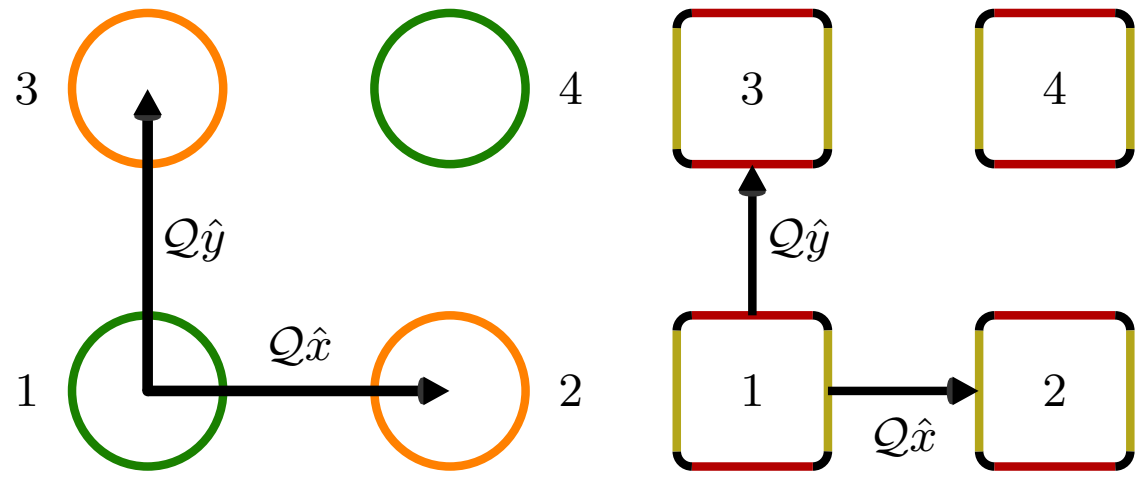
\begin{figure}
    \begin{tikzpicture}
    
        \draw[ultra thick,goodgreen] (-4,0.5) circle (0.5cm);
        \draw[ultra thick,orange] (-4,2.5) circle (0.5cm);
        \draw[ultra thick,orange] (-2,0.5) circle (0.5cm);
        \draw[ultra thick,goodgreen] (-2,2.5) circle (0.5cm);
        \node at (-4.8,0.5){$1$};
        \node at (-4.8,2.5){$3$};
        \node at (-1.2,0.5){$2$};
        \node at (-1.2,2.5){$4$};
        \draw[-{Circle[sep=-.9pt,width=5.04pt,length=2pt,black!80]Triangle[width=5pt,length=5pt,black]},line width = 2pt,black] (-4,0.5)--(-2,0.5);
    \draw[-{Circle[sep=-.9pt,width=5.04pt,length=2pt,black!80]Triangle[width=5pt,length=5pt,black]},line width = 2pt,black] (-4,0.5)--(-4,2.5);
    \node at (-3,0.8){$\mathcal{Q}\hat{x}$};
    \node at (-3.7,1.5){$\mathcal{Q}\hat{y}$};

  \draw[ultra thick, goodred] (-0.4,1) -- (0.4,1);
  \draw[ultra thick, black]
    (0.4,1) arc[start angle=90, end angle=0, radius=0.1];
  \draw[ultra thick, goodyellow] (0.5,0.9) -- (0.5,0.1);
  \draw[ultra thick, black]
    (0.5,0.1) arc[start angle=0, end angle=-90, radius=0.1];
  \draw[ultra thick, goodred] (0.4,0) -- (-0.4,0);
  \draw[ultra thick, black]
    (-0.4,0) arc[start angle=-90, end angle=-180, radius=0.1];
  \draw[ultra thick, goodyellow] (-0.5,0.1) -- (-0.5,0.9);
  \draw[ultra thick, black]
    (-0.5,0.9) arc[start angle=180, end angle=90, radius=0.1];
  \draw[ultra thick, goodred] (1.6,1) -- (2.4,1);
  \draw[ultra thick, black]
    (2.4,1) arc[start angle=90, end angle=0, radius=0.1];
  \draw[ultra thick, goodyellow] (2.5,0.9) -- (2.5,0.1);
  \draw[ultra thick, black]
    (2.5,0.1) arc[start angle=0, end angle=-90, radius=0.1];
  \draw[ultra thick, goodred] (2.4,0) -- (1.6,0);
  \draw[ultra thick, black]
    (1.6,0) arc[start angle=-90, end angle=-180, radius=0.1];
  \draw[ultra thick, goodyellow] (1.5,0.1) -- (1.5,0.9);
  \draw[ultra thick, black]
    (1.5,0.9) arc[start angle=180, end angle=90, radius=0.1];
  \draw[ultra thick, goodred] (-0.4,3) -- (0.4,3);
  \draw[ultra thick, black]
    (0.4,3) arc[start angle=90, end angle=0, radius=0.1];
  \draw[ultra thick, goodyellow] (0.5,2.9) -- (0.5,2.1);
  \draw[ultra thick, black]
    (0.5,2.1) arc[start angle=0, end angle=-90, radius=0.1];
  \draw[ultra thick, goodred] (0.4,2) -- (-0.4,2);
  \draw[ultra thick, black]
    (-0.4,2) arc[start angle=-90, end angle=-180, radius=0.1];
  \draw[ultra thick, goodyellow] (-0.5,2.1) -- (-0.5,2.9);
  \draw[ultra thick, black]
    (-0.5,2.9) arc[start angle=180, end angle=90, radius=0.1];
  \draw[ultra thick, goodred] (1.6,3) -- (2.4,3);
  \draw[ultra thick, black]
    (2.4,3) arc[start angle=90, end angle=0, radius=0.1];
  \draw[ultra thick, goodyellow] (2.5,2.9) -- (2.5,2.1);
  \draw[ultra thick, black]
    (2.5,2.1) arc[start angle=0, end angle=-90, radius=0.1];
  \draw[ultra thick, goodred] (2.4,2) -- (1.6,2);
  \draw[ultra thick, black]
    (1.6,2) arc[start angle=-90, end angle=-180, radius=0.1];
  \draw[ultra thick, goodyellow] (1.5,2.1) -- (1.5,2.9);
  \draw[ultra thick, black]
    (1.5,2.9) arc[start angle=180, end angle=90, radius=0.1];
 \node at (0,0.5){$1$};
        \node at (0,2.5){$3$};
        \node at (2,0.5){$2$};
        \node at (2,2.5){$4$};
    \draw[-{Circle[sep=-.9pt,width=5.04pt,length=2pt,black!80]Triangle[width=5pt,length=5pt,black]},ultra thick,black] (0.5,0.5)--(1.5,0.5);
    \draw[-{Circle[sep=-.9pt,width=5.04pt,length=2pt,black!80]Triangle[width=5pt,length=5pt,black]},ultra thick,black] (0,1)--(0,2);
    \node at (1,0.2){$\mathcal{Q}\hat{x}$};
    \node at (0.3,1.5){$\mathcal{Q}\hat{y}$};
                \fill (-4,0.5) circle (1pt);
\end{tikzpicture}
\caption{\textbf{Left)} Sketch of the Fermi surface of one of the example systems that host shielded order. The Fermi surface consists of four connected components, 2 electron pockets (green) with dispersion $\xi + \eta\xi$, and 2 hole pockets (orange) with dispersion $-\xi + \eta \xi$. Correlated insulator phases are possible both through correlations with momentum along the horizontal direction, connecting pocket 1 with 2 and 3 with 4, and along the vertical, connecting pocket 1 with 3 and 2 with 4. \textbf{Right)} Fermi surface with four electron pockets, or four hole pockets, that exhibit spin-density wave correlations whose strength depends strongly on the direction along the Fermi surface pocket. The spin-density wave is formed using those parts of the Fermi surface that satisfy the nesting condition, indicated using yellow and red for spin-density waves along the $x$ and $y$ direction respectively.}
    \label{fig:Fermisurfaces}
\end{figure}

To introduce the concept of shielded orders, we study a toy model in which the noninteracting Fermi surface has four distinct components, 2 electron-like (labeled 1 and 4) and 2 hole-like (2 and 3), as illustrated in Fig.~\ref{fig:Fermisurfaces}(left).

In such a system, close to the Fermi surface, the Hamiltonian has band degrees of freedom, denoted using Pauli matrices $\mu_i,\nu_i$, $i = 0,1,2,3$, in addition to the spin ($\sigma_i$) degree of freedom. Moreover, to be able to describe superconductivity, we use Nambu space, with Pauli matrices denoted as $\tau_i$. As discussed in Supplementary Material (SM) Sec.~\ref{sec:Construction}, in this space the Hamiltonian can be approximated as
\begin{align}
    H = \Psi^{\dagger}\Big(\xi(k) \tau_3 \mu_3\nu_3 + \eta\xi(k) \tau_3\Big)\Psi  + H_{\text{int}}\;,\label{eq:Hamiltonian}
\end{align}
where $\xi(k)$ contains the band dispersion,  $H_{\text{int}}$ all interactions, and $\Psi$ is a 16 component spinor containing the electron operators with band, Nambu and spin degrees of freedom. For illustrative purposes we assume that the electron and hole dispersions are almost similar (up to a sign), and include the parameter $\eta\ll 1$ to describe their asymmetry.

The interaction part of the Hamiltonian prevents exact diagonalization. A robust approach for describing correlated interacting systems is via path integrals and effective actions. There, the interaction is decoupled using a Hubbard-Stratonovich transformation that introduces auxiliary fields, which are identified with possible  order parameters for different phases. For example, denoting expectation values by $\langle\cdot \rangle$,  for ferromagnetism the order parameter is $\text{tr}\langle\Psi^{\dagger}\vec{\sigma}\Psi\rangle$, and for superconductivity $\text{tr}\langle \Psi^{\dagger}e^{i\phi\tau_3}\tau_2\Psi \rangle$. Moreover, there are order parameters for charge density waves (CDWs) along $x$ ($\text{tr}\langle\Psi^{\dagger} e^{i\phi_{\text{cx}}\mu_3}\mu_2 \Psi\rangle$) and $y$ ($\text{tr}\langle \Psi^{\dagger}e^{i\phi_{\text{cy}}\nu_3}\nu_2\Psi\rangle$) directions, as well as spin-density waves (SDWs) ($\text{tr}\langle\Psi^{\dagger}e^{i\phi_{\text{sx}}\mu_3}\mu_2 \vec{\sigma}\Psi\rangle$ and $\text{tr}\langle\Psi^{\dagger}e^{i\phi_{\text{sy}}\nu_3}\nu_2 \vec{\sigma}\Psi\rangle$). For superconductivity, the phase $\phi$ couples to the EM field and describes supercurrents, while the phases $\phi_{c(s)x,y}$ are connected to the position of the peaks of the charge (spin) density waves.

At the saddle-point level, typically one of these phases dominates for a given set of parameters, while the others are suppressed.

This mean-field theory generally works well in three dimensions. However, for two-dimensional materials the range of coherence is important, and there is a regime in which the phase coherence of the dominant order is lost completely, and only local order remains, but with a rapidly varying phase. As an example, let us assume this dominant order to be a spin-density wave (SDW) order with an order parameter $\Delta_{\rm SDW}(r)$. Its amplitude is coherent, but it is phase disordered, which can be described via correlators 
\begin{subequations}
\begin{align}
    \langle\Delta_{\text{SDW,x}}(r)\Delta_{\text{SDW,x}}(r')\rangle \sim |\Delta^{0}_{\text{SDW}}|^2\xi_{\text{V}}^2\delta(r-r')\label{eq:nostiffnessmain1}\;,\\
    \langle\Delta_{\text{SDW,y}}(r)\Delta_{\text{SDW,y}}(r')\rangle \sim |\Delta^{0}_{\text{SDW}}|^2\xi_{\text{V}}^2\delta(r-r')\label{eq:nostiffnessmain2}\;.
\end{align}
\label{eq:nostiffnessmain}
\end{subequations}

Here $\xi_{\text{V}}$ is the short-range scale over which coherence is lost, $\Delta_{\text{SDW}}^0$ is the typical amplitude of the remaining short-range correlations and $\delta(r)$ is a representation of the Dirac delta functional. We assume $\xi_{\text{V}}$  to be large compared to the coherence length of the spin-density wave $\xi_{\text{SDW}}$, so that its amplitude and temporal coherence remain well defined, but very small compared to the minimum relevant length scales $\xi_{\text{sub}}$ set by the geometry and the subdominant interactions, i.e. $\xi_{\text{SDW}}\ll \xi_{\text{V}}\ll \xi_{\text{sub}}$. The density wave thus forms a glass \cite{edwards1975theory}. 

The rapid and random spatial dependence of phase of the order parameter makes the description of the system for a specific realization computationally very expensive. However, when the system is self-averaging, we can describe the properties of a correlated sample much larger than $\xi_{\text{V}}$ via the ensemble average over those phase configurations. In path integrals, disorder averaging introduces a new interaction channel, which again needs to be decoupled via Hubbard-Stratonovich transformation, as detailed in SM Sections~\ref{sec:Construction} and \ref{sec:Actiondetails}. We denote the ensuing self-consistent field $Q$ \cite{wegner1979mobility,efetov1980interaction,finkelshtein1983effect}. The $Q$-field describes the disorder-averaged, universal, properties of the system. This comes with a cost, as the disorder averaging cannot be done in the band-Nambu-spin space, but an additional matrix degree of freedom is needed. We use the Keldysh formulation \cite{feigelman2000keldysh,levchenko2009transport,kamenev2023field}, in which case this additional degree of freedom contains the retarded, advanced and Keldysh components.  Therefore, unlike the order parameters for ferromagnetism, superconductivity or density waves, the $Q$-field is a 32 by 32 matrix field in the resulting Keldysh-band-Nambu-spin space. 

Because of the construction of the Nambu space, which artificially doubles the space, $Q$ must obey a charge conjugation symmetry \cite{kokkeler2025universal}. Moreover, when the normal-state bandwidth of the electron system is much larger than the energy scales relevant for the correlations, it is possible to restrict the $Q$-field. In this case low-energy degrees of freedom are described by those $Q$-matrices that satisfy $Q^2 = \mathbf{1}$ \cite{wegner1979mobility,efetov1980interaction,finkelshtein1983effect}. The $Q$-fields that obey these two constraints are called soft modes. On this soft-mode manifold, the effective action takes a simple form, namely that of a diffusion theory \cite{wegner1979mobility,efetov1980interaction,finkelshtein1983effect}. Because of the non-linear normalization constraint, this action is called the non-linear sigma model (NLSM) action.

Similar to conventional scalar disorder, also the phase disorder described by Eq.~\eqref{eq:nostiffnessmain} allows for the construction of a $Q$-field. There is however a key distinction between the soft-mode manifold created by scalar disorder and the soft-mode manifold for systems with short-range order. Scalar disorder affects all states equally. On the other hand, short-range spin-density waves have a non-trivial structure in band-Nambu-spin space. Indeed, the remaining short-scale dominant order opens up a gap between different bands, and only those components of the $Q$-field which do not mix states on opposite sides of the gap are unaffected. They correspond to components that commute with the short-range order parameters; they are preserved upon scattering by the short-range phase disorder. Similar situation was found in the case of magnetic impurities \cite{marchetti2002tail}.

Due to this, as elaborated in SM Sec.~\ref{sec:Actiondetails}, next to the normalization and charge conjugation constraints, the soft-mode manifold is characterized by additional constraints that can be specified using matrices $B$:
\begin{align}
    Q = BQB\;.
\end{align}
Here $B$ depends on the band-Nambu-spin structure of the dominant short-range order. For  spin-density waves $B \in \{\mu_{1,2}\tau_3\sigma_z, \nu_{1,2}\tau_3\sigma_z\}$. These constraints significantly reduce the dimensionality of the soft-mode manifold and the terms that are allowed in the NLSM action. In fact, the soft-mode manifold has the same structure as the standard single-band one, but with modified Pauli matrices $\tilde{\tau},\tilde{\sigma}$ defined as:
\begin{subequations}
\begin{align}
    (\tilde{\tau}_{0,3},\tilde{\sigma}_{0,3}) &= (\tau_{0,3},\sigma_{0,3})\;,\\
    (\tilde{\tau}_{1,2},\tilde{\sigma}_{1,2}) &= (\tau_{1,2}\mu_3\nu_3,\sigma_{1,2}\mu_3\nu_3)\;.\label{eq:tausigmaoffdiag}
\end{align}
\end{subequations}

This connection allows us to identify and describe the allowed orders in complete analogy with the standard procedure \cite{kamenev2023field}.
Including only terms relevant at the Fermi surface level, the resulting action that includes all possible shielded orders reads 
\begin{align}
    iS&[Q,\Delta_{\text{d}},\vec{h}] = \frac{\pi\nu}{2}\text{Tr}\Bigg[\Big(\eta\hat{\omega}_{t,t'}\tilde{\tau}_3 + i\vec{h}\cdot\vec{\tilde{\sigma}}\tilde{\tau}_3 + \Delta_{\text{d}}e^{i\phi_{\text{d}} \tilde{\tau}_3}\tilde{\tau_2}\Big)Q \nonumber\\&-D_{\Delta}(\nabla Q)^2+ \frac{1}{\lambda_{\text{d}}}\hat{\Delta}_{\text{d}}\rho_1 \hat{\Delta}_{\text{d}}^{\dagger}-\sum_{i = x,y,z}\frac{1}{\lambda_i}h_i\rho_1h_i \Bigg]+ iS_{\text{r}}\;.\label{eq:ActionShielded}
\end{align}
Here $\nu$ is the density of states per band per spin, $\hat{\omega}_{t,t'} = \delta(t-t')\partial_t$, $D_{\Delta} =\frac{1}{2} v_F^2\tau_{\Delta}$ is the effective diffusion constant of the system, $v_F$ is the average magnitude of Fermi velocity and $\tau_{\Delta}^{-1} =2\pi\nu |\Delta^{0}_{\text{SDW}}|^2\frac{\text{ln} [(8/e)k_F\xi_V]}{k_F^3\xi_V}$, while we introduced the shorthand notation $\hat{\Delta}_{\text{d}} = \Delta_{\text{d}}e^{i\phi_{\text{d}}\tilde{\tau}_3}\tilde{\tau}_2$ and used $\rho_1$ for the first Pauli matrix in Keldysh space. The notation Tr denotes a trace over Keldysh-Nambu-band-spin space as well as spatiotemporal integration.

The two possible sub-dominant order parameters consistent with the soft-mode manifold are $\vec{h}$ and $\Delta_{\text{d}}$. They describe magnetic and superconducting phases. Their appearance requires non-zero values of the corresponding interaction strengths that we denote by $\lambda_{\text{d}}$ and $\lambda_{x,y,z}$. $iS_{\text{r}}$ describes relaxation mechanisms, which contains interband scattering as well as spin-relaxation through spin-orbit coupling and magnetic impurities.

Variation of Eq.~(\ref{eq:ActionShielded}) with respect to $Q$ gives the effective diffusion equation, called the Usadel equation \cite{usadel1970generalized}. The $Q$-field satisfying this equation corresponds to the quasiclassical Green's function $g\sim \langle \Psi\Psi^{\dagger}\tau_3\mu_3\nu_3\rangle$, where the $\tau_3\mu_3\nu_3$ stems from the $\xi$-term in the Hamiltonian, Eq.~(\ref{eq:Hamiltonian}), see SM Sec.~\ref{sec:Actiondetails} for a discussion. Variation with respect to the order parameters gives their self-consistency equations. 

There are several conclusions that we can draw from the action in Eq.~(\ref{eq:ActionShielded}). First of all, there is only one type of superconductivity that survives the short-range dominant mode, $\Delta_{\text{d}} e^{i\phi_{\text{d}} \tilde{\tau}_3}\tilde{\tau}_2$. This however, is not conventional \textit{s}-wave superconductivity, as the sign of the order parameter depends on the band index. Specifically, it has one sign on Fermi pockets 1 and 4, and the opposite sign on Fermi pockets 2 and 3. It therefore has \textit{d}-wave symmetry. 

The \textit{d}-wave superconductivity competes with magnetic interactions, which contribute the $\vec{h}\cdot\vec{\tilde{\sigma}}\tilde{\tau}_3$ term in the action. One of them has spins collinear with the dominant SDW order, has opposite signs on pockets 1 and 4 compared to pockets 2 and 3, and  corresponds to itinerant \textit{d}-wave altermagnetism \cite{smejkal2022emerging}, while the other two are ferromagnetic orders that have spins perpendicular to the dominant order.

In the following, we assume that superconductivity dominates over the remaining magnetic orders.  A crucial aspect of Eq.~(\ref{eq:ActionShielded}) is that the time-derivative term contains a suppression factor $\eta$, which allows pair amplitudes to extend to much larger energies than in conventional superconductors. For small pair relaxation through interband scattering or magnetic impurity scattering, the critical temperature can be calculated by exploiting  a natural cut-off $\tau_{\Delta}^{-1}$ for the attractive interaction (see SM Sec.~\ref{sec:Tc} for details)
\begin{align}
    k_{\text{B}}T_{\text{c}} = \hbar\tau_{\Delta}^{-1}e^{-\frac{\eta}{\lambda_{\text{d}}}}\;.
\end{align}
We conclude that the critical temperature for the \textit{d}-wave superconductivity is much higher than the BCS prediction due to the renormalization of the exponential suppression factor via $\eta \ll 1$. Thus, the gap induced by the dominant order enhances the critical temperature of the subdominant superconducting order well beyond what is expected from its bare interaction strength. We therefore call the subdominant order a shielded order. 

Another interesting feature of Eq.~(\ref{eq:ActionShielded}) is that without mismatch ($\eta = 0$), and without subdominant correlations, the action is constant over the entire soft-mode manifold. The existence of a continuum of saddle points indicates that in this regime the dynamics is dominated by phase disorder. We associated this to a pseudogap state. In summary, the NLSM action derived from the short-ranged dominant order naturally gives way to both a pseudogap state and shielded orders that have a much larger critical temperature and excitation gap than expected based on their own interaction strength.

An important requirement for shielded long-range order to be visible in experiment is that it is not reduced to short-range correlations itself. This puts additional restrictions on which orders can appear as a shielded order. For example, if the dominant order is a spin-density wave, a subdominant different spin-density wave typically also loses spatial coherence and is therefore hard to measure as a shielded order. On the other hand, the combination of density waves and superconductivity is a good candidate for shielded order. The phase of density waves is tied to the position of the peak of the density wave and thus affected by dopants, while the phase of the superconductor governs currents and is thus instead affected by electromagnetic fields. The phase coherence of these orders is therefore governed by entirely different processes and therefore superconductivity can retain long-range order when spin-density waves are phase disordered.

\section{Specific examples}
Having introduced the concept of shielded order, we turn our attention to a class of materials that potentially realizes them, the cuprates. In cuprates, depending on the filling, the material realizes a state with magnetic order, that is usually referred to as a strongly correlated antiferromagnetic Mott insulator.

Here, we treat it as a spin-density wave order. While the theory is inherently quasiclassical and thus assumes interaction-induced perturbations to be smaller than the single-electron bandwidth, we expect non-quasiclassical corrections to change the behavior only quantitatively.  
At large enough dopant concentration, cuprates display \textit{d}-wave superconductivity, whose origin and pairing symmetry have been debated \cite{anderson1987resonating,vanharlingen1995phase,scalapino1999cuprate,tsuei2000pairing,keimer2015quantum} since their discovery \cite{bednorz1986possible}. In previous literature, the superconductivity has been linked to the magnetic order via the proposition that spin fluctuations might serve as an unconventional pairing mechanism \cite{schmalian1998pairing,chubukov2003spin}. Our model provides an alternative point of view that does not require an exotic pairing mechanism to explain the high critical temperature. There, cuprate superconductivity can be considered as a shielded order, in which the remnant short-range spin-density wave does not act as a pairing mechanism, but shields it from quasiparticle excitations.

There are a few modifications that need to be made to the model introduced above for the application to cuprates. In cuprates, there is only one type of pocket, no electron and hole pockets. Only those parts of the Fermi surface  that satisfy the nesting condition contribute to the spin-density waves, see Fig.~\ref{fig:Fermisurfaces} (right). As detailed in SM  Sec~\ref{sec:NLSMwithangledependence},  two distinct $Q$-fields can be derived for electrons participating in the SDW order along the $x$ and $y$ directions. The actions for these $Q$-fields are similar to the NLSM discussed above, but all coefficients are calculated using weighted integrals over the Fermi surface. The weights $f_{1,2}(\theta)$ peak at the parts of the Fermi surface that satisfy the nesting condition and  are normalized such that $\int_{0}^{2\pi}\frac{d\theta}{2\pi}\; f_{1,2}^2(\theta) = 1$. Because of this weighting, the two $Q$-matrices $Q_{1,2}$ do not encode the density of states on their own anymore. Instead, as elaborated in SM Sec.~\ref{sec:NLSMwithangledependence}, the excitation gap depends on the direction of the momentum relative to the center of the pocket. Parameterizing its angle as $\theta$, we have
\begin{align}
    E_{\text{g}}(\theta) = \tau_{\Delta}^{-1}\big(f^2_1(\theta)+f^2_{2}(\theta)\big)\;.
\end{align}
Thus, the variation of the SDW strength on the Fermi surface directly translates to a variation of the excitation gap, resulting in a density of states that does not have a single hard gap, but rather a gradually increasing density of states, which in experiment is observed as a V-shape. The exact shape depends on the form of $f_{1,2}$, which are material dependent. The smallest gap appears near the corners of the Fermi surfaces, for which the nesting condition is not satisfied, in agreement with experiment \cite{ding1996angle}. However, the \textit{d}-wave superconductivity that is realized has the same sign within a connected component and the sign changes happen only between them. Therefore, the quasiparticle excitation gap is governed by the SDW-strength rather than nodes of the pair potential.

Let us consider the scenario in which \textit{d}-wave order in cuprates originates as a shielded order protected by the AFM/SDW order that appears at zero doping. A remaining question is when it does so. As discussed in Ref.~\cite{lee2003two}, dopants create spatial inhomogeneity in such systems. These inhomogeneities lead to scattering of short-range fluctuations \cite{sadovskii2001pseudogap}, exactly what is needed for shielded orders. Here we provide a simple model for the transition between a phase coherent and a phase disordered state based on a U(1) phase theory. 

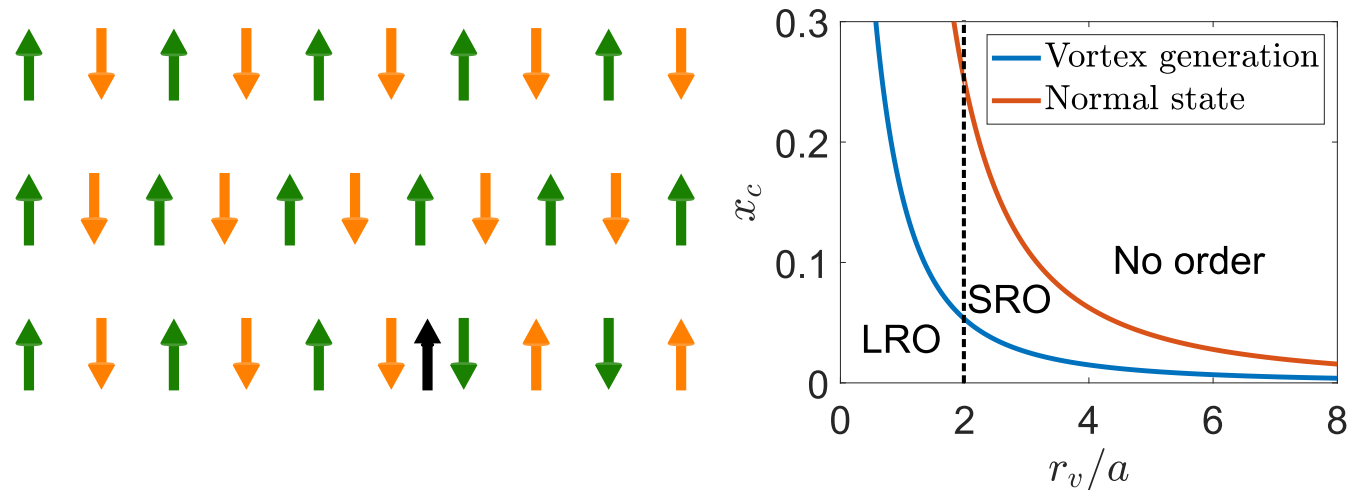
\begin{figure}
\resizebox{0.5\textwidth}{!}{
    \begin{tikzpicture}
        \draw[-{Circle[sep=-0.9pt,width=10.3pt,length=2pt,goodgreen!80]Triangle[width=10pt,length=10pt,goodgreen]} ,line width = 4pt,goodgreen](0,-0.5,0)--(0,0.5,0);
        \draw[-{Circle[sep=-0.9pt,width=10.3pt,length=2pt,orange!80]Triangle[width=10pt,length=10pt,orange]} ,line width = 4pt,orange](1,0.5,0)--(1,-0.5,0);
        \draw[-{Circle[sep=-0.9pt,width=10.3pt,length=2pt,goodgreen!80]Triangle[width=10pt,length=10pt,goodgreen]} ,line width = 4pt,goodgreen](2,-0.5,0)--(2,0.5,0);
        \draw[-{Circle[sep=-0.9pt,width=10.3pt,length=2pt,orange!80]Triangle[width=10pt,length=10pt,orange]} ,line width = 4pt,orange](3,0.5,0)--(3,-0.5,0);
        \draw[-{Circle[sep=-0.9pt,width=10.3pt,length=2pt,goodgreen!80]Triangle[width=10pt,length=10pt,goodgreen]} ,line width = 4pt,goodgreen](4,-0.5,0)--(4,0.5,0);
        \draw[-{Circle[sep=-0.9pt,width=10.3pt,length=2pt,orange!80]Triangle[width=10pt,length=10pt,orange]} ,line width = 4pt,orange](5,0.5,0)--(5,-0.5,0);
        \draw[-{Circle[sep=-0.9pt,width=10.3pt,length=2pt,goodgreen!80]Triangle[width=10pt,length=10pt,goodgreen]} ,line width = 4pt,goodgreen](6,-0.5,0)--(6,0.5,0);
        \draw[-{Circle[sep=-0.9pt,width=10.3pt,length=2pt,orange!80]Triangle[width=10pt,length=10pt,orange]} ,line width = 4pt,orange](7,0.5,0)--(7,-0.5,0);
        \draw[-{Circle[sep=-0.9pt,width=10.3pt,length=2pt,goodgreen!80]Triangle[width=10pt,length=10pt,goodgreen]} ,line width = 4pt,goodgreen](8,-0.5,0)--(8,0.5,0);
        \draw[-{Circle[sep=-0.9pt,width=10.3pt,length=2pt,orange!80]Triangle[width=10pt,length=10pt,orange]} ,line width = 4pt,orange](9,0.5,0)--(9,-0.5,0);
        \draw[-{Circle[sep=-0.9pt,width=10.3pt,length=2pt,goodgreen!80]Triangle[width=10pt,length=10pt,goodgreen]} ,line width = 4pt,goodgreen](0,-2.5,0)--(0,-1.5,0);
        \draw[-{Circle[sep=-0.9pt,width=10.3pt,length=2pt,orange!80]Triangle[width=10pt,length=10pt,orange]} ,line width = 4pt,orange](0.9,-1.5,0)--(0.9,-2.5,0);
        \draw[-{Circle[sep=-0.9pt,width=10.3pt,length=2pt,goodgreen!80]Triangle[width=10pt,length=10pt,goodgreen]} ,line width = 4pt,goodgreen](1.8,-2.5,0)--(1.8,-1.5,0);
        \draw[-{Circle[sep=-0.9pt,width=10.3pt,length=2pt,orange!80]Triangle[width=10pt,length=10pt,orange]} ,line width = 4pt,orange](2.7,-1.5,0)--(2.7,-2.5,0);
        \draw[-{Circle[sep=-0.9pt,width=10.3pt,length=2pt,goodgreen!80]Triangle[width=10pt,length=10pt,goodgreen]} ,line width = 4pt,goodgreen](3.6,-2.5,0)--(3.6,-1.5,0);
        \draw[-{Circle[sep=-0.9pt,width=10.3pt,length=2pt,orange!80]Triangle[width=10pt,length=10pt,orange]} ,line width = 4pt,orange](4.5,-1.5,0)--(4.5,-2.5,0);
        \draw[-{Circle[sep=-0.9pt,width=10.3pt,length=2pt,goodgreen!80]Triangle[width=10pt,length=10pt,goodgreen]} ,line width = 4pt,goodgreen](5.4,-2.5,0)--(5.4,-1.5,0);
        \draw[-{Circle[sep=-0.9pt,width=10.3pt,length=2pt,orange!80]Triangle[width=10pt,length=10pt,orange]} ,line width = 4pt,orange](6.3,-1.5,0)--(6.3,-2.5,0);
        \draw[-{Circle[sep=-0.9pt,width=10.3pt,length=2pt,goodgreen!80]Triangle[width=10pt,length=10pt,goodgreen]} ,line width = 4pt,goodgreen](7.2,-2.5,0)--(7.2,-1.5,0);
        \draw[-{Circle[sep=-0.9pt,width=10.3pt,length=2pt,orange!80]Triangle[width=10pt,length=10pt,orange]} ,line width = 4pt,orange](8.1,-1.5,0)--(8.1,-2.5,0);
        \draw[-{Circle[sep=-0.9pt,width=10.3pt,length=2pt,goodgreen!80]Triangle[width=10pt,length=10pt,goodgreen]} ,line width = 4pt,goodgreen](9,-2.5,0)--(9,-1.5,0);
        \draw[-{Circle[sep=-0.9pt,width=10.3pt,length=2pt,goodgreen!80]Triangle[width=10pt,length=10pt,goodgreen]} ,line width = 4pt,goodgreen](0,-4.5,0)--(0,-3.5,0);
        \draw[-{Circle[sep=-0.9pt,width=10.3pt,length=2pt,orange!80]Triangle[width=10pt,length=10pt,orange]} ,line width = 4pt,orange](1,-3.5,0)--(1,-4.5,0);
        \draw[-{Circle[sep=-0.9pt,width=10.3pt,length=2pt,goodgreen!80]Triangle[width=10pt,length=10pt,goodgreen]} ,line width = 4pt,goodgreen](2,-4.5,0)--(2,-3.5,0);
        \draw[-{Circle[sep=-0.9pt,width=10.3pt,length=2pt,orange!80]Triangle[width=10pt,length=10pt,orange]} ,line width = 4pt,orange](3,-3.5,0)--(3,-4.5,0);
        \draw[-{Circle[sep=-0.9pt,width=10.3pt,length=2pt,goodgreen!80]Triangle[width=10pt,length=10pt,goodgreen]} ,line width = 4pt,goodgreen](4,-4.5,0)--(4,-3.5,0);
        \draw[-{Circle[sep=-0.9pt,width=10.3pt,length=2pt,orange!80]Triangle[width=10pt,length=10pt,orange]} ,line width = 4pt,orange](5,-3.5,0)--(5,-4.5,0);
        \draw[-{Circle[sep=-.45pt,width=5.2pt,length=2pt,black]Triangle[width=10pt,length=10pt,black]} ,line width = 4pt](5.5,-4.5,0)--(5.5,-3.5,0);
        \draw[-{Circle[sep=-0.9pt,width=10.3pt,length=2pt,goodgreen!80]Triangle[width=10pt,length=10pt,goodgreen]} ,line width = 4pt,goodgreen](6,-3.5,0)--(6,-4.5,0);
        \draw[-{Circle[sep=-0.9pt,width=10.3pt,length=2pt,orange!80]Triangle[width=10pt,length=10pt,orange]} ,line width = 4pt,orange](7,-4.5,0)--(7,-3.5,0);
        \draw[-{Circle[sep=-0.9pt,width=10.3pt,length=2pt,goodgreen!80]Triangle[width=10pt,length=10pt,goodgreen]} ,line width = 4pt,goodgreen](8,-3.5,0)--(8,-4.5,0);
        \draw[-{Circle[sep=-0.9pt,width=10.3pt,length=2pt,orange!80]Triangle[width=10pt,length=10pt,orange]} ,line width = 4pt,orange](9,-4.5,0)--(9,-3.5,0);
        \node (pdf) at (14,-2.5,0) {\includegraphics[width=8.6cm]{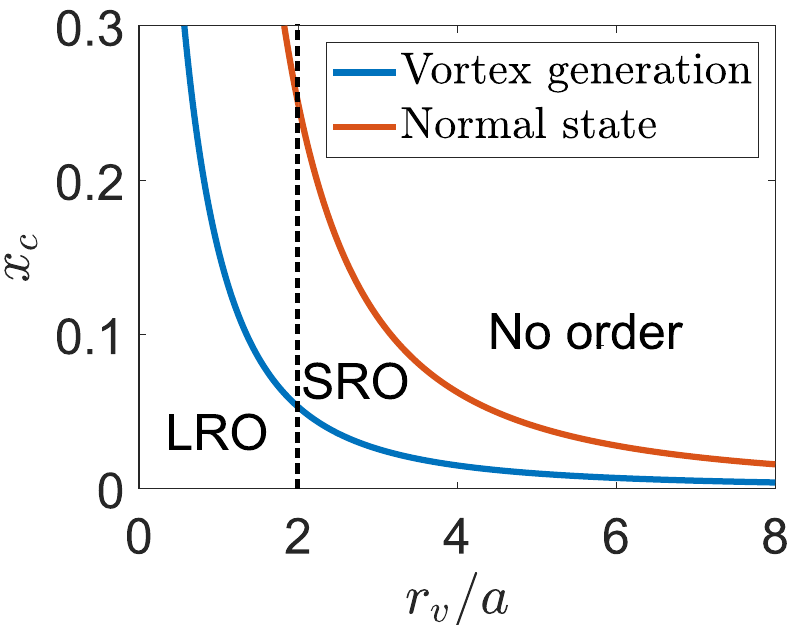}};
    \end{tikzpicture}}  
    \caption{\textbf{Left)} \textit{Upper row:} 1D illustration of SDW order without doping. \textit{Middle row:} Wigner crystal with modified distance between electrons. \textit{Bottom row:} Wigner crystal with unmodified distance. The additional impurity in 1D is a phase slip, that is, a topological excitation. \textbf{Right)} Phase diagram of cuprates including the long-range ordered (LRO) and short-range ordered (SRO) dominant spin-density wave order, and the normal state. The boundary between long-range and short-range order is determined by $F_{\text{p}} = F_{\text{v}}$, where $F_{\text{p,v}}$ are defined in Eqs. (\ref{eq:Fp},\ref{eq:Fv}), while the boundary with the normal state is determined by the overlap condition $x = \frac{a^2}{r_{\text{v}}^2}$. Dashed line indicates $r_{\text{v}}$ consistent with \cite{keimer2015quantum}.}
    \label{fig:WhyBKTzero}
\end{figure}

Without dopants, the electrons can form a perfect crystal, with lattice constant set by the nesting vector $\mathcal{Q}$. Upon doping $x$ extra electrons per unit cell, the material can accommodate them in two different ways, which are illustrated in Fig.~\ref{fig:WhyBKTzero}(left). In the first way all the additional electrons are accommodated by altering the lattice constant, which in SDW theory corresponds to a constant phase gradient $q= x\frac{\pi}{a}$. In this state, long-range coherence is preserved. Its free energy cost is
\begin{align}
    F_{\text{p}} = \frac{\pi^2 W_{\text{s}} A}{2a^2} x^2\;,\label{eq:Fp}
\end{align}
where $W_{\text{s}}$ is the stiffness of the SDW order, $a$ is the unit cell size and $A$ is the sample area.
The second option is that the additional electrons appear as topological defects. In 2D this corresponds to vortices, which destroy long-range coherence. The energy of a configuration with $x\frac{A}{a^2}$ vortices with winding $\pm 1$ and spacing of order $\frac{a}{\sqrt{x}}$ is, as elaborated in SM Sec.~\ref{sec:Dopants}
\begin{align}
    F_{\text{v}} = \frac{\pi W_{\text{s}} A}{2a^2}x\text{ln}\frac{a^2}{4r_{\text{v}}^2 x}\;,\label{eq:Fv}
\end{align}
where we introduced the vortex core radius $r_{\text{v}}$. Figure \ref{fig:WhyBKTzero} illustrates the competition between $F_{\text{p}}$ and $F_{\text{v}}$, along with the normal state, to which the system transitions if the distance between vortices equals the core radius, i.e., $x = \frac{a^2}{r_{\text{v}}^2}$. A vertical cut along a given value of $r_{\text{v}}$ thus provides a zero-temperature phase diagram vs doping $x$, with transitions between long-range dominant order below the blue line, to the subdominant order between the blue and red lines, and to the normal state without order above the red line. 

For many cuprates, the pseudogap and superconductor state persist up to $x\approx 0.25$, indicating that $r_{\text{v}} \approx 2a$. For such $r_{\text{v}}$ (dashed line in Fig.~\ref{fig:WhyBKTzero}] the model predicts a transition to the short-range ordered state $x \approx 0.05$, in good agreement with experimental observations \cite{keimer2015quantum}.  
Between these $x$-values, the short-range ordered spin-density wave order shields the  \textit{d}-wave superconductivity whose long-range phase correlations are not directly affected by the distance between electron spins.

We note that just above $x\approx 0.05$, the dominant correlations decay exponentially, but on a very large length scale. The shielded superconducting order only appears once this correlation length decays below the superconductor coherence length. Therefore, close to $x_{c} = 0.05$ there is a window in which neither SDW nor superconductivity appear. The size of this window is non-universal and strongly depends on the details of the underlying order. 
Moreover, at finite temperatures thermal effects also contribute, so this window widens with increasing temperature.

\begin{table*}[ht!]
    \centering
    \begin{tabular}{|c|c|c|}
    \hline
      Dominant order  & Shielded SC & Shielded Magnetism\\
         \hline
        CDW ($\mu_{1,2}$ or $\nu_{1,2}$) & \textit{s}-wave ($\tau_{1,2}$) & \textit{d}-wave AM \\
        SDW ($\mu_{1,2}\tau_3\sigma_z$ or $\nu_{1,2}\tau_3\sigma_z$) & \textit{s}$_{\pm}$/\textit{d}-wave ($\tau_{1,2}\mu_3\nu_3$) & FM ($\perp$), \textit{d}-wave AM ($\parallel$)\\
       T-IVC ($\eta_3\kappa_{1,2}$) & \textit{s}-wave ($\tau_{1,2}$) & -  \\ 
        K-IVC ($\tau_3\kappa_{1,2}$)  & \textit{f}-wave ($\tau_{1,2}\kappa_3\sigma_z$) & - \\
        Singlet IX ($\upsilon_{1,2}$) &Layer-even \textit{s}-wave ($\tau_{1,2}$) & Layer AFM\\
        Triplet IX ($\upsilon_{1,2}\tau_3\sigma_z$)& Layer-odd \textit{s}-wave  ($\tau_{1,2}\mu_3$)& Layer AFM ($\parallel$),  FM($\perp$)\\
    \hline
    \end{tabular}
    \caption{Summary of dominant orders and shielded orders that can appear in materials with different types of Fermi surface topology. For magnetic orders shielded by a magnetic dominant order, $\parallel,\perp$ indicate whether the spins of the dominant and shielded orders are parallel or perpendicular to each other. Pauli matrices appearing in the intervalley coherent (IVC) and interlayer exciton (IX) order describe valley / layer degrees of freedom, as discussed in SM Sec.~\ref{sec:OtherFermisurfacetopologies}.}
    \label{tab:shieldedordersummary}
\end{table*}
Our theory is not limited to cuprate superconductors, as spin-density waves and superconductivity can also be present in other materials, such as iron pnictides \cite{mazin2009pairing,paglione2010high,hirschfeld2011gap}, in which case the involvement of the $\Gamma$-pocket implies the shielded order should be classified as \textit{s}$_{\pm}$ rather than \textit{d}-wave. The \textit{s}$_{\pm}$ was shown before to be the superconducting order in pnictides that is most compatible with the SDW state \cite{fernandes2010competing}. Our results indicate that the critical temperature of the previous can even be enhanced by the latter.

The main effect of doping is to introduce extra electrons, which means that also gating can induce loss of spatial SDW phase coherence. This makes shielded order a candidate mechanism for superconductivity in twisted multilayer graphene or TMD based systems \cite{lu2019superconductors}, in which the phase diagram harbors superconducting domes next to correlated insulator states, and are thus candidates for shielded order. 

The resulting type of superconductivity depends strongly on which pairing channel is compatible with  the dominant insulating order, that is, which one commutes with it. As we have shown, spin-density waves in materials with square lattices and an easy axis favor \textit{d}-wave superconductivity. Also easy-plane materials shield \textit{d}-wave superconductivity. For such materials, two of the shielded magnetic orders become massive, leaving only competition of superconductivity with ferromagnetic order along the hard axis. Charge density waves on the other hand do not break time-reversal symmetry, and  shield conventional \textit{s}-wave superconductivity, consistent with Anderson's theorem \cite{anderson1959theory}. 

In materials with hexagonal or triangular lattices, the Fermi surface is separated into two different valleys. In such systems insulating states due to intervalley interactions with a U(1) gauge have been predicted, the time-reversal preserving (T) and Kramers (K) intervalley coherent (T-IVC and K-IVC) states. Like charge-density waves, the T-IVC state does not break time-reversal symmetry, and thus shields \textit{s}-wave superconductivity. The K-IVC state on the other hand breaks time-reversal and is thus incompatible with an isotropic order parameter. Instead, it shields valley-odd \textit{f}-wave superconductivity. In many such materials though, the bandwidth of the single-electron excitations is small, which means that the quasiclassical approximation  might fail. In such case it is not possible to restrict $Q$ only to the soft-mode manifold, and thus a more general action, see Eq.~(\ref{eq:QeqTrln}) in SM, needs to be considered. Moreover, shielded orders can also appear in layered materials, in which interlayer excitons (IX) can shield subdominant order when they lose coherence.

\section{Discussion}
We have introduced the concept of shielded orders, orders that are subdominant at the mean-field level, but can prevail in materials if the dominant order loses spatial phase coherence, and therefore appears only as a short-range order. We have shown that in this regime, the absence of long-range dominant order allows subdominant order to arise. Moreover, the remaining short-range dominant order creates a large quasiparticle excitation gap that shields the subdominant order. Through this, shielded orders survive up to much higher temperatures than expected from  their pairing strength.

Shielded orders can appear in any material that hosts multiple correlation induced orders. Our theory allows for predicting the pairing symmetries of superconductors by identifying which orders are shielded by the dominant orders. Table \ref{tab:shieldedordersummary}, whose entries are derived in SM Sec.~\ref{sec:Actiondetails} and \ref{sec:OtherFermisurfacetopologies} summarizes several compatible orders. In these examples, we have taken the correlated insulators to be dominant and superconductors to reflect that in cuprates and van der Waals materials the critical temperatures of the former are usually higher. However, in principle their roles can be reversed, superconductors with phase disorder shield correlated insulator orders. 

In cuprates spin-density waves favor \textit{d}-wave superconductivity in which the gap magnitude changes as a function of angle, but its sign is the within a Fermi surface pocket. Because the sign is different on each component, such superconductivity is not distinguishable from nodal \textit{d}-wave superconductors through the density of states or phase shifts in junctions. They can, however, be probed by techniques that are sensitive to the phase shift upon a small change of momentum. For this purpose,  angle resolved photo-emission-based techniques are promising. We note that in cuprates the state at optimal doping is usually considered to be a strongly correlated Mott insulator \cite{keimer2015quantum}. The intricacies of strong correlations that come along with this \cite{lee2006doping} are not fully covered by our model. Nevertheless, we think that shielded orders provide a potential route towards explaining superconductivity in these materials.

Besides this, our theory predicts that in the search for high-temperature superconductors, it is not required to focus on strong pairing mechanisms. Rather, for shielded orders the pairing mechanism can be weak and its critical temperature is determined mainly by the critical temperature of the dominant order. We therefore suggest that to maximize the critical temperature of superconductivity, it is useful to look for materials that maximize the SDW state, that is, have strong magnetic interactions and good nesting. To this state, dopants should be added that add electrons to the conduction band, but only weakly induce scattering between the different components of the Fermi surface, since interband scattering suppresses \textit{d}-wave superconductivity. This criterion can serve as a descriptor for automated machine learning searches for superconductors \cite{mustaf2026machine}.

\section{Acknowledgements}
We thank P. Virtanen, S. Ili\'c, R. Ojajärvi, A. Dunbrack and J. Schmalian for useful discussions. The Authors acknowledge the help of AI for direction towards relevant literature and for refereeing the paper, and take full responsibility for the contents of the manuscript.

The work of T.~K. and T.~T.~H. was supported by the Research Council of Finland through DYNCOR, Project Number 354735,  through the Finnish Quantum Flagship, Project Number 359240 and Jane and Aatos Erkko Foundation and Keele Foundation as part of the SuperC collaboration. This work is part of the Finnish Centre of Excellence in Quantum Materials (QMAT). F.~S.~B. acknowledges financial support from the the Spanish MCIN/AEI/10.13039/501100011033 through the grant PID2023-148225NB-C31,  the European Union’s Horizon Europe research and innovation program under grant agreement No.~101130224 (JOSEPHINE), and InstituteQ under the Chair of Excellence programme. 
\bibliography{sources}

\appendix
\clearpage
\onecolumngrid\
\section*{Appendices}
In the main text, we have introduced shielded orders, long-range orders whose critical temperature and excitation gap are strongly enhanced by a dominant short-range order, and have identified several candidates for dominant and shielded orders, such as spin-density waves and \textit{d}-wave or \textit{s}$_{\pm}$ superconductivity, charge density waves and \textit{s}-wave superconductivity, Kramers degenerate intervalley coherent (K-IVC) states and \textit{f}-wave superconductivity, and time-reversal preserving intervalley coherent (T-IVC) states and \textit{s}-wave superconductivity.
In this supplementary material, we provide technical steps that have been omitted in the main text. To improve readability, we not only present those technical steps, but maintain a narrative. For this reason, we will repeat some of the arguments made in the main text.

The supplemental material is structured as follows. The first part of the Supplemental Material focuses on the derivation of the low energy theory and shielded orders given that there exists a dominant mode that only has short-range coherence. In Sec.~\ref{sec:Construction} we discuss the construction of the spinors for the non-linear sigma model, the form of the charge-conjugation constraint in band-Nambu-spin space and the construction of an action from the Hamiltonian. Thereafter, in Sec.~\ref{sec:Actiondetails}, we show the details behind the Hubbard-Stratonovich transformations that yield the $Q$-matrix, verify the normalization condition and expand the action on the soft-mode manifold. In Sec.~\ref{sec:Tc} we show the derivation of the critical temperature from the saddle point equations. After this, in Sec.~\ref{sec:NLSMwithangledependence} we generalize the derivation of the nonlinear sigma model to phase disorder that affect certain angles stronger than others. 

The next part of the Supplemental Material, Sec.~\ref{sec:Dopants}, is devoted to a description of how orders can lose phase coherence, while retaining amplitude and temporal coherence. We focus on the explicit example of spin-density waves and derive the free energy cost of phase gradients and vortices that are introduced by dopants, and use these expressions to derive an expression for a transition between long-range and short-range spin-density wave order. 

In Sections \ref{sec:Layeredandconcentric} and \ref{sec:IVC} we discuss a few examples of systems that host shielded order, and discuss the structure of the low-energy manifold and the allowed shielded orders in those systems. We conclude the Supplemental Material with a discussion of the comparison of loss of spatial coherence of the amplitude and phase in Sec. \ref{sec:Phasevsamplitudeinhomogeneities}.

\section{Spinors, interactions and order parameters}\label{sec:Construction}
Before we discuss the physics of shielded order, we first provide the mathematical setting in which they can be derived. In this section we introduce the action formalism that we use to describe shielded orders, define the spinors and discuss the order parameters that we consider. It serves as the basis on which the derivations in later sections build. To make the steps explicit, we consider here a specific example, using a Fermi surface with 4 pockets and spin-density wave and superconducting instabilities. We emphasize that this is not the only type of system in which shielded order appears, the derivation for other systems, like those with valley or layer degree of freedom, follows a similar construction. We provide a discussion of the basic setup of a few more examples near the end of the Supplemental Material, in Sec. \ref{sec:OtherFermisurfacetopologies}.

In condensed matter, states whose energy is far from the Fermi surface compared to interaction strengths and temperature do not play a role in the phase transitions and transport features of a material. Therefore, we can describe these materials considering only the energy levels near the Fermi surface. Moreover, in many materials, the distance from the band edge to the Fermi surface is large compared to interaction scales. In such materials, the dispersion can be considered to be almost linear, and therefore the material can be described within the quasiclassical limit.

\begin{figure}
    \begin{tikzpicture}
        \draw[ultra thick,goodgreen] (-1,-1) circle (0.7cm);
        \draw[ultra thick,orange] (-1,1) circle (0.7cm);
        \draw[ultra thick,orange] (1,-1) circle (0.7cm);
        \draw[ultra thick,goodgreen] (1,1) circle (0.7cm);
        \node at (-1,-2){$1$};
        \node at (-1,2){$3$};
        \node at (1,-2){$2$};
        \node at (1,2){$4$};
        \draw[-{Circle[sep=-.45pt,width=2.04pt,length=2pt,black]Triangle[width=5pt,length=5pt,black]},line width = 2,black] (-1,-1)--(1,-1);
    \draw[-{Circle[sep=-.45pt,width=2.04pt,length=2pt,black]Triangle[width=5pt,length=5pt,black]},line width = 2,black] (-1,-1)--(-1,1);
    \node at (0,-0.7){$\mathcal{Q}\hat{x}$};
    \node at (-0.7,0){$\mathcal{Q}\hat{y}$};
    \fill (-1,-1) circle (1pt);
    \end{tikzpicture}
    \caption{Illustration of a Fermi surface with four connected componets, two of which are electron-like and two of which are hole-like, as indicated using color. For illustrative purposes, the Fermi surfaces have been drawn as circles. Their shape however depends on the dispersion chosen.}
    \label{fig:FermisurfaceSM}
\end{figure}
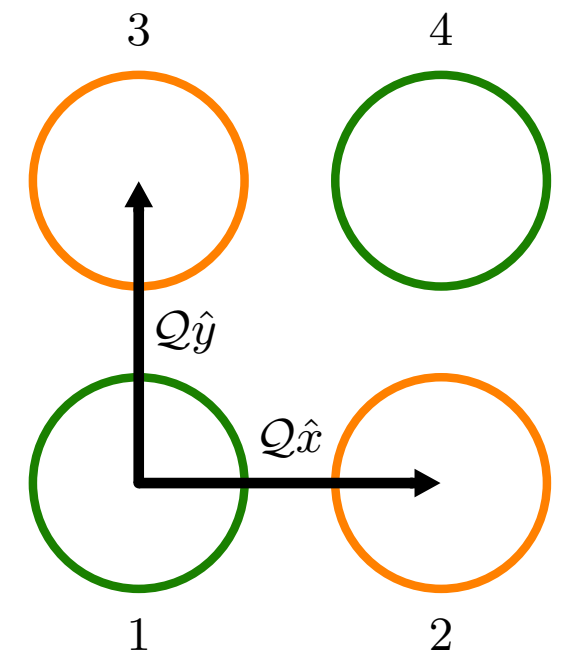

Using this formalism, we want to describe a system in which interactions can induce both superconductivity and spin-density waves. Both of these interactions can naturally appear if the Fermi surface has multiple components. We consider the minimal model that captures the concept of shielded order. The Fermi surface has four connected components, as illustrated in Fig.~\ref{fig:FermisurfaceSM}. Of these four components. We assume two are electron-like, while the other two are hole-like, with a dispersion that is almost exactly opposite.

\subsection{Keldyh-Nambu-band-spin space and their Pauli matrices}\label{sec:spaces}
Before introducing the Hamiltonian and the action, we first need to introduce the space in which we consider the action. Within the standard theory of conventional superconductors on a simple Fermi surface, the system is described in a 4 by 4 matrix degree of freedom in the spin-Nambu space that is a function of momentum. The first degree of freedom corresponds to the spin of the electron, described using Pauli matrices $\sigma_i$, $i = 1,2,3$, while the second degree of freedom describes the electron ($\psi$) and hole ($\psi^{\dagger}$), which is needed to describe the anomalous correlators $\langle\psi\psi\rangle$ and $\langle\psi^{\dagger}\psi^{\dagger}\rangle$. The Pauli matrices in Nambu degree of freedom are denoted by $\tau_i$.

It should be noted that there are several ways to construct this Nambu-space, such as keeping the same spin structure for electrons and holes $(\psi_{\uparrow},\psi_{\downarrow}, \psi_{\uparrow}^{\dagger}, \psi_{\downarrow}^{\dagger})$ or a construction with time-reversed partners $(\psi_{\uparrow},\psi_{\downarrow}, \psi_{\downarrow}^{\dagger}, -\psi_{\uparrow}^{\dagger})$. While the former is often used in BdG Hamiltonians, the latter has significant advantages in the construction of kinetic equations, especially in the presence of time-reversal symmetry breaking. In this basis, time-reversal odd terms are proportional to $\tau_3$ in Nambu space \cite{kokkeler2025quantum}, while time-reversal even terms are proportional to $\tau_0$. Moreover, it makes sure that singlet superconductivity appears proportional to $\tau_{1,2}$, while triplet superconductivity appears proportional to $\tau_{1,2}\vec{d}\cdot\vec{\sigma}$. We therefore keep this latter convention, and in the next subsection we construct Nambu space in the presence of multiple Fermi pockets following a similar procedure, making sure that time-reversal odd terms are proportional to $\tau_3$, and \textit{s}-wave superconductivity to $\tau_{1,2}$.

The Fermi surface in Fig.~\ref{fig:FermisurfaceSM} contains four connected components, the four pockets. It is most convenient to describe those pockets with an additional 4 by 4 matrix degree of freedom. We will parameterize matrices in this space using two sets of Pauli matrices $\mu_{i}, \nu_i$, where $\mu_3$ has opposite sign on pockets with even or odd number in Fig.~\ref{fig:Fermisurfaces}, while $\nu_3$ has opposite sign on pockets with number above or below $2.5$ in Fig.~\ref{fig:Fermisurfaces}. Off-diagonal terms in these spaces correspond to density-waves. Indeed, because even and odd numbered pockets have a momentum difference $\mathcal{Q}\hat{x}$, correlations that are off-diagonal in $\mu$-space correspond to density waves with wavenumber $\mathcal{Q}\hat{x}$, while correlations that are off-diagonal in $\nu$ correspond to density waves with wavenumber $\mathcal{Q}\hat{y}$. The flexibility between $\mu_{1,2}$ or $\nu_{1,2}$ indicates a U(1) phase of these density waves, and a shift $\delta\phi$ of such phase corresponds to a shift $\frac{\delta\phi}{2\pi}\mathcal{Q}^{-1}$ of the spin-density wave in real space \cite{gruner1994dynamics}.

The band-Nambu-spin space construction however, is  not enough to derive the phase disordered quasiclassical action. The reason for this is that under time-evolution, even in equilibrium the spinors pick up a time-dependent phase, which depends on the phase disorder realization. This makes disorder averaging in band-Nambu-spin space extra-ordinarily difficult. We therefore need a technique to eliminate this equilibrium phase dependence, which is done by introducing extra degrees of freedom. Historically, for scalar disorder three methods have been used, the replica trick \cite{wegner1979mobility}, supersymmetry \cite{efetov1980interaction}  and the Keldysh formalism \cite{feigelman2000keldysh,kamenev2023field}. Here we choose the latter one,  which also allows for describing non-equilibrium problems. The Keldysh contour has two branches, one running from $t = -\infty$ to $t= \infty$ and the other from $t=\infty$ to $t=-\infty$. Due to this, in equilibrium the time-dependent phases of the two tracks cancel out, and thus disorder averaging can be performed. The Keldysh degree of freedom can be described using another 2 by 2 matrix degree of freedom.  We denote the Pauli matrices in Keldysh space using $\rho_i$. The structure of the theory is simplified via a rotation in Keldysh space, $\cdot \rightarrow \frac{1}{2} (1-i\rho_2) \cdot (1+i\rho_2)$, which ensures that all the Green's functions of all causal systems are upper triangular in Keldysh space, and contain retarded, advanced and Keldysh components \cite{kamenev2023field}. 

With this, we conclude that the theory requires a 32 by 32 matrix space, the Keldysh-Nambu-band-spin space.
\subsection{Spinors and bispinors}\label{sec:Spinors}
Next, we consider the bispinors on which the matrices act. Bispinors are vectors containing all of the band and spin degrees of freedom, as well as the Keldysh and Nambu degrees of freedom. Their entries are Grassmann variables, anticommuting numbers that describe the fermionic degrees of freedom. Before constructing these bispinors, let us first ignore the Keldysh and Nambu degree of freedom and consider the spinors in band-spin space, which can be constructed as $\psi = (\psi_{1\uparrow},\psi_{1\downarrow},\psi_{2\uparrow},\psi_{2\downarrow},\psi_{3\uparrow},\psi_{3\downarrow},\psi_{4\uparrow},\psi_{4\downarrow})$.
When considering the states close to the Fermi surface, the non-interaction Hamiltonian can be written using this basis as:
\begin{align}
    H_0 = \psi^{\dagger}\Big(\xi(\vec{k}) \mu_3\nu_3 + \tilde{\xi}(\vec{k})\Big)\psi\;,\label{eq:HamiltonianSM}
\end{align}
Here $\xi(\vec{k})$ is the component of the dispersion that is opposite in the electron pockets compared to the hole pockets, while $\tilde{\xi}(k)$ is the component that is the same in the two, and thus leads to an asymmetry in size or shape. To simplify momentum integrations and obtain an analytical expression, we assume that $\tilde{\xi}$ takes the particular form $\tilde{\xi}(k) = \eta\xi(k)$, where $\eta\ll 1$ gives the deviation from the perfect antisymmetry between the pockets. A deviation from this approximation only qualitatively changes the results as long as close to the Fermi surface $\tilde{\xi}(\vec{k}) \ll \xi(\vec{k})$, the calculation of the correct prefactor in the final result however becomes more intricate.

When treating systems within the quasiclassical formalism, the dominant dispersion is assumed to be the largest energy scale of the system. To find the low-energy degrees of freedom, we follow a strategy similar to the one used in Ref.~\cite{kamenev2023field} for Nambu space and write the action in a form in which $\xi(k)$ comes with the identity matrix in band-Nambu-spin space. To achieve this, we introduce a new spinor $\bar{\psi} = \psi^{\dagger}\mu_3\nu_3 = (\psi^{\dagger}_{1\uparrow},\psi^{\dagger}_{1\downarrow},-\psi^{\dagger}_{2\uparrow},-\psi^{\dagger}_{2\downarrow})$. In terms of $\psi, \bar{\psi}$, we may write
\begin{align}
    H_0 = \bar{\psi}\Big(\xi(k) + \eta\xi(k) \mu_3\nu_3 \Big)\psi\;,\label{eq:HamiltonianSM2}
\end{align}
From a non-interacting Hamiltonian perspective this transformation might seem an unnecessary complication. Indeed, from Eq.~(\ref{eq:HamiltonianSM}) the energies can be extracted as the eigenvalues of the matrix, which is not possible using Eq.~(\ref{eq:HamiltonianSM2}). However, we are using a mesoscopic approach, and therefore we are not interested in these eigenvalues, but rather in the generating function that allows us to extract all observables. To obtain the generating function, we need to integrate over all configurations of the Grassmann fields, and thus both representations give equivalent results. As we show in the next section, the derivation of the soft-mode manifold and the subsequent expansion in the quasiclassical limit take a particularly simple form when using the representation in Eq.~(\ref{eq:HamiltonianSM2}), because the dispersive term $\xi(k)$ comes with an identity matrix in band-Nambu-spin space.

Following this strategy, we are ready to construct the bispinors in Nambu space and write the action in terms of bispinors that ensure the dispersion is proportional to the identity term.  This is achieved by using $\Psi$ and $\bar{\Psi} = \Psi^{\dagger}\tau_3\mu_3\nu_3$ instead of $\Psi$ and $\Psi^{\dagger}$, where the additional $\tau_3$ is required because the electron-hole conjugated system has opposite dispersion by construction.

Moreover, as discussed in Sec.~\ref{sec:spaces} it is convenient to construct the Nambu space using time-reversed partners, where the time-reversal operator is $\mathcal{T} = i\mu_1\nu_1\sigma_2\mathcal{K}$ and $\mathcal{K}$ denotes complex conjugation.

The exploitation of time-reversal partners ensures that time-reversal odd mechanisms can be identified using $\tau_3$ matrices, while \textit{s}-wave superconductivity, for which $\langle \psi_{1\uparrow}\psi_{4\downarrow}-\psi_{1\downarrow}\psi_{4\uparrow} +\psi_{2\uparrow}\psi_{3\downarrow}-\psi_{2\downarrow}\psi_{3\uparrow}+\psi_{3\uparrow}\psi_{2\downarrow}-\psi_{3\downarrow}\psi_{2\uparrow} +\psi_{4\uparrow}\psi_{1\downarrow}-\psi_{4\downarrow}\psi_{1\uparrow} \rangle\neq 0$ comes with a matrix proportional to $\tau_{1,2}$ but the identity matrix in other spaces.
Thus, we use the following basis of 16 component spinors: 
\begin{align}
\Psi & = (\psi^T, -i\bar{\psi}\mu_1\nu_1\sigma_2)^T = (\psi_{1\uparrow},\psi_{1\downarrow},\psi_{2\uparrow},\psi_{2\downarrow},\psi_{3\uparrow},\psi_{3\downarrow},\psi_{4\uparrow},\psi_{4\downarrow},\psi^{\dagger}_{4\downarrow},-\psi^{\dagger}_{4\uparrow},-\psi^{\dagger}_{3\downarrow},\psi^{\dagger}_{3\uparrow},-\psi^{\dagger}_{2\downarrow},\psi^{\uparrow}_{2\downarrow},\psi^{\dagger}_{1\downarrow},-\psi^{\dagger}_{1\uparrow})^{T}\;,\\
\begin{split}
\Bar{\Psi} & =  \Psi^{\dagger}\tau_3\mu_3\nu_3= (\bar{\psi},i\psi^T\sigma_2\mu_1\nu_1)  \\ &=(\psi^{\dagger}_{1\uparrow},\psi^{\dagger}_{1\downarrow},-\psi^{\dagger}_{2\uparrow},-\psi^{\dagger}_{2\downarrow},-\psi^{\dagger}_{3\uparrow},-\psi^{\dagger}_{3\downarrow},\psi^{\dagger}_{4\uparrow},\psi^{\dagger}_{4\downarrow},-\psi_{4\downarrow},\psi_{4\uparrow},-\psi_{3\downarrow},\psi_{3\uparrow},-\psi_{2\downarrow},\psi_{2\uparrow},-\psi_{1\downarrow},\psi_{1\uparrow})\;.
\end{split}
\end{align}

This is the structure of the bispinors in band-Nambu-spin space that we use to derive the quasiclassical action of the system. Incorporating the Keldysh degree of freedom in the standard way, $\Psi$ and $\bar{\Psi}$ are promoted to 32 component spinors. 

\subsection{Charge conjugation}
As mentioned in the introduction of Nambu space, the electron-hole doubling puts constraints on the allowed configurations because the description of a system in terms of electrons is the same as its description in terms of holes. Therefore, not all 32 by 32 configurations can be achieved. Indeed, $\Psi$ and $\bar{\Psi}$ are not independent, but related by
\begin{align}
    \Psi = i\tau_1\mu_1\nu_1 \sigma_2\bar{\Psi}^{T}\;.
\end{align}
On the time contour this implies
\begin{align}
    \langle\Psi(t) \bar{\Psi}(t')\rangle = \tau_1\mu_1\nu_1 \sigma_2\langle\Psi(t') \bar{\Psi}(t)\rangle^T\tau_1\mu_1\nu_1 \sigma_2\;,
\end{align}
which is the charge conjugation symmetry. As derived in Ref.~\cite{kokkeler2025quantum}, in the usual Keldysh space this means
\begin{align}\label{eq:ChargeConjugationPsibarPsi}
    \langle\Psi(t) \bar{\Psi}(t')\rangle  = \rho_1\tau_1\mu_1\nu_1 \sigma_2\langle\Psi(t') \bar{\Psi}(t)\rangle^T\sigma_2\nu_1\mu_1\tau_1\rho_1\;.
\end{align}
With this, we have setup the spinors. We now turn our attention to the order parameters that can be constructed with the help of these spinors.

\subsection{Order parameters}
We are interested in the behavior of this system in the presence of interactions. Due to the existence of four components of the Fermi surface interactions can induce not only superconductivity or magnetism, but also charge density waves and spin-density waves along both the $x$ and $y$ directions. For all of these orders we can define order parameters, that indicate which components of the correlator are nonzero. For charge density waves, we obtain the following complex order parameters: 
\begin{align}
     \Delta_{\text{cx}}e^{i\phi_{\text{cx}}}\sim -i\langle[\psi_{1\uparrow},\psi_{2\uparrow}^{\dagger}]+[\psi_{1\downarrow},\psi_{2\downarrow}^{\dagger}]-[\psi_{3\uparrow},\psi_{4\uparrow}^{\dagger}]-[\psi_{3\downarrow},\psi_{4\downarrow}^{\dagger}]\rangle^K = \text{Tr}\langle\rho_1\frac{\mu_2-i\mu_1}{2} \Psi \bar{\Psi} \rangle\;,\\
          \Delta_{\text{cy}}e^{i\phi_{\text{cy}}}\sim -i\langle [\psi_{1\uparrow},\psi_{3\uparrow}^{\dagger}]+[\psi_{1\downarrow},\psi_{3\downarrow}^{\dagger}]-[\psi_{2\uparrow},\psi_{4\uparrow}^{\dagger}]-[\psi_{2\downarrow},\psi_{4\downarrow}^{\dagger}]\rangle^K = \text{Tr}\langle\rho_1\frac{\nu_2-i\nu_1}{2} \Psi \bar{\Psi} \rangle\;,
\end{align}
where $\Delta_{\text{cx,y}}$ and $\phi_{\text{cx,y}}$ are real numbers and the notation $\langle\cdot\rangle$ is introduced to express expectation values and the superscript $K$ denotes the Keldysh component. The amplitude of the order parameter gives the excitation gap, while its phase corresponds to the position of the peak of the density wave within the unit cell of the density wave. Like for superconductors, a global phase shift does not alter the system, but a position dependent phase shift does. The gradients of $\phi_{\text{cx,y}}$ however do not correspond to charge currents, but rather to a deviation of the distance between peaks from the optimum charge density wave periodicity.

Next to charge density waves, we may also consider spin-density waves, in which the spin-density oscillates as a function of $x$ or $y$. For simplicity of presentation, we choose the spin-direction to be in the z-direction. The order parameters of the spin density wave are defined as
\begin{align}
     \Delta_{\text{sx}}e^{i\phi_{\text{sx}}}&\sim -i\langle[\psi_{1\uparrow},\psi_{2\uparrow}^{\dagger}]-[\psi_{1\downarrow},\psi_{2\downarrow}^{\dagger}]-[\psi_{3\uparrow},\psi_{4\uparrow}^{\dagger}]+[\psi_{3\downarrow},\psi_{4\downarrow}^{\dagger}]\rangle^K = \text{Tr}\langle\rho_1\frac{\mu_2-i\mu_1}{2}\sigma_z \Psi \bar{\Psi} \rangle\;,\\
          \Delta_{\text{sy}}e^{i\phi_{\text{sy}}}&\sim -i\langle [\psi_{1\uparrow},\psi_{3\uparrow}^{\dagger}]-[\psi_{1\downarrow},\psi_{3\downarrow}^{\dagger}]-[\psi_{2\uparrow},\psi_{4\uparrow}^{\dagger}]+[\psi_{2\downarrow},\psi_{4\downarrow}^{\dagger}]\rangle^K = \text{Tr}\langle\rho_1\frac{\nu_2-i\nu_1}{2}\sigma_z \Psi \bar{\Psi} \rangle\;.
\end{align}
Just as for the charge density wave order parameters, the amplitude of the order parameter corresponds to the excitation gap, while the phase corresponds to the relative position within the unit cell, so that phase gradients correspond to a deviation of the periodicity from the optimal one.

Next to density wave order parameters, we also consider superconductor order parameters. Our main focus is on two types of order parameters, \textit{s}-wave and \textit{d}-wave, which are defined as
\begin{align}
    \Delta_{\text{s}} e^{i\phi_{\text{s}}} \sim \langle \psi_{1\uparrow}\psi_{4\downarrow}-\psi_{1\downarrow}\psi_{4\uparrow} +\psi_{2\uparrow}\psi_{3\downarrow}-\psi_{2\downarrow}\psi_{3\uparrow}+\psi_{3\uparrow}\psi_{2\downarrow}-\psi_{3\downarrow}\psi_{2\uparrow} +\psi_{4\uparrow}\psi_{1\downarrow}-\psi_{4\downarrow}\psi_{1\uparrow} \rangle^K &= \text{Tr}\langle\rho_1\frac{\tau_2-i\tau_1}{2}\Psi \bar{\Psi}\rangle\;,\\
    \Delta_{\text{d}} e^{i\phi_{\text{d}}} \sim \langle \psi_{1\uparrow}\psi_{4\downarrow}-\psi_{1\downarrow}\psi_{4\uparrow} -\psi_{2\uparrow}\psi_{3\downarrow}+\psi_{2\downarrow}\psi_{3\uparrow}-\psi_{3\uparrow}\psi_{2\downarrow}+\psi_{3\downarrow}\psi_{2\uparrow} +\psi_{4\uparrow}\psi_{1\downarrow}-\psi_{4\downarrow}\psi_{1\uparrow} \rangle^K &= \text{Tr}\langle\rho_1\frac{\tau_2-i\tau_1}{2}\mu_3\nu_3\Psi \bar{\Psi}\rangle\;.
\end{align}

Again, the amplitude of the order parameter indicates the excitation gap. Phase gradients of superconductors correspond to dissipationless charge currents.

Lastly, we consider the magnetic order parameters. Because of the four components, we can include two types of order parameters with spin-split bands, ferromagnetism and \textit{d}-wave altermagnetism.
Introducing $\vec{\psi}_{i} = (\psi_{i\uparrow},\psi_{i\downarrow})^{T}$, the order parameters $\vec{m}$ of the ferromagnet and $\vec{N}$ of the altermagnet correspond to
\begin{align}
    \vec{m} &\sim \langle \vec{\psi}^{\dagger}_1\sigma\vec{\psi_{1}}\rangle^K +\langle \vec{\psi}^{\dagger}_2\sigma\vec{\psi_{2}}\rangle^K+\langle \vec{\psi}^{\dagger}_3\sigma\vec{\psi_{3}}\rangle^K+\langle \vec{\psi}^{\dagger}_4\sigma\vec{\psi_{4}}\rangle^K  = \text{Tr}\langle\rho_1\tau_3\vec{\sigma}\mu_3\nu_3\Psi\bar{\Psi}\rangle\;,\\
    \vec{N} &\sim  \langle \vec{\psi}^{\dagger}_1\sigma\vec{\psi_{1}}\rangle^K -\langle \vec{\psi}^{\dagger}_2\sigma\vec{\psi_{2}}\rangle^K-\langle \vec{\psi}^{\dagger}_3\sigma\vec{\psi_{3}}\rangle^K+\langle \vec{\psi}^{\dagger}_4\sigma\vec{\psi_{4}}\rangle^K  = \text{Tr}\langle\rho_1\tau_3\vec{\sigma}\Psi\bar{\Psi}\rangle\;.
\end{align}
Note that in the basis chosen, the \textit{s}-wave superconductor order parameter that is isotropic is proportional to the identity in band space, while isotropic magnetic order is proportional to $\mu_3\nu_3$, due to the construction of the bispinors. 

\subsection{From Hamiltonian to action}

Having defined the bispinors and the types of order parameters, we are ready to start constructing the action. For simplicity, we first focus on the case in which the dispersion in the hole pockets is exactly opposite to that of the electron pockets, i.e. $\eta =  0$. The action for the non-interacting system with Hamiltonian given in Eq.~(\ref{eq:HamiltonianSM2})  can be written as
\begin{align}
    Z &=\int \mathcal{D}\Psi\int\mathcal{D}\bar{\Psi}\;e^{iS}\;,\nonumber\\ iS &= \text{Tr}  i\Bar{\Psi}(\tau_3\mu_3\nu_3 \hat{\omega}_{t,t'} + \xi)\Psi \;,
\end{align}
where $\hat{\omega}_{t,t'} = \delta(t-t')\partial_t$. 
The time-derivative comes with all third Pauli matrices because we defined $\bar{\Psi} = \Psi^{\dagger}\tau_3\mu_3\nu_3$ to make the dispersion term proportional to the identity in band-Nambu-spin space. Moreover, we use the notation $\int \mathcal{D}\cdot $ to denote path integrals and $\text{Tr}$ to indicate tracing over matrix degrees of freedom as well as spatiotemporal integration.

Now, let us incorporate the interactions. Specifically, we are interested in interactions leading to SDW and superconductivity. We denote the easy axis for SDWs by the z-axis. In this case, for SDW, the relevant interaction in momentum space in terms of the non-pocket projected electron operators $c$ is
\begin{align}
    H_{\text{int}} = \sum_{k,k',q} V(q)c^{\dagger}_{k+q,\uparrow}c_{k,\uparrow}c^{\dagger}_{k'-q,\downarrow}c_{k',\uparrow}\;,
\end{align}
where $V(q)$ is a momentum dependent potential whose momentum ($q$) dependence is determined by the electron-electron interactions. We have to separate this into $q$'s that 
have a magnitude much smaller than the distance between Fermi surfaces and $q$'s that have a magnitude close to the distance $Q$ between Fermi surfaces. The resulting interaction Hamiltonian can be written in terms of the spinors $\psi$ with pocket degree of freedom as
\begin{align}
    H_{\text{int}} &= \sum_{k,k',q \ll \mathcal{Q}} V(q)\sum_{a,b = 1}^{4} \psi^{\dagger}_{k+q,a,\uparrow}\psi_{k,a,\uparrow}\psi^{\dagger}_{k'-q,b,\downarrow}\psi_{k',b,\downarrow}\nonumber\\&+\sum_{k,k',q \sim \mathcal{Q}\hat{x}} V(q)\sum_{a,b \in \{1,3\}} \psi^{\dagger}_{k+q,a+1,\uparrow}\psi_{k,a,\uparrow}\psi^{\dagger}_{k'-q,b,\downarrow}\psi_{k',b+1,\downarrow}\nonumber\\&+\sum_{k,k',q \sim -\mathcal{Q}\hat{x}} V(q)\sum_{a,b \in \{2,4\}} \psi^{\dagger}_{k+q,a-1,\uparrow}\psi_{k,a,\uparrow}\psi^{\dagger}_{k'-q,b,\downarrow}\psi_{k',b-1,\downarrow}\nonumber\\&+\sum_{k,k',q \sim \mathcal{Q}\hat{y}} V(q)\sum_{a,b \in \{1,2\}} \psi^{\dagger}_{k+q,a+2,\uparrow}\psi_{k,a,\uparrow}\psi^{\dagger}_{k'-q,b,\downarrow}\psi_{k',b+2,\downarrow}\nonumber\\&+\sum_{k,k',q \sim -\mathcal{Q}\hat{y}} V(q)\sum_{a,b \in \{3,4\}} \psi^{\dagger}_{k+q,a-2,\uparrow}\psi_{k,a,\uparrow}\psi^{\dagger}_{k'-q,b,\downarrow}\psi_{k',b-2,\downarrow}
    \nonumber\\&+\sum_{k,k',q \sim \mathcal{Q}(\hat{x}+\hat{y})} \sum_{\sigma = \uparrow,\downarrow} V(q) \psi^{\dagger}_{k+q,4,\sigma}\psi_{k,1,\sigma}\psi^{\dagger}_{k'-q,1,\bar{\sigma}}\psi_{k',4,\bar{\sigma}}\nonumber\\&+\sum_{k,k',q \sim -\mathcal{Q}(\hat{x}-\hat{y})} \sum_{\sigma = \uparrow,\downarrow}V(q) \psi^{\dagger}_{k+q,2,\sigma}\psi_{k,3,\sigma}\psi^{\dagger}_{k'-q,3,\bar{\sigma}}\psi_{k',2,\bar{\sigma}}\;.\label{eq:Interactionsdecomposedinmomenta}
\end{align}

Here $\bar{\sigma}$ is used to denote the spin opposite to $\sigma$.
Such four-fermion terms can be decoupled using Hubbard-Stratonovich decoupling, which reduces it to a four-fermion term at the cost of integration over an additional variable and its complex conjugate. This additional variable is a boson, and its expectation value is an order parameter. Here we focus on SDW with wave factor $\mathcal{Q}\hat{x}$ or $\mathcal{Q}\hat{y}$ and superconductivity. Because their main contributions come from different $q$, there exist separate Hubbard-Stratonovich transformations for all of them. We assume that the spin-density waves are dominant and thus first focus only on them, and decouple the interaction with the fields $\hat{\Delta}_{\text{SDW},x} = \Delta_{\text{SDW,x}}e^{i\phi_{\text{SDW,x}}\mu_3}\mu_2$ for $q \sim \pm \mathcal{Q}\hat{x}$ and $\hat{\Delta}_{\text{SDW},y} = \Delta_{\text{SDW,y}}e^{i\phi_{\text{SDW,y}}\nu_3}\nu_2$ for $q\sim \mathcal{Q}\hat{y}$. For now, we ignore the subdominant interactions with $q\ll \mathcal{Q}$ for simplicity of presentation. We consider them them again in Sec. \ref{sec:manifoldstructure}. The resulting two fermion action reads
\begin{align}
   Z &= \int \mathcal{D}\hat{\Delta}_{\text{SDW,x}}\int \mathcal{D}\hat{\Delta}_{\text{SDW,y}}\int \mathcal{D}\Psi\int \mathcal{D}\bar{\Psi}\;e^{iS_i}\\
   iS_i &=  i\Bar{\Psi}(\tau_3\mu_3\nu_3 \hat{\omega}_{t,t'} + \xi + i\Delta _{\text{SDW},x} e^{i\phi_{\text{SDW},x}\mu_3}\tau_3\sigma_z\mu_2 + i\Delta _{\text{SDW},y} e^{i\phi_{\text{SDW},y}\nu_3}\tau_3\sigma_z\nu_2)\Psi \nonumber\\& + \frac{i\nu}{\lambda_{\text{sx}}}\hat{\Delta}_{\text{SDW,x}}^{\dagger}\rho_1\hat{\Delta}_{\text{SDW,x}} + \frac{i\nu}{\lambda_{\text{sy}}}\hat{\Delta}_{\text{SDW,y}}^{\dagger}\rho_1\hat{\Delta}_{\text{SDW,y}}\;, \label{eq:SDWaction}
\end{align}
where $\rho_1$ is the first Pauli matrix in Keldysh space, $\lambda_{\text{sx}} = \nu V(q = \mathcal{Q}\hat{x})$ and $\lambda_{\text{sy}} = \nu V(q = \mathcal{Q}\hat{y})$ and $\nu$ is the density of states per Fermi surface, while $\hat{\Delta}_{\text{SDW,x(y)}} = \Delta _{\text{SDW},x(y)} e^{i\phi_{\text{SDW},x(y)}\mu_3}\tau_3\sigma_z\mu_2$ is the spin-density wave potential, with $\Delta _{\text{SDW},x(y)}$ providing the excitation gap, while $\phi_{\text{SDW}}\in[0,2\pi)$ denotes the relative position of the peak of the spin-density wave. With this, we have introduced the setting and the orders that are relevant for the system under consideration. In what follows we consider a system in which the dominant order parameter $\hat{\Delta}_{\text{SDW}}$ loses spatial coherence of its phase, so that spin-density wave correlations exist locally, but decay quickly.

\section{Action formalism in the presence of a short-range dominant order}\label{sec:Actiondetails}
Now that we have introduced the relevant spaces and constructed the microscopic action for systems with four Fermi pockets, we turn our attention to the description of the physics determined by the states near the Fermi level, and identify the low-energy degrees of freedom. We focus on the state in which the dominant order has lost its spatial phase coherence. Because system responses are governed by states that are close in energy to the Fermi surface, we identify the low-energy degrees of freedom.
\subsection{Treatment of the phase disorder}\label{sec:Decouplingthings}
In the following,  we derive the nonlinear sigma model for a material with a dominant order that is short-ranged, that is, has no phase stiffness, but the local order parameter persists. If we consider phenomena that appear on length scales much longer than the characteristic short length scale $\xi_{\text{V}}$, we can approximate\footnote{In this formulation we treat $\Delta_{\text{SDW}}$ as Gaussian distributed to remain as close as possible to the standard disorder formulation. In Sec. \ref{sec:Phasevsamplitudeinhomogeneities} we show that this choice does not affect the structure of the low-energy theory.}
\begin{align}
    \langle\Delta_{\text{SDW,x}}(r)\Delta^*_{\text{SDW,x}}(r')\rangle \sim |\Delta^{0}_{\text{SDW}}|^2e^{-\frac{|r-r'|}{\xi_V}}\;.\label{eq:verylowstiffness}
\end{align}
When considering the slow diffusion modes, we consider scales much larger than $\xi_{\text{V}}$, in which case we may treat $\Delta_{\text{SDW}}$ as local. 

\begin{align}
    \langle\Delta_{\text{SDW,x}}(r)\Delta^*_{\text{SDW,x}}(r')\rangle \sim \frac{1}{\pi\nu\tau_{\Delta}}\delta(r-r')\;.\label{eq:nostiffness}
\end{align}
Here $\tau_{\Delta}^{-1}$ is the rate over which the direction of momentum is randomized. It therefore corresponds to the transport scattering time, defined as \cite{kamenev2023field}:
\begin{align}
    \tau_{\Delta}^{-1} &= \int_{\text{F.S.}} \frac{d\hat{k}}{A_{\text{F.S.}}}\tilde{\tau}_{\Delta}(\vec{k})\;,\\
    \tilde{\tau_{\Delta}}^{-1}(\vec{k}) &= \int dk' \langle\Delta_{\text{SDW,x}}\Delta^*_{\text{SDW,x}}(k-k')\rangle (1-\vec{k}\cdot\vec{k}') =  \int_{\text{F.S.}} \frac{d\hat{k}'}{A_{\text{F.S.}}} \nu(k')\langle\Delta_{\text{SDW,x}}\Delta^*_{\text{SDW,x}}\rangle(k-k') (1-\vec{k}\cdot\vec{k}')\;.
\end{align}
Here $\langle\Delta_{\text{SDW,x}}\Delta^*_{\text{SDW,x}}(k-k')\rangle$ is the Fourier transform of Eq.~(\ref{eq:verylowstiffness}), while $A_{\text{F.S.}}$ is defined such that $\int_{\text{F.S.}} \frac{d\hat{k}}{A_{\text{F.S.}}} 1 = 1$, that is, it corresponds to an average over Fermi surface.

For a spherical Fermi surface, all directions are equivalent, so that $\tilde{\tau}_{\Delta}(\vec{k}) = \tau_{\Delta}$ for all $\vec{k}$, and it equals
\begin{align}
    \tau_{\Delta}^{-1} = 2\pi \nu|\Delta^{0}_{\text{SDW}}|^2 \xi_V^2 \int \frac{d\theta}{2\pi} \frac{1-\cos\theta}{(1+2k_F^2\xi_V^2(1-\cos\theta))^{\frac{3}{2}}}\;. 
\end{align}
This integral can be solved analytically. 
For $k_{\text{F}}\xi_{\text{V}}\gg 1$, it reduces to
\begin{align}
    \tau_{\Delta}^{-1} = \pi\nu |\Delta^{0}_{\text{SDW}}|^2\frac{\text{ln}[ (8/e)k_F\xi_V]}{k_F^3\xi_V}\;.
\end{align}
Like for scalar disorder, for energies below the transport scattering time the electronic properties are almost isotropric, and an NLSM can be constructed.

In this case the dominant order has a structure very similar to the one known for scalar disorder. For scalar disorder, the low energy degrees of freedom can be derived with the help of a nonlinear sigma model (NLSM) \cite{kamenev2023field}, which produces a theory that is valid on length scales much larger than the mean free path of the system. Therefore, we start from the action Eq.~(\ref{eq:SDWaction}) but treat $\Delta_{\text{SDW,x}}$ similarly to disorder and derive the resulting NLSM. Averaging over this effective disorder term, we obtain an effective theory valid at length scales much larger than the phase disorder length $\xi_{\text{V}}$:
\begin{align}
     Z &= \int \mathcal{D}\hat{\Delta}_{\text{SDW,x}}\int \mathcal{D}\hat{\Delta}_{\text{SDW,y}}\int \mathcal{D}\Psi\int\mathcal{D}\bar{\Psi}\;e^{iS_2}\\
   iS_2 &=  \int dt i\Bar{\Psi}(\tau_3\mu_3\nu_3 \hat{\omega}_{t,t'} + \xi + i\Delta _{\text{SDW},x} e^{i\phi\mu_3}\tau_3\sigma_z\mu_2 + i\Delta _{\text{SDW},y} e^{i\phi\nu_3}\tau_3\sigma_z\nu_2)\Psi \nonumber\\& - \pi\nu\tau_{\Delta}|\Delta_{\text{SDW,x}}|^2 + \frac{i\nu}{\lambda_{\text{sy}}}\hat{\Delta}_{\text{SDW,y}}^{\dagger}\rho_1\hat{\Delta}_{\text{SDW,y}}\Big\}\;. \label{eq:SDWactions}
\end{align}
 The spin-density wave $\Delta_{\text{SDW}}$ appears as an uncorrelated disorder potential, however, with the distinction that it has a structure in Keldysh-Nambu-band-spin space: it contains both $\mu_2 \tau_3\sigma_z$ and $\mu_1\tau_3\sigma_z$. We therefore use a similar treatment as used in Ref.~\cite{kamenev2023field} for scalar disorder, but take into account the non-trivial structure of the phase disorder. 
 
 First, we want to eliminate the rapidly and randomly fluctuating term from the action and therefore integrate out $\Delta_{\text{SDW}}$ using Gaussian integration rules. By doing this, we obtain a four fermion term. However, unlike for scalar disorder, not all combinations can appear in this four-fermion term. Like for magnetic disorder \cite{marchetti2002tail}, correlations that are not preserved upon scattering off the phase disorder, do not contribute. Formally, only those terms $\Psi\bar{\Psi}$ that commute with both $\mu_2\tau_3\sigma_3$ and $\mu_1\tau_3\sigma_z$ contribute to the Gaussian integrations over both real and imaginary parts. Because the action appears in an exponential, it is safe to ignore all contributions that do not satisfy this condition. These other components do not contribute to the low-energy theory. With this, we obtain
\begin{align}
    Z &= \int \mathcal{D}\hat{\Delta}_{\text{SDW,x}}\int \mathcal{D}\hat{\Delta}_{\text{SDW,y}}\int \mathcal{D}\Psi\int\mathcal{D}\bar{\Psi} \;\text{Exp}\Big\{\int dt i\Bar{\Psi}(\tau_3\mu_3\nu_3 \hat{\omega}_{t,t'} + \xi + i\Delta _{\text{SDW},y} e^{i\phi\nu_3}\tau_3\sigma_z\nu_2)\Psi \nonumber\\& + \frac{i\nu}{\lambda_{\text{sy}}}|\Delta_{\text{SDW,y}}|^2\Big\}\;\text{Exp}\{-\sum_m\frac{1}{4\pi\nu\tau_{\Delta}}\int dt\int dt'\bar{\Psi}(t)P_m\Psi(t) \bar{\Psi}(t')P_m\Psi(t')\}\;, \label{eq:SDWaction2}
\end{align}
where $P_m$, $m = 1...N_e$ projects on the $N_e$ space of matrices that commute with $\tau_3\sigma_z \mu_{1,2}$. Similar to the derivation for scalar disorder \cite{kamenev2023field}, we proceed by noting that due to anticommutation relations $\bar{\Psi}(t)P_m\Psi(t)\bar{\Psi}(t')P_m\Psi(t') = -\bar{\Psi}(t)P_m\Psi(t') \bar{\Psi}(t')P_m\Psi(t)$. Like for the interactions introduced in the Sec.~\ref{sec:Construction}, we want to reduce this term to a two-fermion term using a Hubbard-Stratonovich decoupling. This interchange of $\Psi$'s allows us to do this decoupling with a $Q$-field that has two time-coordinates instead of one and still contains a structure in Keldysh-Nambu-band-spin space: 
\begin{align}
    \text{Exp}\{\sum_m\frac{1}{4\pi\nu\tau_{\Delta}}\int dt \int dt' \bar{\Psi}(t)P_m\Psi(t) \bar{\Psi}(t')P_m\Psi(t')\} =  \int \mathcal{D}Q\; \text{Exp}\{\text{Tr}-\frac{\pi\nu}{4\tau_{\Delta}}Q^2 +\frac{i}{\tau_{\Delta}} \bar\Psi Q\Psi\}\;, \label{eq:decoupled}
\end{align}
where $\nu$ is the density of states per spin per Fermi pocket and $Q$ contains the long-range correlations of the fermions, while $\text{Tr}$ denotes matrix traces as well as spatiotemporal integration. We note that the choice of the prefactor in front of $Q^2$ in Eq.~(\ref{eq:decoupled}) is arbitrary; Choosing a different prefactor merely rescales the $Q$-field. The prefactor here is chosen so the resulting soft-mode manifold is characterized by the simplest possible constraint. 

With this, we obtain again an action with only two fermion terms, 
\begin{align}
    Z &= \int \mathcal{D}Q\int \mathcal{D}\hat{\Delta}_{\text{SDW,x}}\int \mathcal{D}\hat{\Delta}_{\text{SDW,y}}\int \mathcal{D}\Psi\int\mathcal{D}\bar{\Psi}\; \text{Exp}\Big\{\text{Tr}\Big(i\Bar{\Psi}(\tau_3\mu_3\nu_3 \hat{\omega}_{t,t'} + \xi + i\Delta _{\text{SDW},y} e^{i\phi\nu_3}\tau_3\sigma_z\nu_2 +\frac{i}{\tau_{\Delta}}Q)\Psi \nonumber\\& + \frac{i\nu}{\lambda_{\text{sy}}}|\Delta_{\text{SDW,y}}|^2\Big)\Big\}\;\text{Exp}\Big\{-\frac{\pi\nu}{4\tau_{\Delta}}\text{Tr}\Big(Q^2\Big)\Big\}\;, \label{eq:SDWaction3}
\end{align}

which can be integrated out to give
\begin{align}
    Z = \int \mathcal{D}Q \;e^{iS[Q]}\;,
\end{align}
where $S[Q]$ is given by
\begin{align}\label{eq:iSQ}
    iS[Q] = -\frac{\pi\nu}{4\tau_{\Delta}}\text{Tr}(Q^2) + \text{Tr ln}\Big((\tau_3\mu_3\nu_3 \hat{\omega}_{t,t'} + \xi + i\Delta _{\text{SDW},y} e^{i\phi\nu_3}\tau_3\sigma_z\nu_2 +\frac{i}{\tau_{\Delta}}Q)\Big)\;.
\end{align}

The integration over the $Q$-matrix is an integration over a restricted part of the Keldysh-band-Nambu-spin space. Because $Q$ is related to $\Psi \bar{\Psi}$ via Hubbard-Stratonovich decoupling, it inherits its charge conjugation symmetry, introduced in Eq.~(\ref{eq:ChargeConjugationPsibarPsi}), that is, $Q$ satisfies
\begin{align}
    Q(t,t') = \rho_1\tau_1\mu_1\nu_1 \sigma_2Q(t',t)^T\sigma_2\nu_1\mu_1\tau_1\rho_1\;.\label{eq:ChargeConjugationQ}
\end{align}

Moreover, because the four-fermion terms only contains $\Psi$'s that are in the shared eigenspace of $\tau_3\sigma_z\mu_{1,2}$, the integration has to be restricted to those $Q$ that satisfy
\begin{align}
    Q = \tau_3\sigma_z \mu_1 Q\mu_1\sigma_z\tau_3 = \tau_3\sigma_z \mu_2 Q\mu_2\sigma_z\tau_3\;.\label{eq:BQBlike}
\end{align}
It is this restriction that distinguishes the phase disorder from scalar disorder, the $Q$-matrix integration is restricted to a smaller soft-mode manifold.

With this, we have gone from a two-fermion term, via a four-fermion term, back to a two-fermion term. However, via successive transformations, we have interchanged integration over realizations of the rapidly spatially varying field $\Delta_{\text{SDW}}$ that is a function of a single time-coordinate, with an integration over the 32 by 32 matrix field $Q$ that depends on two time coordinates. The main advantage of this procedure is that similar to the case of scalar disorder, the averaged fields $Q$ do not vary rapidly over space, and they describe the phase disorder averaged low-energy modes.

\subsection{Normalization}
The action in Eq.~(\ref{eq:iSQ}) gives a saddle point equation that can be solved for a general class of materials. This resulting saddle point equation is in many cases hard to solve. However, for many materials, the Fermi energy is by far the largest relevant energy scale in the system. In those materials, the $Q$-matrix can be restricted to the so-called soft-mode manifold, which captures the low energy degrees of freedom. On this soft-mode manifold, Eq.~(\ref{eq:iSQ}) can be simplified significantly. In this Subsection, we show that the soft-mode manifold for the materials under consideration is characterized by $Q^2 = \mathbf{1}$, like in the case of scalar disorder. To this end, we exploit the quasi-classical limit, in which dispersion is linearized near the Fermi surface, $\xi \approx \vec{v}_{\text{F}}(\vec{\hat{k}}_{\text{F}})\cdot(\vec{k}-\vec{k}_{\text{F}})$, where $\vec{v}_{\text{F}}(\vec{\hat{k}}_{\text{F}})$ is the Fermi velocity at $\vec{k}_F$.

Taking the variation of Eq.~(\ref{eq:iSQ}) with respect to the $Q$-matrix, we obtain
\begin{align}
    0 = \pi^2\nu^2 \xi_{\text{V}}^2|\Delta^{0}_{\text{SDW}}|^2\Big(-Q + \frac{i}{\pi \nu}((\tau_3\mu_3\nu_3 \hat{\omega}_{t,t'} + \xi + i\Delta _{\text{SDW},y} e^{i\phi\nu_3}\tau_3\sigma_z\nu_2 +\frac{i}{\tau_{\Delta}}Q))^{-1}\Big)\;,
\end{align}
so that
\begin{align}
    Q(r) = \frac{i}{\pi \nu}((\tau_3\mu_3\nu_3 \hat{\omega}_{t,t'} + \xi + i\Delta _{\text{SDW},y} e^{i\phi\nu_3}\tau_3\sigma_z\nu_2 +\frac{i}{\tau_{\Delta}}Q))^{-1}_{r' = r}\;.
\end{align}
To evaluate this relation, we use the same technique as for scalar disorder \cite{kamenev2023field}, and we make a transformation from time domain to energy domain. Moreover, instead of real space, we exploit momentum space, in which evaluation of $r' = r$ yields an integral over all momenta.

It is the performance of this momentum integral at which the use of $\bar{\Psi}$ instead of $\Psi^{\dagger}$ proves useful; the momentum dependent term $\xi$ comes with an identity matrix in space, which allows us to choose a single eigenbasis, the eigenbasis of of $-i\epsilon\tau_3\mu_3\nu_3 + \Delta_{\text{SDW,y}}^2 +\frac{i}{\tau_{\Delta}}$, which is independent of momentum. In this eigenbasis, the integrals are easily evaluated. Within the quasiclassical approximation they yield $\pm 1$ because the dispersive term dominates \cite{kamenev2023field}.  Therefore, the eigenvalues of $Q$ are $\pm 1$, indicating that the soft-mode manifold is given by
\begin{align}
    Q^2 = \mathbf{1}\;.
\end{align}
This is the same normalization condition as found for scalar disorder \cite{kamenev2023field}, which allows us to exploit the same techniques. Its nonlinear character implies that the action restricted to the soft-mode manifold is a nonlinear sigma model (NLSM).
\subsection{Expansion}\label{sec:quasiclassicalexpansion}
Now that we have identified the soft-mode manifold, we expand the action within the quasiclassical limit on this manifold. The procedure is in many ways similar to the standard one \cite{kamenev2023field}, but with one important distinction. Namely, we first note that the constraints imposed by the phase disorder, Eq.~\ref{eq:BQBlike}, imply that
\begin{align}
    &\text{Tr ln}\Big((\tau_3\mu_3\nu_3 \hat{\omega}_{t,t'} + \xi + i\Delta _{\text{SDW},y} e^{i\phi\nu_3}\tau_3\sigma_z\nu_2 +\frac{i}{\tau_{\Delta}}Q)\Big)\nonumber\\ &= \text{Tr ln}\Big((\tau_3\mu_3\nu_3 \hat{\omega}_{t,t'} + \xi + i\Delta _{\text{SDW},y} e^{i\phi\nu_3}\tau_3\sigma_z\nu_2 +\frac{i}{\tau_{\Delta}}\tau_3\mu_2\sigma_zQ\tau_3\mu_2\sigma_z)\Big) \nonumber\\
    & = \text{Tr ln}\Big((-\tau_3\mu_3\nu_3 \hat{\omega}_{t,t'} + \xi + i\Delta _{\text{SDW},y} e^{i\phi\nu_3}\tau_3\sigma_z\nu_2 +\frac{i}{\tau_{\Delta}}Q)\Big)\;.
\end{align}
From this we conclude that to first order, the time-derivative term $\hat{\omega}_{t,t'}$, which in Matsubara space corresponds to the Matsubara frequency, drops out of the equations, and it can only appear in the next, beyond quasiclassical, order. Such terms carry a factor $T/E_{\text{F}}$ and are thus neglected in the following. 

The procedure for the spatial derivative term and the $\Delta_{\text{SDW}}$ term is entirely similar as in the case of scalar disorder \cite{kamenev2023field} but with $\tau\rightarrow\tau_{\Delta}$, so that we end up with
\begin{align}
    iS[Q] = -\frac{\pi\nu}{4}\text{Tr}\Big(\frac{D_{jk}}{4}\partial_j Q \partial_k Q + \Delta_{\text{SDW,y}}e^{i\phi\nu_3}\tau_3\sigma_z\nu_2 Q\Big)\;,
\end{align}
where $D_{jk} = \int_{0}^{2\pi} \frac{d\theta}{2\pi} v_{\text{F},j}(\theta)v_{\text{F},k}(\theta)\tau_{\Delta}$ and $v_{\text{F}}(\theta)$ is the magnitude of the Fermi velocity in the non interacting state.

The absence of the usual energy term implies that in the bulk the saddle point of this action is $e^{i\phi}\tau_3\sigma_z\nu_2$ for all energies, indicating that at any temperature for which the SDW along $x$ has local correlations only, the saddle point corresponds to SDW correlations with period along $y$. 

Moreover, because the action contains no time-derivative, within the quasiclassical approximation the saddle point is independent of energy. Thus, the density of states vanishes for any energy for which the low-energy theory can be used, and the excitation gap is defined by the maximum energy for which it can be used, $\tau_{\Delta}^{-1}$.
Thus, because the spin-density wave along the x-direction loses its long-range order, the two are not competing anymore. Instead, the local $\Delta_{\text{SDW,x}}$ correlations help the $\Delta_{\text{SDW,y}}$ correlations to develop, by shielding it from the excitations of quasiparticles.

However, since this order is also a spin-density wave, inhomogeneity of its phase is determined by the same mechanisms as for the dominant order.
Thus, in a more realistic treatment, we should start from an action in which also this type of order loses its mean-field characteristics and satisfies
\begin{align}
    \langle \Delta_{\text{SDW,y}}(r)\Delta_{\text{SDW,y}}(r')\rangle \sim |\Delta^{0}_{\text{SDW}}|^2e^{-\frac{|r-r'|}{\xi_{\text{V}}}} \approx \pi\nu\tau_{\Delta}\delta(r-r')\;.
\end{align}
This is what we do in the next section.

\subsection{Two short-range orders}\label{sec:Twoshortranged}
With two disorder terms we can follow a similar procedure as before. Again, we can ignore any four-fermion terms that do not contribute to one of the Hubbard-Stratonovich terms, so that we obtain
\begin{align}
    Z &= \int \mathcal{D}\hat{\Delta}_{\text{SDW,x}}\int \mathcal{D}\hat{\Delta}_{\text{SDW,y}}\int \mathcal{D}\Psi\int\mathcal{D}\bar{\Psi} \;\text{Exp}\Big\{i\Bar{\Psi}(\tau_3\mu_3\nu_3 \hat{\omega}_{t,t'} + \xi + \Delta _{\text{SDW},y} e^{i\phi\nu_3}\tau_3\sigma_z\nu_2)\Psi \nonumber\\& + \frac{i\nu}{\lambda_{\text{sy}}}|\Delta_{\text{SDW,y}}|^2\Big\}\;\text{Exp}\{-\sum_m\frac{1}{2\pi\nu\tau_{\Delta}}\bar{\Psi}(t)\tilde{P}_m\Psi(t) \bar{\Psi}(t')\tilde{P}_m\Psi(t')\}\;, \label{eq:extraSDWaction}
\end{align}
where $\tilde{P} = 1...\tilde{N}_e$ are now the projectors on the shared eigenspaces of $\mu_{1,2}\tau_3\sigma_z, \nu_{1,2}\tau_3\sigma_z$.

We now decouple this using a $Q$-matrix in the same way as before, but where $Q$ is now constrained further, because long-range modes are now required to be preserved on scattering off the phase disorder of both types of spin-density waves. The soft-mode manifold therefore now has the restriction.
\begin{align}
    Q = \mu_{1}\tau_3\sigma_zQ\mu_{1}\tau_3\sigma_z= \mu_{2}\tau_3\sigma_zQ\mu_{2}\tau_3\sigma_z = \nu_{1}\tau_3\sigma_zQ\nu_{1}\tau_3\sigma_z= \nu_{2}\tau_3\sigma_zQ\nu_{2}\tau_3\sigma_z\;.\label{eq:ManifoldConstraints}
\end{align}

After integrating out the fermions using the Gaussian integration rules for Grassmann variables, the action now reads
\begin{align}
    Z = \int \mathcal{D}Q \;e^{iS_P[Q]}\;,
\end{align}
where $S_P[Q]$ is given by
\begin{align}
    iS_P[Q] = -\frac{\pi\nu}{2\tau_{\Delta}}\text{Tr}(Q^2) + \text{Tr ln}\Big((\tau_3\mu_3\nu_3 \hat{\omega}_{t,t'} + \xi +2\frac{i}{\tau_{\Delta}}Q)\Big)\;.\label{eq:QeqTrln}
\end{align}
As was the case for Eq.~(\ref{eq:iSQ}), this action is valid for a large range of materials, but in many cases is utilize as such, and therefore we proceed in the quasiclassical limit. 
Again, we find the normalization by taking the variation of Eq.~(\ref{eq:QeqTrln}) with respect to $Q$, obtaining
\begin{align}
    0 = \frac{\pi\nu}{\tau_{\Delta}}\Big(-Q + \frac{i}{\pi\nu}(\tau_3\mu_3\nu_2\hat{\omega}_{t,t'} + \xi + \frac{2i}{\tau_{\Delta}} Q)^{-1}\Big)_{r' = r}\;.
\end{align}
This is the same type of expression as before, so that the soft-mode manifold still is characterized by
\begin{align}
    Q^2 = \mathbf{1}\;.\label{eq:Normalizationfinalmanifold}
\end{align}

Like before, the time-derivative term vanishes from the first order expansion due to the symmetries of $Q$, which means that in similarity with the approach in Sec.~\ref{sec:Decouplingthings}, but recognizing that $\tau_{\Delta}^{-1}\rightarrow 2\tau_{\Delta}^{-1}$, the action in the absence of any additional interactions becomes
\begin{align}
    iS[Q] = -\frac{\pi\nu D_{\Delta}}{4}\text{Tr}(\nabla Q)^2\;.
\end{align}

An interesting feature of this action is that for bulk systems, the action is independent of $Q$, so that all $Q$-matrices on the soft-mode manifold are saddle points. Thus, without any further instabilities, the saddle point approximation cannot be used, and there is no long-range order. We interpret this as a pseudo-gap phase.

\subsection{Structure of the low-energy modes and instabilities}\label{sec:manifoldstructure}
As we have seen in the treatment of the subdominant spin density wave, the remaining local order helps the long-range order by shielding it from quasiparticle excitations. This also happens to the phase that develops from the interactions with $|q|\ll \mathcal{Q}$ in Eq.~(\ref{eq:Interactionsdecomposedinmomenta}).

To this end, we first consider, which phases can develop. In Sec.~\ref{sec:Twoshortranged} we have identified the soft-mode manifold describing the low-energy degrees of freedom via the constraints 
$Q^2 = \mathbf{1}$ (Eq.~\ref{eq:Normalizationfinalmanifold})),
the charge conjugation symmetry defined in Eq.~(\ref{eq:ChargeConjugationQ}) , and $Q = BQB$, where following Eq.~(\ref{eq:ManifoldConstraints}) $B\in\{\mu_{1,2}\tau_3\sigma_z, \nu_{1,2}\tau_3\sigma_z\}$ describe the dominant short-range spin-density wave degrees of freedom. Any low-energy mode must preserve those constraints.

Exploiting the restrictions imposed by the short-range dominant modes, we see that the constrained manifold can be described using four by four matrices squaring to the identity matrix and parameterized by the Pauli matrices
\begin{align}
    \tilde{\tau}_{0,3} &:= \tau_{0,3}\;,\\
    \tilde{\tau}_{1,2} &:= \tau_{1,2}\mu_3\nu_3\;,\\
    \tilde{\sigma}_{0,3} &:= \sigma_{0,3}\;,\\
    \tilde{\sigma}_{1,2} & := \sigma_{1,2}\mu_3\nu_3\;.\label{eq:newPauli}
\end{align}
Inserting these matrices in the charge conjugation constraint, Eq.~(\ref{eq:ChargeConjugationQ}), we obtain the charge constraint on the matrices $\tilde{Q}$ in this space:
\begin{align}
    \tilde{Q} = \rho_1 \tilde{\tau}_1\tilde{\sigma}_2 \tilde{Q}^{T}\tilde{\sigma}_2\tilde{\tau_1}\rho_1\;.
\end{align}
This is exactly the same charge conjugation constraint that appears in the usual Nambu-spin sigma model. Therefore, the structure of the soft-mode manifold, and thus the low-energy theory, is exactly the same as for the usual Usadel equation. This means that there are only a few types of orders that can appear at the zeroth order of quasiclassics, that is without a derivative, or without being odd in energy:
\begin{align}
    &\vec{h}\cdot\vec{\tilde{\sigma}}\tau_3Q\;,&\qquad
    \Delta_{\text{d}} e^{i\phi_{\text{d}} \tilde{\tau_3}}\tilde{\tau}_2\;.
\end{align}
The difference with the standard manifold is that these terms do not correspond to the usual exchange field and \textit{s}-wave superconductivity, because the modified Pauli matrices $\tilde{\tau},\tilde{\sigma}$ have a non-trivial structure in band space, following Eq.~(\ref{eq:newPauli}), while also the spinors have a non-trivial structure, as defined in Sec.~\ref{sec:Spinors}. Indeed, we have
\begin{align}
    \frac{1}{2}\text{tr}\langle\tau_3\sigma_3 \Psi\bar\Psi\rangle = \text{tr}(\frac{\mathbf{1} + \tau_3\sigma_3}{2}) = \langle \psi^{\dagger}_{1\uparrow}\psi_{1\uparrow}-\psi^{\dagger}_{1\downarrow}\psi_{1\downarrow}\rangle - \langle \psi^{\dagger}_{2\uparrow}\psi_{2\uparrow}-\psi^{\dagger}_{2\downarrow}\psi_{2\downarrow}\rangle -
\langle \psi^{\dagger}_{3\uparrow}\psi_{3\uparrow}-\psi^{\dagger}_{3\downarrow}\psi_{3\downarrow}\rangle+\langle \psi^{\dagger}_{4\uparrow}\psi_{4\uparrow}-\psi^{\dagger}_{4\downarrow}\psi_{4\downarrow}\rangle\;,\end{align}
which means that $h_z$ corresponds to a momentum-even spin splitting that is opposite on one pair of Fermi surfaces compared to the other. Such state has the symmetry of an altermagnet \cite{smejkal2022emerging} and a spin-dependent \textit{d}-wave Pomeranchuk instability \cite{pomeranchuk1958stability}.

On the other hand, 
\begin{align}
    \frac{1}{2}\text{tr}\langle\tau_3\sigma_x\mu_3\nu_3\Psi\bar\Psi\rangle = \sum_{i = 1}^4 \langle \psi^{\dagger}_{i\uparrow}\psi_{i\downarrow}+\psi^{\dagger}_{i\downarrow}\psi_{i\uparrow}\rangle\;,
\end{align}
indicating that $h_{x}$, and by similarity also $h_y$, correspond to ferromagnetic terms.

Lastly, there is a superconducting order, which corresponds to
\begin{align}
    \frac{1}{2}\text{tr}\langle(\tau_1-i\tau_2)\Psi\bar{\Psi}\rangle = \langle\psi_{1\uparrow}\psi_{4\downarrow}-\psi_{1\downarrow}\psi_{4\uparrow}\rangle + \langle\psi_{4\uparrow}\psi_{1\downarrow}-\psi_{4\downarrow}\psi_{1\uparrow}\rangle -\langle\psi_{2\uparrow}\psi_{3\downarrow}-\psi_{2\downarrow}\psi_{3\uparrow}\rangle-\langle\psi_{3\uparrow}\psi_{2\downarrow}-\psi_{3\downarrow}\psi_{2\uparrow}\rangle\;.
\end{align}
The minus signs within each expectation value indicate that the order parameter is even-parity, singlet. However, the minus sign in the last two terms indicates it has opposite sign on the two pairs of Fermi surfaces. This means it has \textit{d}-wave symmetry. 

Thus, the resulting soft-mode manifold does not contain \textit{s}-wave superconductivity. Instead, it hosts \textit{d}-wave superconductivity, as one of its only few instabilities. In the next part we show how to describe this \textit{d}-wave superconductivity within the NLSM.
\subsection{\textit{d}-wave superconductivity}
Thus, the only type of superconductivity compatible with a  SDW order without phase coherence is \textit{d}-wave superconductivity. This possibility is intriguing for  different   reasons. First of all, there is experimental evidence of \textit{d}-wave superconductivity appearing near spin-density wave order in cuprates. Secondly, it is sensitive to different correlations than SDW order, since for SDW orders the main contributions are from $q = \pm \mathcal{Q}\hat{x}$ and $q = \pm \mathcal{Q}\hat{y}$, while for superconductivity, the main contributions are from $k = -k'$, without restriction on $q$. We can thus still consider these contributions from the four-fermion term and its decoupling without it being absorbed as a renormalization of the SDW parameter. Thirdly, we are mostly interested in an order that, despite losing to the spin-density wave order in the initial mean field, maintains phase coherence better than the SDW, and superconductivity is a perfect candidate for this.

Indeed, the spin-density wave order is sensitive to the location of the spin within the unit cell:  its phase corresponds to this position. A single dopant shifts this position and thus  leads to a vortex of the SDW order. The superconductor however, does not depend on where in the unit cell the electrons are located, only on the electrons forming a pair. For this reason, the presence of dopants does not immediately lead to a vortex of the superconductor order. It does reduce the amplitude of the superconductor order parameter if  it leads to interband scattering. Formally, the U(1) gauge field  that couples to the phase of the spin-density waves and thus to its vortices, is proportional to $\mu_3$ in the band-Nambu-spin-Keldysh space, and thus commutes with the \textit{d}-wave order parameter $\Delta_{\text{d}} e^{i\phi\tau_3}\tau_2\mu_3\nu_3$.
In fact, vortices in superconductors arise via magnetic flux, a completely different mechanism.

There is also a momentum space picture that can be used to understand the difference in reduction of the phase coherence. The SDW order depends on the coherence between the different components of the Fermi surface, and is thus strongly affected by anything that disturbs this coherence. On the other hand, \textit{d}-wave superconductivity does not require coherence between different components, it only requires them to be different.

Lastly, superconductors are charged, so that the dynamics of the phase of the pair potential couple to the Maxwell equations, and the Goldstone mode gets pushed to the plasmon frequency. The SDW mode however is charge-neutral, and is therefore much easier excited than that of a charged superconductor.

Now, let us consider the case in which the four-fermion term corresponds to an attractive interaction for $k \approx -k'$, which appears for example due to phonon attractions, and let us focus on its \textit{d}-wave component. Such \textit{d}-wave channel is always present if the material does not have $C_4$ symmetry.

We can decouple this interaction through the \textit{d}-wave superconductivity channel similar to how it is usually done for the \textit{s}-wave case \cite{kamenev2023field}, and the resulting action reads
\begin{align}
    Z &= \int\mathcal{D}(\Delta_{\text{d}} e^{i\phi_{\text{d}}})\int\mathcal{D}Q\; e^{iS_{\text{s}}[Q,\Delta_{\text{d}} e^{i\phi_{\text{d}}}]}\;,\\
    iS_{\text{s}} &= -\frac{\pi\nu}{2\tau_{\Delta}}\text{Tr}(Q^2) + \text{Tr ln}\Big((\tau_3\mu_3\nu_3 \hat{\omega}_{t,t'} + \xi + i\Delta_{\text{d}} e^{i\phi_{\text{d}}}\tau_2\mu_3\nu_3 +2\frac{i}{\tau_{\Delta}}Q)\Big) + \frac{i\nu}{\lambda_{\text{d}}}\hat{\Delta}_{\text{d}}^{\dagger}\rho_1\hat{\Delta}_{\text{d}}\;,\label{eq:Ss}
\end{align}
where $\lambda_{\text{d}} = \nu\int dq\; V(q) \frac{q_x^2-q_y^2}{q_x^2+q_y^2}$, where the factor $\frac{q_x^2-q_y^2}{q_x^2+q_y^2}$ is used to select \textit{d}-wave symmetry.
Expanding in $Q$ gives
\begin{align}
    iS[Q] = \frac{\pi\nu}{4}\text{Tr}\Big(-\frac{D}{8}(\nabla Q)^2 + \Delta_{\text{d}} e^{i\phi_{\text{d}}\tau_3}\tau_2\mu_3\nu_3Q+ \frac{i\nu}{\lambda_{\text{d}}}\hat{\Delta}_{\text{d}}^{\dagger}\rho_1\hat{\Delta}_{\text{d}}\Big)\;,
\end{align}
which in terms of the reduced matrices reads
\begin{align}
    iS[\tilde{Q}] = \pi\nu\text{Tr}\Big(-\frac{D}{8}(\nabla Q)^2 + \Delta_{\text{d}} e^{i\phi_{\text{d}}\tilde{\tau}_3}\tilde{\tau_2}Q+ \frac{i\nu}{\lambda_{\text{d}}}\hat{\Delta}_{\text{d}}^{\dagger}\rho_1\hat{\Delta}_{\text{d}}\Big)\;,
\end{align}

Like we saw in Sec.~\ref{sec:quasiclassicalexpansion} for the subdominant SDW order, there is no term in the action that competes with the \textit{d}-wave order due to the absence of the usual time-derivative / Matsubara term, indicating that within the mean field approximation, the superconductivity can exist as long as $\Delta_{\text{SDW}}^{0}$ is nonzero, even for very weak coupling.

Thus, a short-range spin-density wave order can greatly enhance the critical temperature of the superconductor by stabilizing a gap for quasiparticles due to their insulating nature. This situation is completely different compared to the original mean field treatments, in which different orders are competing and thus suppress each others critical temperature. The main reason for this is that the long-range correlations of the dominant order parameter are lost, allowing for the subdominant order to be present, while the remaining local order opens a quasiparticle gap and thus shields the subdominant order parameter from thermal excitations.

The different nature of the contributions to depairing and vortices subsequently allows for the superconductor order to retain its long-range order under dopants in contrast to the SDW orders, so that on a mesoscopic scale the superconductor behaves as a \textit{d}-wave superconductor. 

\subsection{Easy-plane spin-density waves}
In this section, we elaborate on the soft-mode manifold where the spin-density wave does not have an easy axis in spin space, but an easy-plane, which we denote as the $xy$-plane. In this case, there are short-range order potentials $\Delta_B B$. Here $B$ describes density waves in the $x$-direction ($\mu_{1,2}$) as well as the y-direction ($\nu_{1,2}$), with spins along the $x$ ($\sigma_x$) or $y$ ($\sigma_y$) direction. This gives eight different possibilities,
\begin{align}
    B \in\{\mu_1\tau_3\sigma_{x},\mu_2\tau_3\sigma_{x},\nu_1\tau_3\sigma_{x},\nu_2\tau_3\sigma_{x},\mu_1\tau_3\sigma_{y},\mu_2\tau_3\sigma_{y},\nu_1\tau_3\sigma_{y},\nu_2\tau_3\sigma_{y}\}\;.
\end{align}
The only contributions to $\Psi\bar{\Psi}$ that survive scattering off all possible short-range orders, that is, all Hubbard-Stratonovich transformations, are those that commutes with all of these $B$-matrices. Thus, next to the usual charge conjugation symmetry, the soft-mode satisfies the eight constraints that can be written as
\begin{align}
    Q = BQB\;.
\end{align}
The only matrices in band-Nambu-spin space that satisfy this property are the linear combinations of $\tau_{0,3}$, $\tau_{1,2}\mu_3\nu_3$,  $\tau_{0,3}\sigma_z \mu_3\nu_3$ and $\tau_{1,2}\sigma_z$. This manifold is identical to the restriction of the easy-axis soft-mode manifold to those matrices commuting with $\tilde{\sigma}_z$, that is, to the combination of two spinless Keldysh-Nambu spaces related through charge conjugation. In almost exact analogy with the easy-axis case, the action containing all interaction instabilities and all possible relaxation mechanisms is
\begin{align}
    iS[Q] = \frac{\pi\nu}{2}\text{Tr}\Big(-D_{\Delta}(\nabla Q)^2 + \Delta_{\text{d}} e^{i\phi\tilde{\tau}_3}\tilde{\tau}_2Q + ih\tilde{\sigma}_z \tilde{\tau_3}-\frac{i}{\lambda_{\text{d}}}\Delta_{\text{d}}^{\dagger}\rho_1 \Delta_{\text{d}} -\frac{i}{\lambda_h}h\rho_1h \Big)+ iS_{\text{r}}\;.
\end{align}
Here $S_{\text{r}}$ contains all relaxation processes. To lowest order, the elastic relaxation processes can be summarized as
\begin{align}
    iS_{\text{r}} = \frac{\pi\nu}{2}\text{Tr}\Big(\Gamma_{\text{inter}}\tilde{\tau}_3Q\tilde{\tau}_3Q + \Gamma_{\text{s,soc}}\tilde{\sigma}_z Q\tilde{\sigma_z}Q + \Gamma_{\text{s,mag}}\tilde{\tau}_3\tilde{\sigma}_zQ\tilde{\tau}_3\tilde{\sigma}_z Q\Big)\;.
\end{align}
$\Gamma_{\text{inter}}$ is the interband scattering rate, $\Gamma_{\text{s,soc}}$ the contribution of spin-orbit coupling to spin-relaxation and $\Gamma_{\text{s,mag}}$ is the contribution of magnetic disorder to spin-relaxation.

The difference compared to the case of easy-axis SDW is that for the easy-plane SDW, \textit{d}-wave superconductivity competes with only one type of magnetic order (a ferromagnetic one), instead of three. Moreover, fluctuations are suppressed because the $Q$-matrix is restricted to commute with $\tilde{\sigma}_Z$, thus suppressing those fluctuations away from the saddle point that do not commute. This reflects a general principle: if the short-ranged order has more degrees of freedom, the soft-mode manifold has less and the remaining shielded orders are more stable.
 
\section{Derivation of critical temperature}\label{sec:Tc}

As derived in Sec.~\ref{sec:Actiondetails}, in the NLSM action for a system with phase disorder the first order time-derivative drops completely out of the equations. However, since the NLSM has a  structure that is equivalent to that of the standard Nambu-spin-Keldysh NLSM, as derived in Sec. \ref{sec:manifoldstructure}, we know that charge conjugation permits a term of the form $\hat{\omega}_{t,t'}\tilde{\tau}_3 Q$ to appear in the NLSM, and it should thus be possible to obtain this term from a microscopic model. To obtain this term, we focus on the effect of $\eta$, introduced in Eq.~(\ref{eq:HamiltonianSM}) describing the deviation from perfect Fermi surface nesting.

For nonzero $\eta$, there is an imbalance between the density of states in the electron pocket and the hole pocket. We show that, even after the formation of the short-range spin density wave, this imbalance results in a first order time-derivative. This drives $Q$ towards the normal state, with carriers of the type that has a larger density of states in the normal state.
Subsequently, we will derive the critical temperature for the superconducting state, and show that it is much higher than expected based on its own pairing strength.

For the quasiclassical approximation to hold, we require $\eta\ll 1$. Adding this to the dispersion in Eq.~(\ref{eq:Ss}), we obtain the generating function that describes the system,
\begin{align}
    Z &= \int\mathcal{D}(\Delta_{\text{d}} e^{i\phi_{\text{d}}})\int\mathcal{D}Q\; e^{iS_{\text{s}}[Q,\Delta_{\text{d}} e^{i\phi_{\text{d}}}]}\;,\\
    iS_{\text{s}} &= -\frac{\pi\nu}{2\tau_{\Delta}}\text{Tr}(Q^2) + \text{Tr ln}\Big((\tau_3\mu_3\nu_3 \hat{\omega}_{t,t'} + \xi + \eta\xi \nu_3\mu_3 + \hat{\Delta}_{\text{d}} e^{i\phi_{\text{d}}}\tau_2\mu_3\nu_3 +2\frac{i}{\tau_{\Delta}}Q)\Big) + \frac{i\nu}{\lambda_{\text{d}}}\hat{\Delta}_{\text{d}}^{\dagger}\rho_1\hat{\Delta}_{\text{d}}\;.\label{eq:Ssren}
\end{align}
Like before, the notation $\int\mathcal{D}\cdot$ is used to denote path integrals, which are separately taken over the $Q$-matrix and the pair potential $\Delta_{\text{d}} e^{i\phi_{\text{d}}}$.

We expand the action up to first order in $\eta$, and derive its contribution to the soft-mode action by exploiting the rotation matrices, i.e. $Q = \mathcal{R}^{-1}\Lambda\mathcal{R}$ \cite{kamenev2023field}, where where $\Lambda$ is a time-independent natural saddle point. The choice of $\Lambda$ will not affect the final result and will thus be kept general.

We are interested in the effect of $\eta$ on the time-derivative term, which is calculated through \cite{kamenev2023field}
\begin{align}
    iS_t =\text{Tr}\Big(\mathcal{G}\mathcal{R}[\hat{\omega}\tau_3\mu_3\nu_3,\mathcal{R}^{-1}]\Big)\;,
\end{align}
where, because all $Q$-matrices commute with $\mu_3\nu_3$ we write $\mathcal{G} \approx \frac{1}{\epsilon-\xi - \frac{2i}{\tau_{\Delta}}}(1+\Lambda) + \frac{1}{\epsilon-\xi + \frac{2i}{\tau_{\Delta}}}(1-\Lambda)+\frac{1}{(\epsilon-\xi-\frac{2i}{\tau_{\Delta}})^2}\mu_3\nu_3(1+\Lambda)+\frac{1}{(\epsilon-\xi+\frac{2i}{\tau_{\Delta}})^2}\mu_3\nu_3(1-\Lambda)$.

We know that for $\eta =  0$, the resulting expression vanishes and therefore focus on the contributions first order in $\eta$. Exploiting that everything commutes with $\mu_3\nu_3$, and that $(\mu_3\nu_3)^2 = \mathbf{1}$, we find
\begin{align}
    iS_t = \eta\Big(\int d\xi\;\frac{\xi}{(\epsilon-\xi+\frac{2i}{\tau_{\Delta}})^2}\Big)\text{Tr}\Big((1+\Lambda)\mathcal{R}[\hat{\omega}\tau_3,\mathcal{R}^{-1}] \Big)+ \eta\Big(\int d\xi\;\frac{\xi}{(\epsilon-\xi-\frac{2i}{\tau_{\Delta}})^2}\Big)\text{Tr}\Big((1-\Lambda)\mathcal{R}[\hat{\omega}\tau_3,\mathcal{R}^{-1}]\Big)\;.
\end{align}
Using that 
\begin{align}
    \int d\xi\frac{\xi}{(\epsilon-\xi+\frac{2i}{\tau_{\Delta}})^2} = -\int d\xi\;\frac{\xi}{(\epsilon-\xi-\frac{2i}{\tau_{\Delta}})^2} = -i\pi \nu\;,
\end{align}
we see that is nothing else than the usual time-derivative in the standard NLSM that appears in Keldysh-Nambu-spin space, with an extra prefactor $-\eta$.

With this, we find that the action reads, in the reduced matrix space
\begin{align}
    iS_r[Q] = \pi \nu \text{Tr}\Big(-\frac{D}{8}(\nabla \tilde{Q})^2 + \eta\hat{\omega}\tilde{\tau}_3\tilde{Q} + \Delta_{\text{d}} e^{i\phi_{\text{d}}\tilde{\tau}_3}\tilde{\tau_2}Q+ \frac{i\nu}{\lambda_{\text{d}}}\hat{\Delta}_{\text{d}}^{\dagger}\rho_1\hat{\Delta}_{\text{d}}\Big)\;.
\end{align}

This action is almost entirely the same as that for a convential single-band \textit{s}-wave superconductor, only with a scalar prefactor in the time-derivative term, and because the structure of the soft-mode manifold is identical to that for the standard Keldysh-Nambu-spin space, the derivation of the critical temperature can proceed in almost identical way. In a bulk superconductor, using the Matsubara representation, $\hat{\omega}\rightarrow \omega_n = (2n+1)\pi T$, the saddle point equation for $Q$ gives the space homogeneous Usadel equation.

\begin{align}
    0 = [\eta\omega_n\tilde{\tau}_3 + \Delta_{\text{d}} \tilde{\tau}_2,Q]\;.
\end{align}

Exploiting the normalization condition $Q^2 = \mathbf{1}$, the solution to this equation is
\begin{align}
    \tilde{Q}_{\text{sp}} = \frac{1}{\sqrt{\left(\eta\omega_n\right)^2+\Delta_{\text{d}}^2}}(\eta\tilde{\tau}_3 + \Delta_{\text{d}}\tilde{\tau}_2)\;,\label{eq:Qsp1}
\end{align}
The saddle point relation for $\Delta_{\text{d}}$ gives us the self-consistency relation, which in Matsubara space takes the standard form
\begin{align}
    \Delta_{\text{d}} = \lambda_{\text{d}}\pi T \sum_{\omega}\text{tr}\Big(\frac{\tau_2-i\tau_1}{2}Q\Big)\;.\label{eq:selfconsistency}
\end{align}
Substituting Eq.~(\ref{eq:Qsp1}) into Eq.~(\ref{eq:selfconsistency}), we find
\begin{align}
    \Delta_{\text{d}} = \lambda_{\text{d}} \pi T \sum_{n = -\frac{\tau_{\Delta}^{-1}}{\pi T_c}}^{\frac{\tau_{\Delta}^{-1}}{\pi T_c}} \frac{\Delta_{\text{d}}}{\sqrt{\left(\eta\omega_n\right)^2+\Delta_{\text{d}}^2}}\;.
\end{align}
Here $\tau_{\Delta}^{-1}$ serves as a natural cut-off frequency, above this energy the high-energy modes start to play an important role and Eq.~(\ref{eq:Qsp1}) cannot be used anymore, and thus the pair amplitudes are strongly suppressed for those energies. If the pairing mechanism has an intrinsic cut-off frequency $\omega_{\text{max}}$, e.g. the Debye frequency, $\tau_{\Delta}^{-1}$ has to be replaced with $\text{min}(\tau_{\Delta}^{-1},\omega_{\text{max}})$. The critical temperature is now determined by
\begin{align}
    1 = \lambda_{\text{d}} \pi T_c \sum_{n = -\Big\lceil\frac{\tau_{\Delta}^{-1}}{\pi T_c}\Big\rceil}^{\Big\lceil\frac{\tau_{\Delta}^{-1}}{\pi T_c}\Big\rceil} \frac{1}{\eta\pi T_c (2n+1)} = \lambda \frac{1}{\eta} \sum_{n = -\Big\lceil\frac{\tau_{\Delta}^{-1}}{\pi T_c}\Big\rceil}^{\Big\lceil\frac{\tau_{\Delta}^{-1}}{\pi T_c}\Big\rceil} \frac{1}{2n+1}\;.
\end{align}
This is entirely similar to the usual BCS relation, bu with an enhanced interaction strength $\lambda\rightarrow \lambda \frac{1}{\eta}$. Due to this, the critical temperature reads
\begin{align}
    T_c &\approx \tau_{\Delta}^{-1} e^{-\frac{\eta}{\lambda_{\text{d}}}}\;.
\end{align}
Thus, the factor in the exponential is a factor $\eta$  smaller than for the standard BCS result.  This leads to a much larger critical temperature  than expected from the pairing interaction $\lambda$ only. This also shows that the superconducting order survives up to the  temperatures of the same order as the remaining short-range density wave order. In fact, for $\eta\ll 1$, $T_c$ is close to $\tau_{\Delta}^{-1}$.

Since, the high critical temperature, and the large excitation gap are determined by the dominant order, they do not necessarily imply the pair potential itself is also large. In the zero temperature limit, the Matsubara sum that appears in the Matsubara sum can be converted into an integral, $2\pi T\sum_{\omega}\xrightarrow{}\int d\omega$, which can be evaluated analytically:
\begin{align}
    1 = \frac{1}{2}\lambda_{\text{d}} \int_{-\tau_{\Delta}^{-1}}^{\tau_{\Delta}^{-1}} d\omega\; \frac{1}{\sqrt{\left(\eta\omega\right)^2+\Delta_{\text{d}}^2}} = \lambda_{\text{d}} \frac{1}{2\eta}\int_{-\frac{\eta\tau_{\Delta}^{-1}}{\Delta_{\text{d}}}}^{\frac{\eta\tau_{\Delta}^{-1}}{\Delta_{\text{d}}}}du\;\frac{1}{\sqrt{1+u^2}} = \lambda_{\text{d}} \frac{1}{\eta}\text{asinh}\frac{\eta\tau_{\Delta}^{-1}}{\Delta_{\text{d}}}\;.
\end{align}
Solving this equation for $\Delta_{\text{d}}$, we find
\begin{align}
    \Delta_{\text{d}} = \eta\tau_{\Delta}^{-1} \frac{1}{\sinh{\frac{\eta}{2\lambda_{\text{d}}}}}\;.
\end{align}
For small $\frac{\eta}{\lambda_{\text{d}}}$ this reduces to 
\begin{align}\label{eq:ZeroTemperaturepotential}
    \Delta_{\text{d}} \approx \lambda_{\text{d}} \tau_{\Delta}^{-1}\;.
\end{align}
Interestingly, because $\lambda_{\text{d}}$ is usually small, $\Delta_{\text{d}}$ is much smaller than $\tau_{\Delta}^{-1}$, and therefore than $k_B T_c$. However, it is still much larger than following BCS predictions because of the absence of an exponential suppression. This cannot be observed directly in spectroscopical experiments, because $\Delta_{\text{d}}$ no longer corresponds to the quasiparticle gap. However, it is an indication that the pair potential is not as robust against relaxation mechanisms as would be expected from the critical temperature.
 
\section{The gapless density of states and the momentum dependence of the SDW state}\label{sec:NLSMwithangledependence}

In the derivation of the NLSM in Sec.~\ref{sec:Actiondetails}, we assumed that the Fermi surface contains two electron-like and two hole-like pockets, so that the entire Fermi surface participates in the formation of the insulating state. However, spin-density waves can also appear in materials with only electron-like or only hole-like pockets, through Fermi surface nesting. Fermi surface nesting has a strong angular dependence, some parts of the Fermi surface obey the nesting conditions, while other do not. 
In this section we introduce an approach to incorporate this angle dependence. We show that it is still possible to write the low-energy modes with the help of a nonlinear sigma model, but with coefficients that are weighted in a way that reflects the angle dependence of the dominant order. 

We show that also momentum dependent potentials can stabilize superconductivity to much higher temperatures than expected based on their attraction, and that it is possible to choose the angle dependence in a way such that the superconducting states are almost gapless. 

The spin-density wave along the x-direction should arise mostly due to those states that have $\vec{k} \sim \pm k \hat{x}$, where $\vec{k}$ is the momentum compared to the center of the pocket. Indeed, those are the only momenta for which $k$ and $k+Q$ or $k-Q$ are both on the Fermi surface. Moreover, this means that we do not need to assume anymore that $\xi(k)$ are opposite. In fact, because all pockets are of the same type, we consider them be the same. 

\begin{figure}
    \begin{tikzpicture}
          \draw[ultra thick, goodred] (-0.4,1) -- (0.4,1);
  \draw[ultra thick, black]
    (0.4,1) arc[start angle=90, end angle=0, radius=0.1];
  \draw[ultra thick, goodyellow] (0.5,0.9) -- (0.5,0.1);
  \draw[ultra thick, black]
    (0.5,0.1) arc[start angle=0, end angle=-90, radius=0.1];
  \draw[ultra thick, goodred] (0.4,0) -- (-0.4,0);
  \draw[ultra thick, black]
    (-0.4,0) arc[start angle=-90, end angle=-180, radius=0.1];
  \draw[ultra thick, goodyellow] (-0.5,0.1) -- (-0.5,0.9);
  \draw[ultra thick, black]
    (-0.5,0.9) arc[start angle=180, end angle=90, radius=0.1];
  \draw[ultra thick, goodred] (1.6,1) -- (2.4,1);
  \draw[ultra thick, black]
    (2.4,1) arc[start angle=90, end angle=0, radius=0.1];
  \draw[ultra thick, goodyellow] (2.5,0.9) -- (2.5,0.1);
  \draw[ultra thick, black]
    (2.5,0.1) arc[start angle=0, end angle=-90, radius=0.1];
  \draw[ultra thick, goodred] (2.4,0) -- (1.6,0);
  \draw[ultra thick, black]
    (1.6,0) arc[start angle=-90, end angle=-180, radius=0.1];
  \draw[ultra thick, goodyellow] (1.5,0.1) -- (1.5,0.9);
  \draw[ultra thick, black]
    (1.5,0.9) arc[start angle=180, end angle=90, radius=0.1];
  \draw[ultra thick, goodred] (-0.4,3) -- (0.4,3);
  \draw[ultra thick, black]
    (0.4,3) arc[start angle=90, end angle=0, radius=0.1];
  \draw[ultra thick, goodyellow] (0.5,2.9) -- (0.5,2.1);
  \draw[ultra thick, black]
    (0.5,2.1) arc[start angle=0, end angle=-90, radius=0.1];
  \draw[ultra thick, goodred] (0.4,2) -- (-0.4,2);
  \draw[ultra thick, black]
    (-0.4,2) arc[start angle=-90, end angle=-180, radius=0.1];
  \draw[ultra thick, goodyellow] (-0.5,2.1) -- (-0.5,2.9);
  \draw[ultra thick, black]
    (-0.5,2.9) arc[start angle=180, end angle=90, radius=0.1];
  \draw[ultra thick, goodred] (1.6,3) -- (2.4,3);
  \draw[ultra thick, black]
    (2.4,3) arc[start angle=90, end angle=0, radius=0.1];
  \draw[ultra thick, goodyellow] (2.5,2.9) -- (2.5,2.1);
  \draw[ultra thick, black]
    (2.5,2.1) arc[start angle=0, end angle=-90, radius=0.1];
  \draw[ultra thick, goodred] (2.4,2) -- (1.6,2);
  \draw[ultra thick, black]
    (1.6,2) arc[start angle=-90, end angle=-180, radius=0.1];
  \draw[ultra thick, goodyellow] (1.5,2.1) -- (1.5,2.9);
  \draw[ultra thick, black]
    (1.5,2.9) arc[start angle=180, end angle=90, radius=0.1];
 \node at (0,0.5){$1$};
        \node at (0,2.5){$3$};
        \node at (2,0.5){$2$};
        \node at (2,2.5){$4$};
    \draw[-{Circle[sep=-.45pt,width=2.04pt,length=2pt,black]Triangle[width=5pt,length=5pt,black]},ultra thick,black] (0.5,0.5)--(1.5,0.5);
    \draw[-{Circle[sep=-.45pt,width=2.04pt,length=2pt,black]Triangle[width=5pt,length=5pt,black]},ultra thick,black] (0,1)--(0,2);
    \node at (1,0.2){$\mathcal{Q}\hat{x}$};
    \node at (0.3,1.5){$\mathcal{Q}\hat{y}$};
    \node at (1,2.5){$Q_1$};
    \node at (2,1.5){$Q_2$};
    \end{tikzpicture}
    \caption{Four-component Fermi surfaces in which all the Fermi surfaces are the same. The shape of the Fermi surface is almost square to enhance nesting, in which large pieces of the Fermi surface are connected by the same momentum difference $\mathcal{Q}\hat{x}$ or $\mathcal{Q}\hat{y}$. $Q_1$ describes those electrons with momentum that satisfies with nesting condition along the $\vec{\hat{x}}$-direction, corresponding to the yellow parts of the Fermi surface, $Q_2$ describes those electrons with momentum that satisfies the nesting condition along the $\vec{\hat{y}}$-direction, corresponding to the red parts of the Fermi surface.}
    \label{fig:4Fermisurfacesallelectron}
\end{figure}
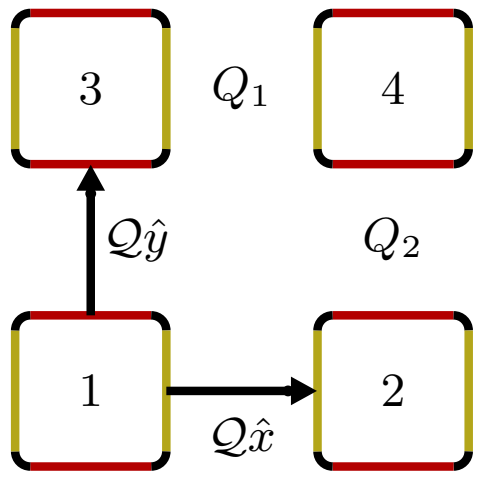

We start with four identical Fermi surfaces, illustrated in Fig.~\ref{fig:4Fermisurfacesallelectron}, and denote $\vec{k}$ as the momentum difference between a state and the center of the pocket, and let us denote the center of the pockets as $\vec{K}_{i}$, $i = 1...4$. Now, suppose the state with momentum $\vec{k}_F$ on Fermi surface 1 and $-\vec{k}_F$ on Fermi surface 2 differ by $\mathcal{Q}$. Then for a state just away from the Fermi surface with momentum $\vec{k}_F + \delta \vec{k}$ in pocket 1, the state that differs by $\mathcal{Q}$ is $-\vec{k}_F + \delta \vec{k} = -\vec{\bar{k}}$, where $\vec{\bar{k}}$ is the mirror image of $\vec{k}_F$ through the Fermi surface.

Since the pairing concerns $\vec{K}_1+\vec{k}$ and $\vec{K}_2-\vec{\bar{k}}$ , and similarly $\vec{K}_1-\vec{k}$ and $\vec{K}_2-\vec{\bar{k}}$ and their counterparts with $(3,4)$ instead of $(1,2)$, where $\vec{\bar{k}}$ is the mirror image of $\vec{k}$ through the Fermi surface, when considering the low-energy degrees of freedom it is most suitable to define the spinors in band space and momentum space as 
\begin{align}\Psi &= \big(\psi_{1}(\vec{k}),\psi_{2}(\vec{\bar{k}}),\psi_3(\vec{\bar{k}}),\psi_4(\vec{k})\big)^{T}\;,\nonumber\\\bar{\Psi} &= \big(\bar{\psi}_1(-\vec{k}),-\bar{\psi}_2(-\vec{\bar{k}}),-\bar{\psi}_3(-\vec{\bar{k}}),\bar{\psi}_4(-\vec{k})\big)\;.
\end{align}
Here we made the notation of the Nambu and spin structure implicit in order to avoid confusion. 

For $\vec{k}\approx \vec{k}_F + \delta \vec{k}$ we have $\bar{k} \approx \vec{k}_F-\delta \vec{k}$ and thus $\xi(\vec{k}) \approx \vec{v}_F\cdot\delta \vec{k} = -\xi(\vec{\bar{k}})$, indicating that for the constructed $\Psi$ and $\bar{\Psi}$, $\xi$  enters as a term proportional to the identity matrix, as is convenient when describing the soft-mode manifold. Moreover, the charge conjugation relation is the same as in Eq.~(\ref{eq:ChargeConjugationPsibarPsi}), while the time-derivative still contains $\tau_3\mu_3\nu_3$ because $\bar{\Psi} = \Psi^{\dagger}\tau_3\mu_3\nu_3$. Therefore, in many aspects, the NLSM is similar to the one discussed before.

However, there is an important difference between the two models. The matching condition $\vec{K}_1 +(-) \vec{k}_F \approx \vec{K}_2-(+)\vec{k}_F$ only holds for specific momenta, and therefore the strength of the spin-density wave potential  depends strongly on the angle on the Fermi surface.

To describe this, we introduce
\begin{align}
    \Delta_{\text{SDW,x}}(\theta) = \Delta_{\text{SDW,1}} f_1(\theta)\;,\\
    \Delta_{\text{SDW,y}}(\theta) = \Delta_{\text{SDW,2}} f_2(\theta)\;,
\end{align}
where $\theta$ is the angle of $\vec{k}$ with respect to the x-axis. The functions $f_{1,2}$ are assumed to be real, even in momentum and positive functions that satisfy a normalization
\begin{align}
    \int_{0}^{2\pi}\frac{d\theta}{2\pi}\; f_{1,2}^2(\theta)  = 1\;,
\end{align}
The functions $f_{1(2)}(\theta)$ are not universal, but depend on the angle between $\mathcal{Q}$ and $\vec{k}-\bar{\vec{k}}$ over the Fermi surface. They are largest at those part of the Fermi surface that satisfy the nesting condition, that is, $f_{1}$ is maximum at momenta in the yellow parts of the Fermi surface in Fig.~\ref{fig:4Fermisurfacesallelectron}, while $f_{2}(\theta)$ is maximum at momenta in the red parts. We assume that no parts of the Fermi surface satisfy two nesting conditions at the same time, that is, $\int_{0}^{2\pi} \frac{d\theta}{2\pi}f_{1}(\theta)f_{2}(\theta)\ll 1$. The calculation of these functions thus requires to study a specific material and go beyond quasiclassics. Here we keep them generic.

To illustrate how we incorporate the momentum dependence, it is instructive to first consider how the random SDW potential discussed in the Sec.~\ref{sec:Actiondetails} appears in momentum space. For an impurity at location $r$, we have
\begin{align}
    \bar{\Psi}(r)\Delta_{\text{SDW}} \tau_3\mu_2\sigma_z \Psi(r) = \int dk \int dk'\; e^{ik r} e^{-ik' r} \bar{\Psi}(k')\Delta_{\text{SDW}} \tau_3\mu_2\sigma_z \Psi(k)\;.
\end{align}
If we descretize momentum space, we obtain,
\begin{align}
    \bar{\Psi}(r)\Delta_{\text{SDW}}(r) \tau_3\mu_2\sigma_z \Psi(r) = \sum_{k,k'}\bar{\Psi}_{k'} \mathcal{P}^s_{\vec{k'k}}\tau_3\mu_2\sigma_z\Psi_{k}\;,\\
    \mathcal{P}^s_{\vec{k'k}}(r) = \frac{1}{N_k} e^{-i\vec{k'}\cdot\vec{r}}e^{i\vec{k}\cdot\vec{r}}\;,
\end{align}
where $N_k$ is the number of points in the discretization. This is a projector. Indeed, $\mathcal{P}^s_{k'k}$ has rank 1, it is the outer product of a vector with itself. Meanwhile, the locality is reflected by the magnitude of the entries not decaying with $|\vec{k}-\vec{k}'|$.

Thus, to describe the random spin-density wave potential whose magnitude depends on momentum, what we need is a projector that encodes a non-trivial momentum dependence.  We can base them on $f_{1,2}$ and define, 
\begin{align}
    \mathcal{P}^{1(2)}_{\vec{k'k}}(r) = \frac{1}{N_k} f_{1(2)}(\theta')e^{-i\vec{k'}\cdot\vec{r}}f_{1(2)}(\theta)e^{i\vec{k}\cdot\vec{r}}\;.
\end{align}
This is still rank 1 as it is still an outer product of two vectors, the normalization of $f_{1,2}$ ensures it is properly normalized, while it encodes the momentum dependence. Moreover, the local character appears because the magnitude only depends on the direction of $\vec{k}$ and $\vec{k}'$, not on their magnitudes.  Let us consider the effect of having a random, spatially uncorrelated potential such a momentum space projection instead of the standard one.

The action for this type of system is
\begin{align}
     Z &= \int \mathcal{D}\hat{\Delta}_{\text{SDW,x}}\int \mathcal{D}\hat{\Delta}_{\text{SDW,y}}\int \mathcal{D}\Psi\int \mathcal{D}\bar{\Psi}\;e^{iS_2}\\
   iS_2 &=  \text{Tr}i\Bar{\Psi}(\tau_3\mu_3\nu_3 \hat{\omega}_{t,t'} + \xi + i\Delta _{\text{SDW},x}\mathcal{P}_1(r) e^{i\phi\mu_3}\tau_3\sigma_z\mu_2 + i\Delta _{\text{SDW},y}\mathcal{P}_2(r) e^{i\phi\nu_3}\tau_3\sigma_z\nu_2)\Psi \nonumber\\& + \frac{1}{4\pi\nu\tau_{\Delta}}\Delta_{\text{SDW,x}}|^2 +  \frac{1}{4\pi\nu\tau_{\Delta}}|\Delta_{\text{SDW,y}}|^2\Big\}\;. \label{eq:SDWactions2}
\end{align}

Now, we can proceed to find the soft-mode manifold and the NLSM using the same strategy as we have used in Sec~\ref{sec:Actiondetails}. Specifically, we do a Hubbard-Stratonovich transformation on the random potentials, after which we have two four-fermion terms
\begin{align}
    Z &= \int \mathcal{D}\hat{\Delta}_{\text{SDW,x}}\int \mathcal{D}\hat{\Delta}_{\text{SDW,y}}\int \mathcal{D}\Psi\int\mathcal{D}\bar{\Psi} \;\text{Exp}\Big\{\text{Tr}\Big(i\Bar{\Psi}(\tau_3\mu_3\nu_3 \hat{\omega}_{t,t'} + \xi)\Psi\Big) \nonumber\\& + \frac{i\nu}{\lambda_{\text{sy}}}|\Delta_{\text{SDW,y}}|^2\Big\}\;\text{Exp}\{-\sum_{m,r_0}\frac{1}{4\pi\nu\tau_{\Delta}}\bar{\Psi}(t)\mathcal{P}_1(r_0)P_m\Psi(t) \bar{\Psi}(t')\mathcal{P}_1(r_0)P_m\Psi(t')\}\nonumber\\&\text{Exp}\{-\sum_{m,r_0}\frac{1}{4\pi\nu\tau_{\Delta}}\bar{\Psi}(t)\mathcal{P}_2(r_0)P_m\Psi(t) \bar{\Psi}(t')P_m\mathcal{P}_2(r_0)\Psi(t')\}\;, \label{eq:SDWaction2Qmatrices}
\end{align}

Thus, written in momentum space, the four fermion term contains projectors not only in the Band-Nambu-spin-space, but also in momentum space. For the isotropic case with $f_{1,2}(\theta) = 1$ this projector simply implies evaluation of $\Psi, \bar{\Psi}$ at a single location $r$, in the generic case it is more convenient to write the projectors explicitly.

As the two four-fermion terms stemming from the two types of spin-density wave project to different momenta, we decouple them using two different $Q$-matrices after we use the usual anticommutation relation of Grassmann variables,
\begin{align}
    &\text{Exp}\{-\sum_m\frac{2|\Delta^{0}_{\text{SDW}}|^2\xi_{\text{V}}^2}{2}\bar{\Psi}(t)\mathcal{P}^{1(2)}\tilde{P}_m\Psi(t) \bar{\Psi}(t')\mathcal{P}^{1(2)}\tilde{P}_m\Psi(t')\} \nonumber\\&=  \int \mathcal{D}Q \;\text{Exp}\{-\frac{\pi\nu_{1(2)}}{4\tau_{\Delta}}Q_{1(2)}^2 - \frac{1}{\tau_{\Delta}}  \bar\Psi \sum_{r_0}Q_{1(2)}(r_0)\mathcal{P}^{1(2)}(r_0)\Psi\} \;.
\end{align}
Here, to simplify notation, we introduced
\begin{align}
    \nu_{1(2)} = \int_{0}^{2\pi} \frac{d\theta}{2\pi}\; \nu(\theta) f_{1(2)}^2(\theta)\;.
\end{align}
For a constant $\nu(\theta) = \nu$, we have $\nu_{1} = \nu_2 = \nu$ because of the normalization, but here we keep the possibility of them being distinct. Like in Sec.~\ref{sec:Actiondetails}, the  $Q$-matrices describe low-energy degrees of freedom. However, because of the angle dependence, they only describe the low-energy degrees of freedom of parts of the Fermi surface. Specifically, $Q_1$ describes the low-energy degrees of freedom for those momenta for which $f_{1}(\theta)$ is nonzero, \textit{i.e.} the yellow parts of the Fermi surface in Fig.~\ref{fig:4Fermisurfacesallelectron}, while $Q_2$ describes the low-energy degrees of freedom for the momenta in the red part of the Fermi surface.

Like before, we can integrate out the fermions, and obtain
\begin{align}
    Z = \int \mathcal{D}Q_1\int \mathcal{D}Q_2\; e^{iS[Q_1,Q_2]}\;,
\end{align}
where $S[Q]$ is given by
\begin{align}\label{eq:iSQ1andQ2}
    iS[Q_{1},Q_{2}] &= -\frac{\pi\nu}{4\tau_{\Delta}}\text{Tr}\Big(\frac{\nu_1^2}{\nu^2} Q_1^2+\frac{\nu_2^2}{\nu^2}Q_{2}^2\Big) + \text{Tr ln}\Big((\tau_3\mu_3\nu_3 \hat{\omega}_{t,t'} + \xi+ i\Delta _{\text{d}} e^{i\phi\tau_3}\tau_2 \nonumber\\&+\frac{i}{\tau_{\Delta}}\sum_{r_0} Q_1(r_0)\mathcal{P}_1(r_0)+\frac{i}{\tau_{\Delta}}\sum_{r_0}Q_2(r_0)\mathcal{P}_2(r_0))\Big)\;.
\end{align}
This is the generic action in terms of the two $Q$-matrices. In the quasiclassical limit, like before we simplify by exploiting that the dispersive term $\xi$ is dominant.
To derive the soft-mode manifold constraint for $Q_1$, we take the variation with respect to $Q_1(r_0)$ and obtain
\begin{align}
    0 &= \frac{\pi\nu_1}{2\tau_{\Delta}}\Bigg(-Q_1(r_0) - \mathcal{P}_1(r_0) \frac{i}{\pi\nu_1}\Big(\tau_3\mu_3\nu_3 \hat{\omega}_{t,t'} + \xi + \Delta _{\text{d}} e^{i\phi\tau_3}\tau_2 + \frac{i}{\tau_{\Delta}}\sum_{r_1}Q_1(r_1)\mathcal{P}^{1}(r_1)\nonumber\\&+ \frac{i}{\tau_{\Delta}}\sum_{r_1}Q_2(r_1)\mathcal{P}^{2}(r_1)\Big)^{-1}\Bigg)\;.\label{eq:NormVar}
\end{align}

The soft-mode manifold can be determined by starting from a spatially homogeneous system, so that $Q_{1(2)}(r) = Q_{1(2)}$ and requiring that the variation, Eq.~(\ref{eq:NormVar}) vanishes. In momentum space, we find
\begin{align}
    \sum_{r_0}Q_{1(2)}(r_0)\mathcal{P}^{1}_{kk'}(r_0) = Q_{1(2)}\sum_{r_0}\mathcal{P}^{1}_{kk'}(r_0) = Q_{1(2)} \frac{1}{N_k}f_{1(2)}(\theta)f_{1(2)}(\theta')\sum_{r_0}e^{i(k-k')r_0} = Q_{1(2)} f_{1(2)}^2(\theta)\frac{1}{N_k}\delta_{\vec{k}\vec{k'}}\;. 
\end{align}
With this we have achieved a form similar to the standard case, and thus exploit the standard procedure. Going to a basis in which $Q_{1}$ is diagonal in band-Nambu-spin basis, and converting the momentum summation to an integral, we find that for energies $\epsilon\ll |\Delta^{0}_{\text{SDW}}|^2\xi_{\text{V}}^2\nu$, $Q_{1}$ can be written as  
\begin{align}
    \lambda = \frac{i}{\pi\nu_{1}}\int d^2\vec{p}\; \frac{1}{ \frac{p^2}{2m}-\mu +\frac{i}{\tau_{\Delta}} f_{1}^2(\theta)\lambda}f_{1}^2(\theta)= \frac{\text{sign}(\lambda)}{\nu_1}\int_{0}^{2\pi} \frac{d\theta}{2\pi}\; \nu(\theta)f_1^2(\theta) = \text{sign}(\lambda)\label{eq:EigenvaluesAnisotropicQ}\;,
\end{align}
where we ignored the $f_{2}^2$ term because $f_{1}$ and $f_{2}$ select different parts of the Fermi surface. We note that for angles such that $f_{1}^2(\theta)\ll \frac{\epsilon}{\pi\nu\xi_{\text{V}} |\Delta^{0}_{\text{SDW}}|^2}$ the Green's function $G$ is not necessarily diagonal in the basis in which $Q_1$ is because the $\epsilon$-term contributes significantly in this range. However, since the contributions of such angles are weighted by $f_{1}^2(\theta)$, corrections to Eq.~(\ref{eq:EigenvaluesAnisotropicQ}) are still $O(\frac{|\epsilon|}{\pi\nu\xi_{\text{V}} |\Delta^{0}_{\text{SDW}}|^2})$.

We conclude that for $|\epsilon|\ll \pi\nu\xi_{\text{V}} |\Delta^{0}_{\text{SDW}}|^2$, the soft-mode manifold consists of $Q$-matrices that satisfy
\begin{align}
    Q_1^2 = \mathbf{1}\;.
\end{align}
By changing $1$ into $2$ in the derivation above we also find that the soft-mode manifold for $Q_2$ is characterized by
\begin{align}
    Q_2^2 = \mathbf{1}\;.
\end{align}

Now, let us consider the additional restrictions on the manifold. Next to the charge conjugation symmetry, they have to  commute with their own spin-density wave. Because spin-density waves involve only those portion of the Fermi surface that are close to the nesting condition that is, that are colored in Fig.~\ref{fig:4Fermisurfacesallelectron}, each part of the Fermi surface can only contribute to one of the spin-density waves, as illustrated in Fig.~\ref{fig:4Fermisurfacesallelectron} by the separation of the yellow and red colors on the Fermi surfaces. Thus, $Q_1$ is diagonal in $\nu$-space while $Q_2$ is diagonal in $\mu$-space.

In this case, 
\begin{align}
    Q_1 &= \tau_3\sigma_z \mu_{1,2} Q_1 \tau_3\sigma_z\mu_{1,2} = \nu_3Q_1\nu_3\;,\\
    Q_2 &= \tau_3\sigma_z \nu_{1,2} Q_2 \tau_3\sigma_z\nu_{1,2} = \mu_3Q_2\mu_3\;.
\end{align}
Thus, both manifolds can be parameterized using an eight by eight Nambu-spin-band basis, spanned by $\tilde{\tau}_{i},\tilde{\sigma}_{i}, i = 0,1,2,3$ as defined before and  now additionally $\tilde{\mu}_i, i = 0,3$, where for band 1 $\tilde{\mu}_i = \mu_i$ and for band 2 $\tilde{\mu}_i = \nu_i$.

The additional degree of freedom allows for extra terms next to the magnetism and the \textit{d}-wave superconductivity, such as
\begin{align}
    &e^{i\phi_{\text{p}}}\tilde{\tau}_{2}\vec{d}\cdot\vec{\tilde{\sigma}}\tilde{\mu}_3\;,\\
    &\tilde{\mu}_3 \vec{b}\cdot\tilde{\sigma}\;,\\
    &\tilde{\mu}_3\tilde{\tau}_3\;.
\end{align}
The first of these terms corresponds to \textit{p}-wave superconductivity, the second to a momentum-odd spin-splitting, and the third to a non-magnetic time-reversal symmetry broken order.

We focus on the \textit{d}-wave superconductivity, as all other terms enter similarly.
First of all we note that because $Q$ still satisfies the commutation relations, the $\hat{\omega}$ term still drops out of the sigma model to linear order. The derivation of the terms is now entirely similar to the usual case, except that each integral has $f_1(\theta)$ and $f_{1}(\theta')$.

We are now ready to do the expansion of the logarithm. As above we are interested in the space-independent case, so that we can drop the derivative term. Moreover, again the commutation relations of $Q_1$ and $Q_2$ imply that there is no $\hat{\omega}Q_1$ or $\hat{\omega} Q_2$ term. Therefore the number of terms is quite limited. Taking the variation we obtain
\begin{align}
    iS_Q&[Q_1+[\delta Q_1,Q_1],Q_2+[\delta Q_2,Q_2]] - iS[Q_1 ,Q_2] \nonumber\\=&\text{Tr ln}\Bigg(\tau_3\mu_3\nu_3\hat{\omega} + \xi + i\Delta_{\text{d}} e^{i\phi\tau_3}\tau_2\mu_3\nu_3  + \frac{i}{\tau_{\Delta}} (Q_1 + [\delta Q_1,Q_1]) f_{1}^2(\theta)  + \frac{i}{\tau_{\Delta}} (Q_2+ [\delta Q_2,Q_2]) f_{2}^2(\theta))\nonumber\\&-\text{Tr ln}(\tau_3\mu_3\nu_3\hat{\omega} + \xi + i\Delta_{\text{d}} e^{i\phi\tau_3}\tau_2 \mu_3\nu_3 + \frac{i}{\tau_{\Delta}} Q_1 f_{1}^2(\theta)  + \frac{i}{\tau_{\Delta}} Q_2 f_{2}^2(\theta)\Bigg)\nonumber\\ =& \text{Tr}\Bigg(\pi\nu  \Big(\frac{\nu_1^2}{\nu^2}\Delta_{\text{d}}([e^{i\phi}\tau_3,Q_1]\delta Q_1 + \frac{\nu_2^2}{\nu^2}[e^{i\phi}\tau_3,Q_2]\delta Q_2)\Big)\Bigg)\;.\label{eq:ActiontwoQs} 
\end{align}
It follows that at the saddle point, where the variation vanishes for any choice of $\delta Q_{1,2}$, we have
\begin{align}
    Q_1 = \pm e^{i\phi\tau_3}\tau_2\;,\\
    Q_2 = \pm e^{i\phi\tau_3}\tau_2\;.
\end{align}

Like in Sec.~\ref{sec:quasiclassicalexpansion}, the absence of a Matsubara frequency term means that this has to be cut-off at $\pi\nu_1\xi_{\text{V}}^2|\Delta_{\text{d}}\Delta_{\text{SDW}}^{0}|^2$, indicating that the momentum dependent phase disorder stabilize superconductivity in the same way as the isotropic model described earlier. With this, we have shown that even if the original spin-density wave is momentum dependent, if they lose phase coherence, the resulting low energy modes can still described by a non-linear sigma model. 

\subsection{Density of states}
Next, we compute the DOS of the superconductor based on the NLSM. We cannot get the LDOS directly from the Q-matrices, since they correspond to a weighted average. Instead, we can look at the angle dependent quasiclassical function. Indeed, from the action we deduce that the microscopic Green's function satisfies
\begin{align}
    G = (\tau_3\mu_3\nu_3 \hat{\omega}_{t,t'} + \xi +\frac{i}{\tau_{\Delta}}Q_1\mathcal{P}_1+\frac{i}{\tau_{\Delta}}Q_2\mathcal{P}_2)^{-1}\;,
\end{align}
so that we define the angle dependent quasiclassical Green's function, which in  can be written as
\begin{align}
    g(\theta) = \frac{i}{\pi\nu(\theta)}\int d\xi\; (-i\epsilon\tau_3\mu_3\nu_3 + \xi +\frac{i}{\tau_{\Delta}}Q_1\mathcal{P}_1+\frac{i}{\tau_{\Delta}}Q_2\mathcal{P}_2)^{-1}\;.
\end{align}
Again, to derive the soft-mode manifold we assume the system is almost homogeneous in space, in which case we have
\begin{align}
    g(\theta) = \frac{i}{\pi\nu(\theta)}\int d\xi\;(-i\epsilon\tau_3\mu_3\nu_3 + \xi +\frac{i}{\tau_{\Delta}}Q_1f_{1}(\phi)^2+\frac{i}{\tau_{\Delta}}Q_2f_{2}(\phi)^2)^{-1}\;.
\end{align}
Now, we exploit again that in the eigenbasis of $g(\phi)$ the quasiclassical limit dictates $g^2(\theta) = \mathbf{1}$. Moreover, this eigenbasis is the eigenbasis of $-i\epsilon\tau_3\mu_3\nu_3 + \frac{i}{\tau_{\Delta}}Q_1f_{1}(\phi)^2+ \frac{i}{\tau_{\Delta}}Q_2f_{2}(\phi)^2$, which indicates that $g$ satisfies:
\begin{align}
    [-i\epsilon\tau_3\mu_3\nu_3 + \frac{i}{\tau_{\Delta}}Q_1f_{1}^2(\theta)+ \frac{i}{\tau_{\Delta}}Q_2f_{2}^2(\theta),g(\theta)] = 0\;.
\end{align}
Thus, clearly, $g$ has an angle dependent gap through the angle dependence of $f_{1}$ and $f_{2}$, it has a quasiparticle excitation gap $E_g$, given by
\begin{align}
    E_g = \frac{1}{\tau_{\Delta}}(f_{1}^2(\theta)+ f_{2}^2(\theta))\;.
\end{align}
 Therefore, suitable choices of $f_{1,2}$ reproduce the density of states of the system.

Notably, for the choice of $\Delta_{\text{d}}$ in Eq.~(\ref{eq:ActiontwoQs}), the sign of $\Delta_{\text{d}}$ is still the same within pockets. If one starts from $\Delta_{\text{d}}(\theta)$, then, because $Q_1$ and $Q_2$ come with different angular averages, their superconductor potential need not be the same. In this case the material settles into a \textit{g}-wave state, with sign changes on each topic, and additionally sign changes when going to the next. However, the \textit{d}-wave is the one that survives the $Q_{1(2)}$ term as well as any additional intraband scattering, which was not considered here, and therefore seems the most robust. To this end, it is important to remark that even though the critical temperature can be high, we found that the pair potential is comparatively weak, Eq.~(\ref{eq:ZeroTemperaturepotential}). It is this scale that has to be compared with the interband disorder.

\subsubsection{Other order parameters}
As discussed above, the enhanced flexibility of the manifold allows for different instabilities. The derivation of these terms in the actions for $Q_{1(2)}$ is entirely similar as for the \textit{d}-wave and \textit{g}-wave superconductivity. Here we discuss their properties.  
First of all, the system is able to shield triplet superconductivity through $\Delta_{\text{p}}e^{i\phi_{\text{p}}}\tilde{\tau}_{2}\tilde{\mu_3}$. For $Q_1$ and $\vec{d}$ along $\vec{\hat{z}}$, or for $Q_2$ and with $\vec{d}\cdot\vec{z} = 0$ this corresponds to a \textit{p}$_{x}$-wave type of superconductivity, while for the other cases it corresponds to a \textit{p}$_y$-wave superconductivity. However, there is no possibility to combine these two with both being nonzero in a way that in each pocket there is only superconductivity of a single sign, indicating that this type of superconductivity is more sensitive against intraband disorder than \textit{d}-wave superconductivity. However, it can spontaneously choose one, if the pairing strength for \textit{p}-wave superconductivity is much stronger than that for \textit{d}-wave superconductivity.

\subsection{Including gradient terms}
Now that we have found the method to determine the bulk $Q$-matrix in systems without any gradients, we turn to the inclusion of gradients. To this end, we note that when expanding the logarithm, we obtain products that themselves still involve convolutions. These convolutions can be expanded using the Moyal product. From the product of the dispersion term  ($\xi$) and the $Q_1$, we thus obtain, to first order in the Moyal product and to second order in the logarithm expansion
\begin{align}
    iS_{\text{D}}[Q_1] = \text{Tr}\Big(\mathcal{P}_1\mathcal{G}\partial_{\vec{p}} \xi \partial_{\vec{r}} Q\mathcal{P}_1\mathcal{G}\partial_{\vec{p'}} \xi \partial_{\vec{r}} Q\Big)\;,
\end{align}
where $\partial_{\vec{p}}\xi \approx \vec{v}_F$ near the Fermi surface, and
\begin{align}
    \mathcal{G}^{-1} &= \tau_3\mu_3\nu_3 \hat{\omega}_{t,t'} + \xi + i\Delta _{\text{d}} e^{i\phi\tau_3}\tau_2 + \frac{i}{\tau_{\Delta}}\sum_{r_1}Q_1(r_1)\mathcal{P}^{1}(r_1)\nonumber\\&+ \frac{i}{\tau_{\Delta}}\sum_{r_1}Q_2(r_1)\mathcal{P}^{2}(r_1)\;.
\end{align}
If $Q_1$ varies sufficiently slowly in-space, that is, it varies on a scale large compared to the real space extend of $\mathcal{P}^{1}$, we can still use the momentum integral and obtain, using the same procedure as usual,
\begin{align}
    iS_{\text{D}}[Q_1] &= \pi\nu_1\text{Tr}\Big(D^1_{jk}\partial_jQ \partial_{k}Q\Big)\;,  \\
    D^1_{jk} &= \int_{0}^{2\pi} \frac{d\theta}{2\pi}\int d\xi\; v_{\text{F},j}(\theta) v_{\text{F},k} (\theta) \frac{f_1^2(\theta)}{\xi + \frac{i}{\tau_{\Delta}}f_1^2(\theta)}\frac{f_1^2(\theta)}{\xi - \frac{i}{\tau_{\Delta}}f_1^2(\theta)}\nonumber\\& = \tau_{\Delta} \int_{0}^{2\pi} \frac{d\theta}{2\pi}\; v_{\text{F},j}(\theta) v_{\text{F},k} (\theta) f_1^{2}(\theta)\; \;.
\end{align}
The diffusion theory for $Q_2$ is found by changing all indices from $1$ to $2$.

With this we have established that even if the dominant order has a strength that varies with the momentum direction on the Fermi surface, shielded orders appear that can be described with the help of a nonlinear sigma model.

\section{Dopants and phase coherence of density waves}\label{sec:Dopants}
Now that we have estabilished that subdominant correlations can be described using a NLSM and attain high critical temperatures in the presence a phase disordered dominant correlation order, we turn our attention to how a dominant order can become phase disordered. To this end, we focus on the example of a density wave.

The purpose of this section is to argue based purely on a free energy for the U(1) phase that the phase coherence in density waves is very sensitive to dopants, with an increasing amount of dopants phase coherence is lost even at zero temperature. We show that the predictions of hte model are consistent with experimental observations in cuprates.

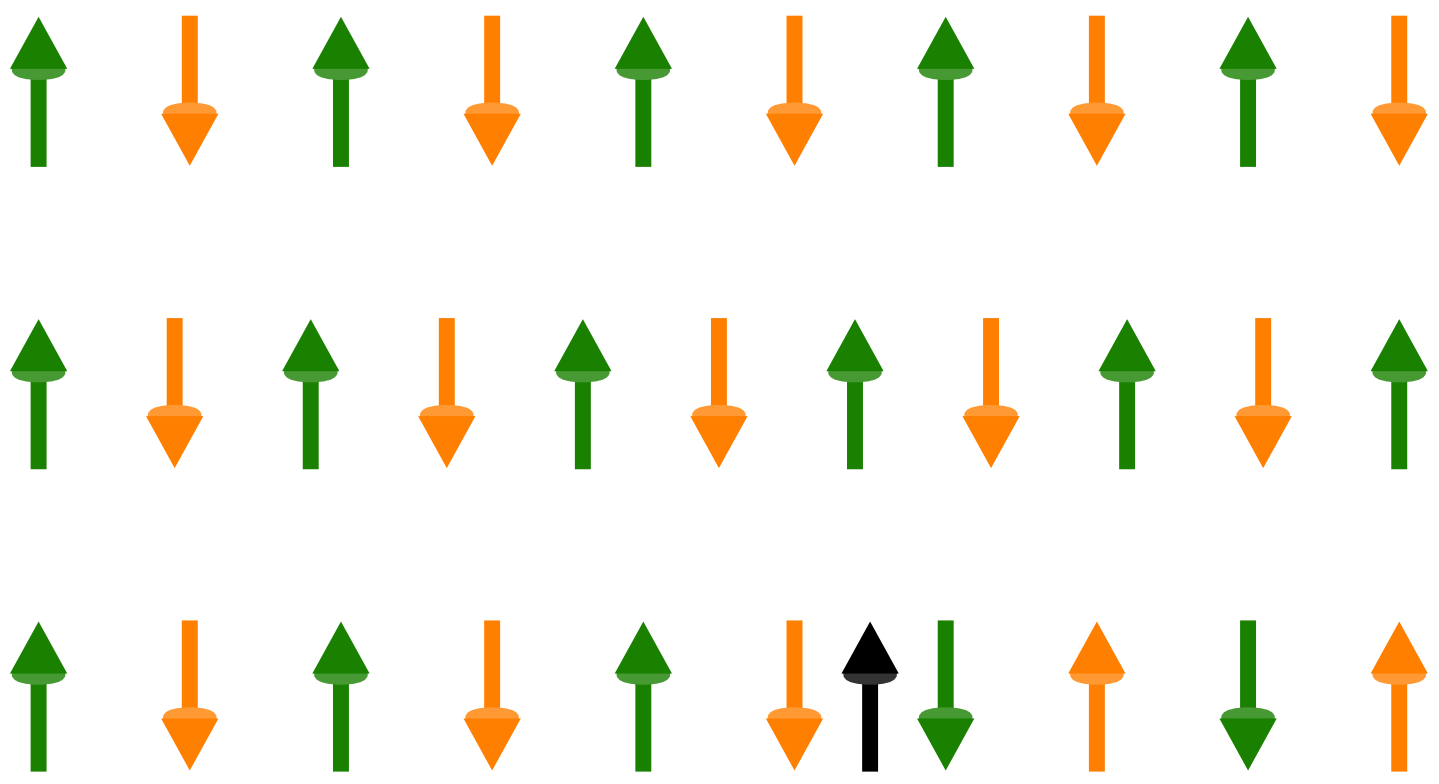
\begin{figure}
    \begin{tikzpicture}
        \draw[-{Circle[sep=-1.9pt,width=10.08pt,length=4pt,goodgreen!80]Triangle[width=10pt,length=10pt,goodgreen]},line width = 3pt,goodgreen](0,-0.5,0)--(0,0.5,0);
        \draw[-{Circle[sep=-1.9pt,width=10.08pt,length=4pt,orange!80]Triangle[width=10pt,length=10pt,orange]},line width = 3pt,orange](1,0.5,0)--(1,-0.5,0);
        \draw[-{Circle[sep=-1.9pt,width=10.08pt,length=4pt,goodgreen!80]Triangle[width=10pt,length=10pt,goodgreen]},line width = 3pt,goodgreen](2,-0.5,0)--(2,0.5,0);
        \draw[-{Circle[sep=-1.9pt,width=10.08pt,length=4pt,orange!80]Triangle[width=10pt,length=10pt,orange]},line width = 3pt,orange](3,0.5,0)--(3,-0.5,0);
        \draw[-{Circle[sep=-1.9pt,width=10.08pt,length=4pt,goodgreen!80]Triangle[width=10pt,length=10pt,goodgreen]},line width = 3pt,goodgreen](4,-0.5,0)--(4,0.5,0);
        \draw[-{Circle[sep=-1.9pt,width=10.08pt,length=4pt,orange!80]Triangle[width=10pt,length=10pt,orange]},line width = 3pt,orange](5,0.5,0)--(5,-0.5,0);
        \draw[-{Circle[sep=-1.9pt,width=10.08pt,length=4pt,goodgreen!80]Triangle[width=10pt,length=10pt,goodgreen]},line width = 3pt,goodgreen](6,-0.5,0)--(6,0.5,0);
        \draw[-{Circle[sep=-1.9pt,width=10.08pt,length=4pt,orange!80]Triangle[width=10pt,length=10pt,orange]},line width = 3pt,orange](7,0.5,0)--(7,-0.5,0);
        \draw[-{Circle[sep=-1.9pt,width=10.08pt,length=4pt,goodgreen!80]Triangle[width=10pt,length=10pt,goodgreen]},line width = 3pt,goodgreen](8,-0.5,0)--(8,0.5,0);
        \draw[-{Circle[sep=-1.9pt,width=10.08pt,length=4pt,orange!80]Triangle[width=10pt,length=10pt,orange]},line width = 3pt,orange](9,0.5,0)--(9,-0.5,0);
        \draw[-{Circle[sep=-1.9pt,width=10.08pt,length=4pt,goodgreen!80]Triangle[width=10pt,length=10pt,goodgreen]},line width = 3pt,goodgreen](0,-2.5,0)--(0,-1.5,0);
        \draw[-{Circle[sep=-1.9pt,width=10.08pt,length=4pt,orange!80]Triangle[width=10pt,length=10pt,orange]},line width = 3pt,orange](0.9,-1.5,0)--(0.9,-2.5,0);
        \draw[-{Circle[sep=-1.9pt,width=10.08pt,length=4pt,goodgreen!80]Triangle[width=10pt,length=10pt,goodgreen]},line width = 3pt,goodgreen](1.8,-2.5,0)--(1.8,-1.5,0);
        \draw[-{Circle[sep=-1.9pt,width=10.08pt,length=4pt,orange!80]Triangle[width=10pt,length=10pt,orange]},line width = 3pt,orange](2.7,-1.5,0)--(2.7,-2.5,0);
        \draw[-{Circle[sep=-1.9pt,width=10.08pt,length=4pt,goodgreen!80]Triangle[width=10pt,length=10pt,goodgreen]},line width = 3pt,goodgreen](3.6,-2.5,0)--(3.6,-1.5,0);
        \draw[-{Circle[sep=-1.9pt,width=10.08pt,length=4pt,orange!80]Triangle[width=10pt,length=10pt,orange]},line width = 3pt,orange](4.5,-1.5,0)--(4.5,-2.5,0);
        \draw[-{Circle[sep=-1.9pt,width=10.08pt,length=4pt,goodgreen!80]Triangle[width=10pt,length=10pt,goodgreen]},line width = 3pt,goodgreen](5.4,-2.5,0)--(5.4,-1.5,0);
        \draw[-{Circle[sep=-1.9pt,width=10.08pt,length=4pt,orange!80]Triangle[width=10pt,length=10pt,orange]},line width = 3pt,orange](6.3,-1.5,0)--(6.3,-2.5,0);
        \draw[-{Circle[sep=-1.9pt,width=10.08pt,length=4pt,goodgreen!80]Triangle[width=10pt,length=10pt,goodgreen]},line width = 3pt,goodgreen](7.2,-2.5,0)--(7.2,-1.5,0);
        \draw[-{Circle[sep=-1.9pt,width=10.08pt,length=4pt,orange!80]Triangle[width=10pt,length=10pt,orange]},line width = 3pt,orange](8.1,-1.5,0)--(8.1,-2.5,0);
        \draw[-{Circle[sep=-1.9pt,width=10.08pt,length=4pt,goodgreen!80]Triangle[width=10pt,length=10pt,goodgreen]},line width = 3pt,goodgreen](9,-2.5,0)--(9,-1.5,0);

        \draw[-{Circle[sep=-1.9pt,width=10.08pt,length=4pt,goodgreen!80]Triangle[width=10pt,length=10pt,goodgreen]},line width = 3pt,goodgreen](0,-4.5,0)--(0,-3.5,0);
        \draw[-{Circle[sep=-1.9pt,width=10.08pt,length=4pt,orange!80]Triangle[width=10pt,length=10pt,orange]},line width = 3pt,orange](1,-3.5,0)--(1,-4.5,0);
        \draw[-{Circle[sep=-1.9pt,width=10.08pt,length=4pt,goodgreen!80]Triangle[width=10pt,length=10pt,goodgreen]},line width = 3pt,goodgreen](2,-4.5,0)--(2,-3.5,0);
        \draw[-{Circle[sep=-1.9pt,width=10.08pt,length=4pt,orange!80]Triangle[width=10pt,length=10pt,orange]},line width = 3pt,orange](3,-3.5,0)--(3,-4.5,0);
        \draw[-{Circle[sep=-1.9pt,width=10.08pt,length=4pt,goodgreen!80]Triangle[width=10pt,length=10pt,goodgreen]},line width = 3pt,goodgreen](4,-4.5,0)--(4,-3.5,0);
        \draw[-{Circle[sep=-1.9pt,width=10.08pt,length=4pt,orange!80]Triangle[width=10pt,length=10pt,orange]},line width = 3pt,orange](5,-3.5,0)--(5,-4.5,0);
        \draw[-{Circle[sep=-1.9pt,width=10.08pt,length=4pt,black!80]Triangle[width=10pt,length=10pt,black]},line width = 3pt](5.5,-4.5,0)--(5.5,-3.5,0);
        \draw[-{Circle[sep=-1.9pt,width=10.08pt,length=4pt,goodgreen!80]Triangle[width=10pt,length=10pt,goodgreen]},line width = 3pt,goodgreen](6,-3.5,0)--(6,-4.5,0);
        \draw[-{Circle[sep=-1.9pt,width=10.08pt,length=4pt,orange!80]Triangle[width=10pt,length=10pt,orange]},line width = 3pt,orange](7,-4.5,0)--(7,-3.5,0);
        \draw[-{Circle[sep=-1.9pt,width=10.08pt,length=4pt,goodgreen!80]Triangle[width=10pt,length=10pt,goodgreen]},line width = 3pt,goodgreen](8,-3.5,0)--(8,-4.5,0);
        \draw[-{Circle[sep=-1.9pt,width=10.08pt,length=4pt,orange!80]Triangle[width=10pt,length=10pt,orange]},line width = 3pt,orange](9,-4.5,0)--(9,-3.5,0);
    \end{tikzpicture}
    \caption{One-dimensional illustration of a dopant acting as a topological excitation of the spin-density wave order parameter. \textit{Top:} SDW system at zero doping, the spins form a perfect lattice. \textit{Middle} Incorporation of the extra spin via a change in the distance between spins, corresponding to a phase gradient of the density wave. \textit{Down:} SDW system at nonzero doping. The dopant creates a $\pi$-shift of the order parameter. As illustrated using colors, on the left of the dopant, the chain is in-phase with the un-perturbed chain, while on the right it is out-of-phase.}
    \label{fig:Dopanttopologicalexcitation}
\end{figure}

We first argue why dopants decrease phase coherence. To this end we first consider the effect of dopants in a 1D chain with an antiferromagnetic order with spin up on sublattice A and spin down on sublattice B, as illustrated in Fig \ref{fig:Dopanttopologicalexcitation}. If we add a dopant to this chain, we see that there are two possibilities, (i) the spin-density wave changes its periodicity to accomodate for the additional spin or (ii) the period is not altered, and the added spin enters between two randomly chosen spins. In the latter case, because also the electron on this atom wants to be aligned opposite to its neighbors, on the right of the atom the order changes to spin down on sublattice A and spin up on sublattice B, which corresponds to a $\pi$ phase slip of the antiferromagnetic / spin-density wave order. 

A change in the periodicity corresponds to a phase gradient of the density wave order parameter. Meanwhile the analogy of the phase slip in 2D is a vortex. Thus, we can model the effect of adding additional electrons in a 2D material with a density wave as either creating uniform phase gradients or vortices of the spin-density wave order.

To understand why phase coherence may be lost even at zero temperature, we consider the system at zero temperature. At zero doping, the electrons form a perfect SDW. Upon addition of $x$ electrons per unit cell, two things can happen, (i) the period of the SDW changes, which corresponds to a phase gradient $q = x\frac{\pi}{a}$ of the SDW order parameter, (ii) $x$ vortices per unit cell are created. Let us consider the costs of these two cases. For the phase gradient, we have
\begin{align}
    E_{\text{nv}} = \frac{1}{2}W_{\text{s}}\int dA\; (\nabla \phi)^2 = \frac{\pi^2W_{\text{s}}A}{2a^2} x^2\;.\label{eq:Env}
\end{align}
where $A_{m}$ is the area of the material and within a BCS-like theorys $W_{\text{s}} = \frac{\pi}{2}\sigma_{\text{2D}}|\Delta_{\text{SDW,bulk}}|$ is the SDW-liquid weight of the material, with $\sigma_{\text{2D}}$ being the 2D conductivity.
Now let us consider the energy of the configuration in which vortices are created. 

In vortices, there are contributions both from amplitude gradients and from phase gradients. Amplitude gradients are localized within a radius $r_{\text{v}}$ from the center of the vortex, while phase gradients extend all the way up to the next vortex, and thus have their main contribution from $r>r_{\text{v}}$.

We estimate the energy per vortex with winding number $\pm 1$ due to phase gradients in a similar manner as in the superconducting case \cite{kosterlitz1973ordering}, which results in
\begin{align}
    E_{\text{v},g} = \frac{1}{2}W_{\text{s}} \int dA\; (\nabla \phi)^2 = \frac{1}{2}W_{\text{s}} \int_{0}^{2\pi} d\theta \int_{r_{\text{v}}}^{r_c}dr\; r\Big(\frac{1}{ r}\Big)^2 = \pi W_{\text{s}} \text{ln}\frac{r_c}{r_{\text{v}}}\;.\label{eq:Evg}
\end{align}
Here $r_{\text{v}}$ is the radius of the vortex core, and $r_c$ is a cut-off radius needed for regularization. In practice, it is determined by the half-distance to the next vortex, $r_c \sim \frac{a}{2\sqrt{x}}$. Since there is a vortex every $x^{-1}$ unit cells, the total number of vortices is $x\frac{A_{m}}{a^2}$.

Near the vortex core, amplitude gradients are usually important, resulting in an additional constant that accompanies the logarithm. However, if the interaction is so strong that the coherence length is of the order of the atomic distance, phase gradients can be achieved without decreasing the amplitude, so that we ignore such term.

To understand if such transition happens, let us consider the dilute limit $x\ll \Big(\frac{a}{2r_{\text{v}}}\Big)^2$ and the dense limit $x\sim \Big(\frac{a}{2r_{\text{v}}}\Big)^2$, that is, close to the destruction of the mean-field spin-density wave order parameter. In the dilute limit $x\ll \Big(\frac{a}{2r_{\text{v}}}\Big)^2$, the logarithm becomes large and we have $x\text{ln}\frac{a^2}{4r_{\text{v}}^2x} \gg x \gg x^2$, so that at small enough $x$ the energy of the configuration with vortices is larger, and hence the state remains ordered.

In the dense limit in which $x$ becomes of the same order as $\frac{a}{2r_{\text{v}}}$, the argument of the logarithm approaches $1$, and the $x^2$ term dominates.  This shows that if $r_{\text{v}}<\pi a$, for large enough $x$ vortices proliferate through the system, leading to a transition. The critical $x_c$ at which the free energy difference vanishes can be detemined from the identity
\begin{align}
    \pi x_c(-\pi x_c + \text{ln}\frac{a^2}{4r_{\text{v}}^2x_c}) = 0\;.
\end{align}
This function is plotted as a function of $\frac{r_{\text{v}}}{a}$ in Fig.
\begin{figure}
    \centering
    \includegraphics[width=8.6cm]{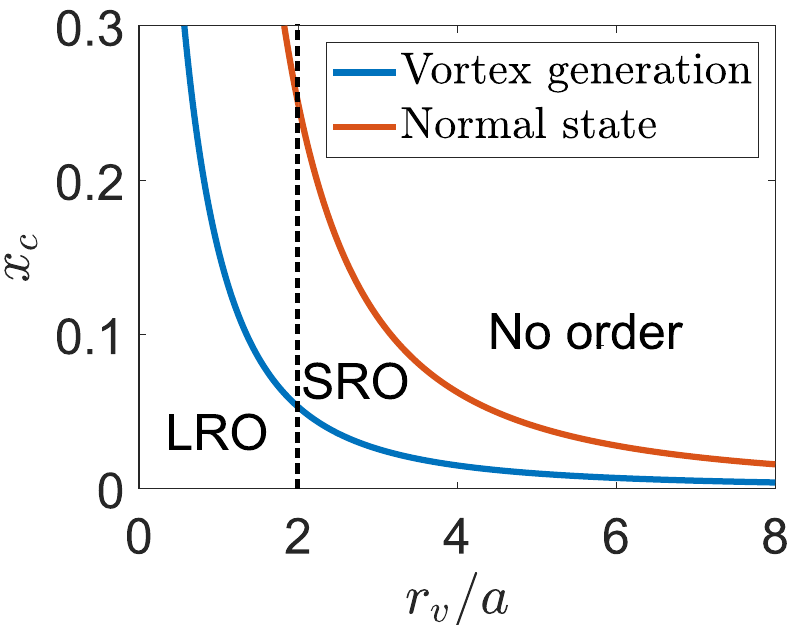}
    \caption{Phase diagram as a function of vortex size $r_{\text{v}}$ and the doping $x$, showing the long-range order phase, the phase in which only local short-range order survives, and the normal phase. The long-range order and short-range order states are of equal energy when $\pi x_c = \text{ln}\frac{a^2}{4r_{\text{v}}^2x_c}$, indicated by the blue line, while the vortices occupy the whole material, resulting in the absence of order for $x_c=\frac{a^2}{r_{\text{v}}^2}$, indicated by the red line.}
    \label{fig:Range}
\end{figure}
Since for $x = \Big(\frac{a}{2r_{\text{v}}}\Big)^2$ the mean-field order parameter vanishes, we can estimate this quantity from the phase diagram. For YBCO, the psuedogap phase extends up to $x\sim 0.25$, indicating that $r_{\text{v}} \sim a$. In this case, the phase coherence is lost at zero temperature for $x\gtrsim 0.06$, which nicely fits the experimental value \cite{keimer2015quantum}.
\subsection{Finite temperature}
Now that we know the behavior at zero temperature, we may turn our focus at finite temperature, and consider how $x_c$ changes. To this end, we exploit that the free energy as a function of the concentration of vortices $x_{\text{V}}$ for a dopant concentration $x$ is
\begin{align}
    F(x_{\text{V}},x) = \frac{\pi W_{\text{s}} A}{2a^2}(-x_{\text{V}} \text{ln}\Big(\lceil\frac{a^2}{4x_{\text{V}} r_{\text{v}}^2}\rceil) - \pi(x-x_{\text{V}})^2)-TS(x_{\text{V}})\;.
\end{align}
Here we modified the logarithm in $E_{\text{v}}$ with the introduction of the ceil function to avoid the logarithm from going negative at large $x$, while not modifying the expression at low $x$. As $x_{\text{V}} \rightarrow 0 $ the previous expression, Eq.~(\ref{eq:Evg}), is obtained. The entropy $S(x_{\text{V}})$ of the configuration with a vortex concentration $x_{\text{V}}$ is, under the assumption that $\frac{A}{a^2}\gg 1$.
\begin{align}
    S(x_{\text{V}}) = k_{\text{B}}\frac{A}{a^2}((1+x_{\text{V}})\text{ln}(1+x_{\text{V}})-x_{\text{V}}\text{ln}(x_{\text{V}}))\;.
\end{align}
We may now define the critical concentration $x_c$ at finite temperatures as the concentration $x$ for which the minimum of $F(x_{\text{V}}, x_c)$ has two degenerate minima. This defines the phase boundary of the dominant SDW order.

For concentrations larger than $x_c$, superconductivity can appear as a shielded order. As derived in this manuscript, its critical temperature depends on the highest energy for which the low-energy theory is valid. The lowest energy state not in the soft-mode manifold in this case is not the normal state, but the ordered SDW state, which for $x\approx 0.05$ is close to the short-range state. In this case the cut-off frequency is not $\tau_{\Delta}^{-1}$, it has to be lowered. The cut-off and thus the critical temperature is determined by $E_{\text{v,g}}-E_{\text{nv}}$, were $E_{\text{nv}}$ and $E_{\text{v,g}}$ are given by Eqs.~(\ref{eq:Env},\ref{eq:Evg}).
With this we have
\begin{align}
    k_{\text{B}}T_{\text{c}} \approx \frac{\pi W_{\text{s}}}{2}x(\pi x-x\text{ln}\frac{x r_{\text{v}}^2}{4a^2}) \;.
\end{align}

\begin{figure}
    \centering
    \includegraphics[width=8.6cm]{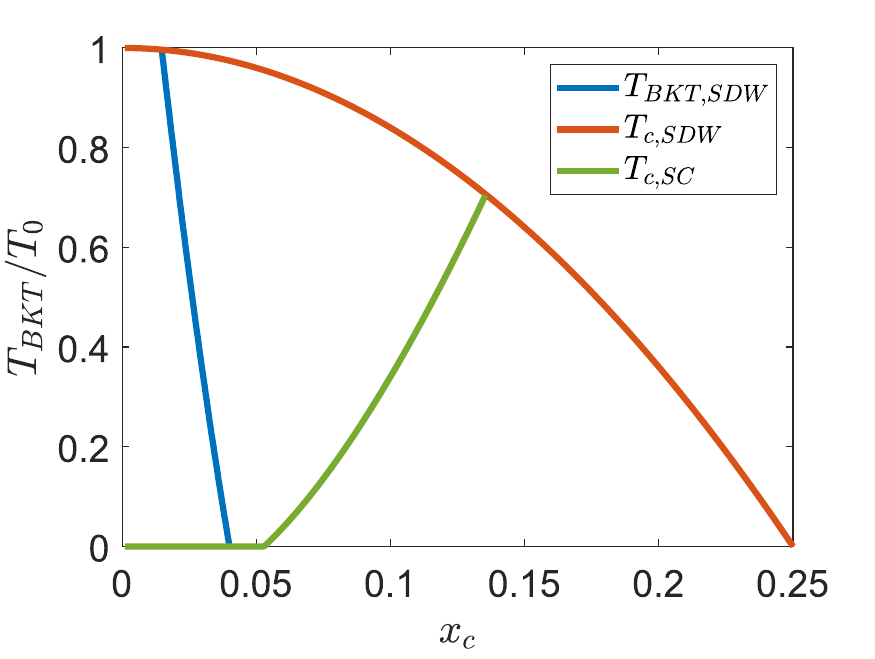}
    \caption{Phase diagram as a function of dopant concentration and temperature of a material with dominant spin-density wave correlations with $r_{\text{v}} = 2a$, and a subdominant shielded \textit{d}-wave superconductor instability. \textit{Note:} the gap between the blue and green lines at $T = 0$ appears because the green line comes from the comparison of the no-phase gradient and no-vortex states, while the blue line allows for any number of vortices. }
    \label{fig:phasediagramcuprates}
\end{figure}

Assuming that the phase stiffness can be approximated by the BCS-like value $W_{\text{s}} = \frac{\pi}{2}\frac{\hbar}{e^2}\sigma_D |\Delta_{\text{SDW}}|$, and using that $r_{\text{v}}\sim 2a$, while $\frac{\hbar}{e^2}\sigma_D \sim 1$ for cuprates, we find the results in Fig.~\ref{fig:phasediagramcuprates}. The resulting phase diagram qualitatively captures the features of cuprate superconductors.
On the superconducting side, we may exploit that the cut-off frequency for the low-energy modes is the free energy difference per unit.
 
\section{Spinors and order parameters for other Fermi surface topologies}\label{sec:OtherFermisurfacetopologies}
Up to this point of the Supplemental Material, we have focused on shielded order for Fermi surfaces with four pockets. However, as mentioned in the main text, Table~\ref{tab:shieldedordersummary}, shielded orders also appear for other Fermi surface topologies. In this section we show consider these other cases. We construct the bispinors based on which the NLSM can be constructed, and derive for several dominant orders which order could arise as a shielded order. Motivated by experimental observation, we always present correlated insulating orders as dominant and superconducting orders as shielded. However, these roles can be reversed for any pair of correlated insulator and superconductivity. 

\subsection{Layered materials and concentric Fermi surfaces}\label{sec:Layeredandconcentric}
First, we consider layered materials and materials with concentric Fermi surfaces. We discuss these two groups together because layer and band indices are both time-reversal even, and therefore their low-energy theories have to satisfy the same charge conjugation constraint. This implies that the soft-mode manifolds have the same structure, only the interpretation of the resulting states changes. Through the derivation of the  shielded orders that arise we refer to layered materials, and discuss their interpretation for concentric Fermi surfaces afterwards.

We consider materials with two layers in which the Fermi velocities are approximately opposite. 
Like we did in Sec.~\ref{sec:Construction} for the four-pocket Fermi surface, we first construct the spinors in band-spin space such that the dispersive term appears proportional to the identity matrix in band-spin space and subsequently the bispinors with the help of time-reversed components.

Because the bands have opposite dispersion, we may write the dispersive term in the Hamiltonian as $H_0 = \psi^{\dagger}\xi(k)\upsilon_3\psi$, where $\upsilon_3$ is the third Pauli matrix in layer space and $\psi = (\psi_{1\uparrow},\psi_{1\downarrow},\psi_{2\uparrow},\psi_{2\downarrow})^T$. This can be written in diagonal form as $H_0 = \xi(k)\bar{\psi}\psi$ if we define $\bar{\psi} = (\psi_{1\uparrow}^{\dagger},\psi_{1\downarrow}^{\dagger},-\psi_{2\uparrow}^{\dagger},-\psi_{2\downarrow}^{\dagger})$.

Because the layer index does not change sign under time-reversal, but the spin does, the spinors should be constructed as
$\Psi = (\psi^T,-i \bar{\psi}\sigma_y)^T = (\psi_{1\uparrow},\psi_{1\downarrow},\psi_{2\uparrow},\psi_{2\downarrow}, \psi_{1\downarrow}^{\dagger},-\psi_{1\uparrow}^{\dagger}, -\psi_{2\downarrow}^{\dagger},\psi_{2\uparrow}^{\dagger})^{T}$ and $\bar{\Psi} = (\bar{\psi},i\psi^T\sigma_y) = \Psi^{\dagger}\tau_3\upsilon_3 = (\psi_{1\uparrow}^{\dagger},\psi_{1\downarrow}^{\dagger},-\psi_{2\uparrow}^{\dagger},-\psi_{2\downarrow}^{\dagger},-\psi_{1\downarrow},\psi_{1\uparrow},-\psi_{2\downarrow},\psi_{2\uparrow})$. The two are related via $\Psi = i\tau_1\sigma_2\bar{\Psi}^{T}$, which means that the charge conjugation is the same as that for standard spin-Nambu space,
\begin{align}
    Q(t,t') = \tau_1\sigma_2 Q^{T}(t',t)\sigma_2\tau_1\;.
\end{align}

To construct shielded orders in analog to the SDW/\textit{d}-wave superconductivity example, we search for correlated insulator states. These must be off-diagonal in layer space, they correspond to interlayer excitons (IX).
We identify two types of IX, singlet
\begin{align}
    \Delta_{\text{IX,s}}e^{i\phi_{\text{IX,s}}} = \langle  [\psi_{1\uparrow},\psi_{2\uparrow}^{\dagger}]+[\psi_{1\downarrow},\psi_{2\downarrow}^{\dagger}]\rangle = \text{Tr}\langle\frac{\upsilon_2-i\upsilon_1}{2}\Psi\bar{\Psi} \rangle\;,
\end{align}
as well as triplet, where without loss of generality we choose spin along the $z$-direction:
\begin{align}
    \Delta_{\text{IX,t}}e^{i\phi_{\text{IX,t}}} = \langle  [\psi_{1\uparrow},\psi_{2\uparrow}^{\dagger}]-[\psi_{1\downarrow},\psi_{2\downarrow}^{\dagger}]\rangle = \text{Tr}\langle\frac{\upsilon_2-i\upsilon_1}{2}\tau_3\sigma_z\Psi\bar{\Psi} \rangle\;.
\end{align}

We take these interlayer excitons as the dominant order, and assume that they are phase disordered. Because due to randomization of momentum the pair potential is constant on each Fermi surface, there exist two candidate types of superconductivity, layer even \textit{s}-wave $\Delta_e$ and layer-odd \textit{s}-wave $\Delta_o$, given by
\begin{align}
    \Delta_{e}e^{i\phi_e} &\sim \langle \psi_{1\uparrow}\psi_{1\downarrow}-\psi_{1\downarrow}\psi_{1\uparrow} + \psi_{2\uparrow}\psi_{2\downarrow}-\psi_{2\downarrow}\psi_{2\uparrow} \rangle = \text{Tr}\langle\frac{\tau_2-i\tau_1}{2}\Psi\bar{\Psi} \rangle\;,\\
    \Delta_{o}e^{i\phi_o} &\sim \langle \psi_{1\uparrow}\psi_{1\downarrow}-\psi_{1\downarrow}\psi_{1\uparrow} - \psi_{2\uparrow}\psi_{2\downarrow}+\psi_{2\downarrow}\psi_{2\uparrow} \rangle = \text{Tr}\langle\frac{\tau_2-i\tau_1}{2}\upsilon_3\Psi\bar{\Psi} \rangle\;.
\end{align}

Because orders can only be shielded if they are preserved upon scattering off the phase disorder, we deduce which exciton shields which superconductor by checking which orders commute. We have for singlet interlayer excitons
\begin{align}
    &[\upsilon_{1,2},\tau_{1,2}] = 0\;,\quad&[\upsilon_{1,2},\tau_{1,2}\upsilon_3]\neq 0\;,
    \end{align}
    while for triplet interlayer excitons
    \begin{align}
    &[\upsilon_{1,2}\tau_3\sigma_z,\tau_{1,2}] \neq 0\;,\quad&[\upsilon_{1,2}\tau_3\sigma_z,\tau_{1,2}\upsilon_3]= 0\;.
\end{align}
From this we conclude that singlet IX shields only layer even \textit{s}-wave superconductivity, while triplet IX shields only layer odd \textit{s}-wave superconductivity. This is entirely analogous with the density wave example in the main text, the time-reversal even dominant order shields conventional \textit{s}-wave superconductivity by Anderson's theorem \cite{anderson1959theory}, while the time-reversal odd dominant order cannot shield conventional superconductivity, but instead preserves an order that has opposite sign on the Fermi pockets that it couples.

Magnetic order parameter that can be constructed are ferromagnetism ($\vec{m}$) and layer antiferromagnetism ($\vec{N}$). Introducing $\vec{\psi}_{i} = (\psi_{i\uparrow},\psi_{i\downarrow})^{T}$, these can be written as
\begin{align}
    \vec{m} &\sim \langle \vec{\psi}^{\dagger}_1\sigma\vec{\psi_{1}}\rangle +\langle \vec{\psi}^{\dagger}_2\sigma\vec{\psi_{2}}\rangle  = \text{Tr}\langle\tau_3\vec{\sigma}\upsilon_3\Psi\bar{\Psi}\rangle\;,\\
    \vec{N} &\sim  \langle \vec{\psi}^{\dagger}_1\sigma\vec{\psi_{1}}\rangle -\langle \vec{\psi}^{\dagger}_2\sigma\vec{\psi_{2}}\rangle  = \text{Tr}\langle\tau_3\vec{\sigma}\Psi\bar{\Psi}\rangle\;.
\end{align}
Evaluating the commutators, we find
\begin{align}
    [\upsilon_{1,2},\tau_3\vec{\sigma}\upsilon_3]\neq 0\;,\quad [\upsilon_{1,2},\tau_3\vec{\sigma}] = 0\;,
\end{align}
indicating that the singlet interlayer exciton shields layer-odd antiferromagnetism with N\'eel vector pointing in arbitrary direction, while ferromagnetic orders are suppressed.

On the other hand, 
\begin{align}
    [\upsilon_{1,2}\tau_3\sigma_z, \tau_3\sigma_a\upsilon_3] \propto \delta_{az}\;,\quad[\upsilon_{1,2}\tau_3\sigma_z,\tau_3\vec{\sigma}] \propto (1-\delta_{az})\;,
\end{align}
which shows that the triplet interlayer exciton shields layer-odd antiferromagnetism that is collinear with it, and ferromagnetism that has spin perpendicular to it. 

In the presence of more than two layers, an analogous procedure can be used. The number of possible pairs of dominant orders and shielded orders increases dramatically in this case because of the increase of the size of the layer space.
\subsubsection{Concentric Fermi surfaces}
For concentric Fermi surfaces with opposite dispersions, time-reversal does not alter the band index. Therefore, the  bispinors and the shielded orders for such system take exactly the same form. They differ however in their interpretation. Indeed, the superconductivity in such system even or odd in bands instead of layer. Therefore, this corresponds to \textit{s}-wave and \textit{s}$_{\pm}$-wave superconductivity. The insulating state now corresponds to an integrate distribution of charge or spin. Simpler charge and spin-density waves arise if one starts from an interaction that depends on momentum direction, akin to the model introduced Sec. \ref{sec:NLSMwithangledependence}. We propose such model for the iron pnictides.
 
\subsection{Systems with a valley degree of freedom}\label{sec:IVC}

Next we consider systems with valley degree of freedom, and show that if they have dominant intervalley coherent (IVC) order, they can host both singlet and triplet superconductivity. 

Also, systems with valley degree of freedom, we have two connected components of the Fermi surface. There is however, a crucial difference with layered systems or systems with concentric Fermi surfaces. The pockets in materials with valley degree of freedom have opposite momentum, so that time-reversal interchanges the two valleys. This significantly alters the structure of the low-energy theory. Moreover, IVC states were introduced for systems that next to the valley degree of freedom have an additional band degree of freedom, the Chern bands, which should be taken into account.

To derive the NLSM for materials that host IVC states, we start from the class of normal state Hamiltonians defined in Ref.~\cite{bultinck2020ground,kolar2023anderson},
\begin{align}
    h_n(k) = \psi^{\dagger}(h_0(k) \kappa_3 + h_x (k) \eta_1 + h_y(k) \eta_2\kappa_3-\mu)\psi\;,
\end{align}
where $\kappa_i,\eta_i$ are Pauli matrices in valley and Chern band space respectively and we write the spinors explicitly incorporating the spin-degree of freedom, $\psi = (\psi_{K,+,\uparrow},\psi_{K,+,\downarrow},\psi_{K,-,\uparrow},\psi_{K,-},\psi_{-K,-,\uparrow},\psi_{-K,-,\uparrow},\psi_{-K,+,\uparrow},\psi_{-K,+,\uparrow})$. Time-reversal in this basis is $\mathcal{T} = \kappa_x i\sigma_y \mathcal{K}$. This basis however is not convenient to derive a low-energy model, because the structure of the Hamiltonian in valley-band state depends on both magnitude and direction of momentum.

We thus need to do basis transformations first. For the first step, we can use the unitary transformation used in Ref.~\cite{kolar2023anderson}, with $\mathcal{U} = \frac{1}{2}(1+\kappa_3) + \frac{1}{2}(1-\kappa_3)i \eta_2$, in which basis, we have $\psi = (\psi_{K,+,\uparrow},\psi_{K,+,\downarrow},\psi_{K,-,\uparrow},\psi_{K,-},\psi_{-K,+,\uparrow},\psi_{-K,+,\uparrow},-\psi_{-K,-,\uparrow},-\psi_{-K,-,\uparrow})$, $\mathcal{T} = i\sigma_y  \eta_2 \kappa_y$ and
\begin{align}
    h_n' = \psi^{\dagger}(h_0(k) + h_1(k) \eta_1 + h_2(k) \eta_2)\kappa_3\psi\;.
\end{align}
Next, we may diagonalize this Hamiltonian using a unitary transform in band ($\eta$) space, which does not affect time-reversal. In this basis we have $\psi = \frac{1}{\sqrt{2}}(\psi_{K,+,\uparrow},\psi_{K,+,\downarrow},\psi_{K,-,\uparrow},\psi_{K,-},\psi_{-K,+,\uparrow},\psi_{-K,+,\uparrow},-\psi_{-K,-,\uparrow},-\psi_{-K,-,\uparrow})^T + \frac{e^{i\varphi(k)}}{\sqrt{2}}((\psi_{K,-,\uparrow},\psi_{K,-,\downarrow},\psi_{K,+,\uparrow},\psi_{K,+},\psi_{-K,-,\uparrow},\psi_{-K,-,\uparrow},-\psi_{-K,+,\uparrow},-\psi_{-K,+,\uparrow}))^T$ where $e^{i\varphi(k)} = \frac{h_1(k) + i h_2(k)}{\sqrt{h_1(k)^2 + h_2(k)^2}}$. Using this basis, the diagonal Hamiltonian reads
\begin{align}
    h_N = \psi^\dagger(h_0(k)\kappa_3 + \sqrt{h_1(k)^2+h_2(k)^2}\kappa_3 \eta_3-\mu)\psi\;.
\end{align}
If $\sqrt{(h_1(k))^2 + (h_2(k))^2}$, dominates the dispersion, the correct soft-mode manifold is thus found by using $\psi$ and $\bar{\psi} = \psi^{\dagger}\kappa_3  \eta_3$.

The construction of Nambu space requires that standard \textit{s}-wave pairing is diagonal in everything except Nambu space, so that we construct the bispinors as $\Psi = (\psi^{T}, -i\bar{\psi}\sigma_2\kappa_2\eta_2)^{T}$ and $\bar{\Psi} = \Psi^{\dagger}\kappa_3\eta_3\tau_3 = (\bar{\psi}, i\psi^{T}\sigma_2\kappa_2\eta_2 )$, which gives the charge conjugation constraint
\begin{align}
    \Psi = i\tau_1\sigma_2\kappa_2\eta_2\bar{\Psi}^T\;.
\end{align}

Thus, in this basis both the valley and the band degree of freedom enter in a manner identical to spin. The large space allows for the construction of many order parameters.

There are now several intervalley coherent states that we may construct, such as the Kramers-degenerate and time-reversal symemtric ones (K-IVC,T-IVC),
\begin{align}
    \Delta_{\text{K-IVC}} &= \text{Tr}\langle \tau_3\frac{\kappa_2-i\kappa_1}{2} \Psi\bar{\Psi}\rangle\;,\\
    \Delta_{\text{T-IVC}} &= \text{Tr}\langle \eta_3\frac{\kappa_2-i\kappa_1}{2} \Psi\bar{\Psi}\rangle\;,
\end{align}
as well as their spin-full components
\begin{align}
    \vec{\Delta}_{\text{spin-IVC}} & = \text{Tr}\langle \tau_3\eta_3\vec{\sigma}\frac{\kappa_2-i\kappa_1}{2} \Psi\bar{\Psi}\rangle\;,\\
    \vec{\Delta}_{\text{T-spin-IVC}} & = \text{Tr}\langle \vec{\sigma}\frac{\kappa_2-i\kappa_1}{2} \Psi\bar{\Psi}\rangle\;.
\end{align}

We consider these intervalley coherent states to be the dominant order with lost spatial phase coherence, and determine the types of superconductivity they can shield. We focus on the intraband pairings, which gives us four different types of allowed order parameters:
\begin{align}
    \Delta_{\text{s}} &= \text{Tr}\langle\frac{\tau_{2}-i\tau_1}{2}\Psi\bar{\Psi}\rangle\;,\\
    \Delta_{\text{f}} \vec{d} &= \text{Tr}\langle\frac{\tau_{2}-i\tau_1}{2}\vec{\sigma}\kappa_3 \Psi\bar{\Psi}\rangle\;,\\
    \Delta_{\text{f}'}\vec{l} &=\text{Tr}\langle\frac{\tau_{2}-i\tau_1}{2}\vec{\eta}\kappa_3 \Psi\bar{\Psi}\rangle\;,\\
    \Delta_{\text{s}'}p_{ab} &=\text{Tr}\langle\frac{\tau_{2}-i\tau_1}{2}\eta_a\sigma_b \Psi\bar{\Psi}\rangle\;.
\end{align}
The first is conventional \textit{s}-wave superconductivity, the second spin-triplet valley odd, i.e. \textit{f}-wave superconductivity, the third orbital triplet \textit{f}-wave superconductivity, and the fourth spin-triplet orbital triplet \textit{s}-wave superconductivity.

We may now consider which superconductor orders can be shielded. To this end, we note that for the K-IVC state
\begin{align}
    [\tau_3\kappa_{1,2},\tau_{1,2}]\neq 0\;\quad [\tau_3\kappa_{1,2},\tau_{1,2}\kappa_3\vec{\sigma}] = 0\;,\quad [\tau_3\kappa_{1,2},\tau_{1,2}\kappa_3\vec{\eta}] = 0\;, [\tau_3\kappa_{1,2},\tau_{1,2}\eta_a\sigma_b]\neq 0\;.
\end{align}
This implies that the K-IVC state shields both \textit{f}-wave order parameters, but forbids \textit{s}-wave pairing to exist.

On the other hand, for the T-IVC state,
\begin{align}
    [\eta_3\kappa_{1,2},\tau_{1,2}]= 0\;\quad [\eta_3\kappa_{1,2},\tau_{1,2}\kappa_3\vec{\sigma}] \neq 0\;,\quad [\eta_3\kappa_{1,2},\tau_{1,2}\kappa_3\eta_a]  \sim\delta_{a3}\;, [\eta_3\kappa_{1,2},\tau_{1,2}\eta_a\sigma_b]\sim (1-\delta_{a3})\;.
\end{align}
Thus, the T-IVC state shields conventional \textit{s}-wave and inter-Chern band \textit{f}-wave superconductivity, as well as the intraorbital triplet \textit{s}-wave. 
For the spin-full states the spin-orientation of the triplets becomes important. We get, for the time-reversal broken variant
\begin{align}
    &[\tau_3\eta_3\sigma_c\kappa_{1,2},\tau_{1,2}]\neq 0,\;\quad 
    [\tau_3\eta_3\sigma_c\kappa_{1,2},\tau_{1,2}\kappa_3\sigma_b] = (1-\delta_{bc})\;,\quad [\tau_3\eta_3\sigma_c\kappa_{1,2},\tau_{1,2}\kappa_3\eta_a] =(1-\delta_{a3})\;,\nonumber\\ &[\tau_3\eta_3\sigma_c\kappa_{1,2},\tau_{1,2}\eta_a\sigma_b] = \delta_{bc}\delta_{a3} + (1-\delta_{bc})(1-\delta_{a3})\;.
\end{align}
and for the time-reversal symmetric variant
\begin{align}
    [\sigma_c\kappa_{1,2},\tau_{1,2}]=0,\;\quad 
    [\sigma_c\kappa_{1,2},\tau_{1,2}\kappa_3\sigma_b] = \delta_{bc}\;,\quad [\sigma_c\kappa_{1,2},\tau_{1,2}\kappa_3\eta_a] \neq 0\;, [\sigma_c\kappa_{1,2},\tau_{1,2}\eta_a\sigma_b] = (1-\delta_{bc})\;.
\end{align}
In the main text we only summarize the spinless K-IVC and T-IVC states.

Besides superconducting phases, in the previous sections we have identified magnetic orders. For ferromagnetism, $\tau_3\kappa_3\eta_3\vec{\sigma}$, we have
\begin{align}
    [\tau_{3}\kappa_{1,2},\tau_3\kappa_3\eta_3\vec{\sigma}]\neq 0\;, \quad [\eta_3\kappa_{1,2},\tau_3\kappa_3\eta_3\vec{\sigma}]\neq 0\;,
\end{align}
 so that neither K-IVC nor T-IVC shields ferromagnetism.

For $f$-wave magnetism ($\tau_3\eta_3\vec{\sigma}$) however, we have
\begin{align}
    [\tau_3\kappa_{1,2},\tau_3\kappa_3\vec{\sigma}] \neq  0\;, \quad [\eta_3\kappa_{1,2},\tau_3\kappa_3\vec{\sigma}] \neq 0\;,
\end{align}
 that is, $\textit{f}$-wave magnetism is not shielded either.

\section{Only phase inhomogeneity}\label{sec:Phasevsamplitudeinhomogeneities}
In the derivation of the NLSM in Sec.~\ref{sec:Actiondetails}, for illustrative purposes we considered a model in which both the real and complex part of $\Delta_{\text{SDW}}$ are Gaussian distributed. Here we take into account that only the phase coherence is lost, while the amplitude of the spin-density wave remains coherent throughout. In this section we consider models in which the difference between phase and amplitude disorder is explicitly taken into account, and show that the resulting low-energy theory is the same as discussed in Sec.~\ref{sec:Actiondetails}.
\subsection{Quartic term}\label{sec:Quartic}
The first manner in which the difference in the amplitude and phase fluctuations can be incorporated, is by making the term quartic, that is, by replacing

\begin{align}
    e^{-\pi\nu\tau_{\Delta}|\Delta_{\text{SDW}}|^2}\rightarrow e^{-\pi\nu\tau_{\Delta}\Big(-\frac{1}{8 |\delta_{\text{SDW}}|^2}|\Delta_{\text{SDW}}|^4 + \frac{|\Delta_{\text{SDW}}^{0}|^2}{4 |\delta_{\text{SDW}}|^2}|\Delta_{\text{SDW}}|^2\Big)}\;.
\end{align}
Here $\Delta_{\text{SDW}}^{0}$ is the strength of the short-range order parameters, while $\delta^{0}_{\text{SDW}}$ characterizes how much it can deviate.
Integration over $\Delta_{\text{SDW}}$ now does not give a simple four-fermion term anymore. In the limit of weak amplitude deviations, $\delta^{0}_{\text{SDW}} \ll \Delta^{0}_{\text{SDW}}$, the quartic integral gives the zeroth order Bessel function, which can be expanded in number of Fermion terms as
\begin{align}
    &\text{Exp}(\Big(-\sum_m \int dt \int dt'\frac{1}{4\pi\nu\tau_{\Delta}}\bar{\Psi}(t)P_m  \Psi(t)\bar{\Psi(t') P_m \Psi(t')}\Big)\nonumber\\& + \frac{1}{128\pi^2\nu^2\tau_{\Delta}^2}\sum_{m,m'}\int dt \int dt'\int dt''\int dt'''\bar{\Psi}(t)P_m  \Psi(t)\bar{\Psi}(t') P_m \Psi(t') \bar{\Psi}(t'')P_m'  \Psi(t'')\bar{\Psi}(t''') P_m' \Psi(t''') + ....)\;.
\end{align}
This reduces to the treatment discussed in Sec~\ref{sec:Actiondetails} in the limit in which we can ignore the higher order in $\Psi$ terms. Because this is a Born-like approximation, we conclude that it is possible in the limit $\Delta_{\text{SDW}}/E_F\ll 1$. If $\Delta_{\text{SDW}}$ becomes larger, and effects akin to Anderson localization may appear. This supports the use of the distribution in Sec.~\ref{sec:Actiondetails}.
\subsection{Gauge transformation}
Another way to achieve a theory with only phase coherence is via a U(1) gauge transformation that completely absorbs its phase. If we now assume that the SDW-phase gradient is Gaussianly distributed, we obtain the following theory:
\begin{align}
    iS[Q] = \bar{\Psi}(\tau_3 \mu_3\nu_3 + \xi(k-A_{\text{SDW}}\mu_3) + |\Delta_{\text{SDW}}|\mu_2\tau_3\sigma_z)\Psi + \pi\nu\tau_{\Delta}v_{\text{F}}^2|A_{\text{SDW}}|^2\;.
\end{align}
An important difference between this theory and the previous is that the disorder now enters in a different way in the action. We show that, none-theless, we obtain an equivalent theory. Indeed, within the quasiclassical approximation, we may expand $\xi(k-A_{\text{SDW}}\mu_3)\approx \xi(k)-\vec{v}_F \cdot\vec{A}_{\text{SDW}}\mu_3$, after which $A_{\text{SDW}}$ can be integrated out to give a four-fermion term. The resulting action reads
\begin{align}
    \text{Tr}\bar{\Psi}(\tau_3 \mu_3\nu_3 + \xi(k) + |\Delta_{\text{SDW}}|\mu_2)\Psi + \text{Tr}\frac{1}{4\pi\nu\tau_{\Delta}} \sum_ m\Bar\Psi \tilde{P}_m \Psi\bar{\Psi}\tilde{P}_m\Psi\;.
\end{align}
where the projectors $\tilde{P}_m$ project on the eigenspaces of $\mu_3$.
We can now decouple this using a $Q$-matrix in the usual way, where because of the projections, $Q$ has to commute with $\rho_3$. This yields the generating functional
\begin{align}
    Z = \int \mathcal{D}Q \int \mathcal{D}\Psi \int\mathcal{D}\bar{\Psi}\; \text{Exp}\Bigg\{\text{Tr}\Big(i\bar{\Psi}(\tau_3\mu_3\nu_3\hat{\omega_{t,t'}} +\xi + |\Delta_{\text{SDW}}|\mu_2 \tau_3\sigma_z+ i\frac{i}{\tau_{\Delta}} Q)\Psi\Big)\Bigg\}\text{Exp}\{-\frac{\pi\nu}{4\tau_{\Delta}}\text{Tr}Q^2\}
\end{align}
As usual, the fermions can be integrated out and give $Z = \int \mathcal{D}Q e^{iS_{A}[Q]}$, where
\begin{align}
    iS_A[Q] = -\frac{\pi\nu}{4\tau_{\Delta}}\text{Tr}Q^2 + \text{Tr ln}(\tau_3\mu_3\nu_3\hat{\omega_{t,t'}} +\xi + |\Delta_{\text{SDW}}|\mu_2\tau_3\sigma_z + \frac{i}{\tau_{\Delta}} Q)\;.
\end{align}
Going to the soft-mode manifold, we notice that there is a large-energy term with non-trivial structure in band-Nambu-spin space in the action, $|\Delta_{\text{SDW}}|$. In such case, the low-energy theory has to be projected to its eigenspaces, as has been done for example in the case of strong Rashba spin-orbit coupling.

Variation with respect to $Q$ now gives the soft-mode constraint
\begin{align}
    0 = \frac{\pi\nu}{2\tau_{\Delta}}\Big(-Q + \frac{i}{\pi\nu}(\tau_3\mu_3\nu_3 \hat{\omega_{t,t'}}+\xi + \Delta_{\text{SDW}}\mu_2\tau_3\sigma_z + i\pi \nu\xi_A^2Q)^{-1}\Big)\;,
\end{align}
and therefore the soft-mode manifold satisfies
\begin{align}
    Q(r) = \frac{i}{\pi\nu}\Big(\tau_3 \mu_3\nu_3\tau_3\sigma_z + \xi + \Delta_{\text{SDW}}\mu_2 + \frac{i}{\tau_{\Delta}}Q\Big)_{r' = r}\;.
\end{align}
Evaluation at $r = r'$ corresponds to an integral over all momenta in momentum space. The integrals are easily evaluated in the eigenbasis of the argument. In the quasiclassical limit, $\xi$ dominates, and therefore the eigenvalues $\lambda$ satisfy $\lambda \pm \lambda$, that is, $\lambda = \pm 1$. Thus,
\begin{align}
    Q^2 = \mathbf{1}\;.
\end{align}
Thus, the soft-mode manifold satisfies, next to the standard charge conjugation constraint, 3 additional constrains, 
\begin{subequations}
\begin{align}
    Q^2 &= \mathbf{1}\;,\label{eq:Constraint1}\\
    Q &= \mu_2 \tau_3\sigma_z Q \mu_2\tau_3\sigma_z \;.\label{eq:Constraint2}\\
    Q &= \mu_3 Q \mu_3\;,\label{eq:Constraint3}
\end{align}
\label{eq:3constraints}
\end{subequations}

Comparing this to the previous treatments in Secs.~\ref{sec:Actiondetails} and \ref{sec:Quartic}, we note that the first two constraints are identical. There, the third constraint was $Q = \mu_1\tau_3\sigma_z Q\mu_2\tau_2\sigma_z$, which on first sight appears to be different. However, since $\mu_1\tau_3\sigma_z = i \mu_2\tau_3\sigma_z \mu_3$, under the provision that constraint Eq.~(\ref{eq:Constraint2}) hold this constraint is equivalent to Eq.~(\ref{eq:Constraint3}). We conclude that the model derived in this section is equivalent to the one derived in Sec.~\ref{sec:Actiondetails}. The fact that equivalent models can be obtained using distinct approximations shows the robustness of the low energy structure of the theory and corroborates the approach used in Sec. \ref{sec:Actiondetails}.
\end{document}